\documentclass[12pt,a4paper]{article}
\usepackage[top=1.2in, left=0.8in, right=0.8in, bottom=1.2in]{geometry}
\usepackage[colorlinks=true,linkcolor=blue,citecolor=magenta,urlcolor=red]{hyperref}
\usepackage{booktabs} 
\usepackage{textgreek} 
\usepackage[ruled]{algorithm2e} 
\usepackage{algorithmic}		
\usepackage{amsmath} 
\usepackage{multirow}
\usepackage[numbers]{natbib}
\usepackage{graphicx}
\usepackage{float}
\usepackage{subfloat}
\usepackage{subfig}	
\usepackage{txfonts}	
\usepackage[T1]{fontenc}

\newcommand{\footremember}[2]{%
    \footnote{#2}
    \newcounter{#1}
    \setcounter{#1}{\value{footnote}}%
}
\newcommand{\footrecall}[1]{%
    \footnotemark[\value{#1}]%
} 

\author{%
   Maria Plakia \footremember{alley}{Foundation for Research and Technology-Hellas, Greece.}%
   \and Evripides Tzamousis  \footremember{trick}{University of Crete and Foundation for Research and Technology-Hellas, Greece.}
   \and Thomais Asvestopoulou  \footrecall{alley}
   \and Giorgos Pantermakis  \footrecall{alley}
   \and Nick Filippakis \footrecall{alley}
   \and Henning Schulzrinne  \footremember{trailer}{Columbia University, United States of America.}%
   \and Yana Kane-Esrig
   \and Maria Papadopouli \footrecall{trick} \footremember{second trick}{Contact author: Maria Papadopouli (\texttt{mgp@ics.forth.gr)}.}
  }
\date{}
\begin{document}
\title{Should I stay or should I go: Analysis of the impact of application QoS on user engagement in YouTube}
\maketitle
\begin{footnotesize}
\noindent
{\bf Abstract}
To improve the quality of experience (QoE), especially under moderate to high traffic demand, it is important to understand the impact of the network and application QoS on user experience. This paper comparatively evaluates the impact of impairments, their intensity and temporal dynamics, on user-engagement in the context of video streaming. The analysis employed two large YouTube datasets. To characterize the user engagement and the impact of impairments, several new metrics were defined. We assessed whether or not there is a {\em statistically significant relationship} between different types of impairments and QoE and user engagement metrics, taking into account not only the characteristics of the impairments but also the covariates of the session (e.g., video duration, mean datarate). After observing the relationships across the entire dataset, we tested whether these relationships also persist under specific conditions with respect to the covariates. The introduction of several new metrics and of various covariates in the analysis are two innovative aspects of this work. We found that the number of negative bitrate changes (BR-) is a stronger predictor of abandonment than rebufferrings (RB). Even positive bitrate changes (BR+) are associated with increases in abandonment.
Specifically, BR+ in low resolution sessions is not well-received. Temporal dynamics of the impairments have also an impact: a BR- that follows much later a RB appears to be perceived as a worse impairment than a BR- that occurs immediately after a RB. These results can be used to guide the design of the video streaming adaptation as well as suggest which parameters should be varied in controlled field studies.
\end{footnotesize}

\section{Introduction}\label{sec:intro}
To provide users with high quality of experience (QoE), especially under moderate to high utilization, it is important to understand the impact of the network and application quality of service (QoS) on user experience. 
%
To effectively adapt and improve a service, it is useful to distinguish those QoS degradations that are the strongest predictors of user behaviors that signal poor QoE (for example, propensity to abandon the session) from those that are weaker predictors. 
The ability to make this distinction makes it possible to address the QoS degradations that truly impact QoE, while avoiding attempts to address less impactful QoS degradation. This distinction can also help us avoid making adaptations that may be more annoying to the users than the problem that they are designed to fix.
%
%
%

In studying the relationship between QoE and QoS, the community has followed two main approaches: (1) out-in-the-wild participatory studies, often through crowd-sourcing systems, that cover diverse conditions in realism, have relatively small costs, are of large scale but unfortunately maintain limited contextual knowledge, and (2) controlled field studies that focus on homogeneous fixed controlled conditions, have larger cost/overheard but are often of limited scale and not under representative/typical contextual conditions. The assessment of QoE with explicit feedback from users
can be intrusive, time-consuming, expensive, and may also introduce bias. Most frequently in empirical studies, there is no explicit quantification of the QoE, but the improvement or deterioration of the QoE needs to be inferred from the users' behavior. 
In general, the design of such QoE-driven empirical studies faces tradeoffs that involve intrusiveness, privacy, reliability, realism, and cost.  


Our work employs two large YouTube datasets collected in the context of an out-in-the-wild participatory study using the YouSlow plugin developed by Nam and Schulzrinne \cite{nam16,nam-rep16}.  The YouSlow has collected over 1,400,000 YouTube views from more than 1,000 PC-based (not mobile) viewers located in more than 110 countries. To assess the QoE, Nam and Schulzrinne focused on the video abandonment rates of these YouTube sessions \cite{nam16,nam-rep16}. 
They analyzed the role of RBs, BR changes and startup delay on the abandonment. The negative impact of RBs and startup delays has been also reported in other studies (e.g. \cite{hoss13,krishnan2013video,plakia2016videoqoe}). These papers agree that the large duration and number of RBs are critical parameters on QoE. 

Our work was motivated by the following questions: 
What are the differences among various types of impairments in their impact on user engagement? How do impairments of different intensity impact the user engagement? Does the time elapsed between impairments impact the user engagement? 
Does the relationship between the impairments and the user engagement vary based on the duration or resolution of the video? 
How do the accumulated impairments affect a user?
%
We also considered an important methodological issue: what methodology should be used to assess the impact of an impairment on viewer engagement, taking into consideration that sessions can be abandoned due to lack of sufficient interest in the content?
%
This paper extends the state-of-the-art by taking advantage of these large YouTube datasets to answer the aforementioned questions and {\em comparatively} evaluate the impact of impairments, their intensity and temporal dynamics on user-engagement, under several metrics and conditions.

Our key findings are included in Table \ref{tbl:key_findings}. Some of our findings are consistent with the results of previous research studies.  The negative role of the number of RB and large RB duration has been
also reported in several other studies \cite{plakia2016videoqoe,hoss13,nam16,krishnan2013video,Casa17,wams16}. Moreover, the negative impact of BR changes has been demonstrated in previous studies (for YouTube sessions in \cite{nam16}, in controlled field studies
with distorted video \cite{seshadrinathan11}).  Other findings in our research were surprising and went counter to our expectations. For example, the number of negative bit rate changes (BR-) is a stronger predictor of abandonment than the number of rebufferings (RB). The stronger impact of BR- compared to RB emerged based on several QoE/user-engagement metrics. We also observed that a BR- that occurs much later after a RB appears to be perceived as a worse impairment than a BR- that occurs immediately after a RB. Another surprising finding was that even positive bit rate changes (BR+) are associated with increases in abandonment under some conditions. Specifically, BR+ in low resolution sessions is not well-received.  

\begin{table}[t!]
\centering 
\begin{tabular}{| p{14.8cm} | c |}
\hline
\hspace{5.5cm} Key Findings &  Section \\
\hline
BR- exhibits  the largest impact on video watching percentage \& abandonment ratio & 3.1\\ 
\hline
BR+ in sessions with low initial resolution is not well-received; Users tolerate the BR+ when the weighted mean data rate is relatively high & 3.1, 3.9\\ 
\hline
RBs of large duration have more prominent impact than BR- & 3.4\\ 
\hline
Compared to startup delay, RBs have a larger impact on the video watching percentage & 3.5\\  
\hline
Features with predictive power for the video watching percentage include the number of RBs, number of BR changes, number of negative BR changes, mean weighted bit rate & 3.6\\  
\hline
An impairment prior to a BR- increases the likelihood of abandonment & 3.8\\ 
\hline
\end{tabular}
\caption{A brief summary of our key findings. The negative role of the number of RB and of the large RB duration has been also reported in other studies \cite{plakia2016videoqoe,hoss13,nam16,krishnan2013video,Casa17,wams16}. The negative impact of the BR changes has been also shown \cite{Moorthy-2012,nam16,seshadrinathan11}. Startup delay has smaller impact on video watching percentage \cite{hossfeld2012initial}. The users' tolerance to BR+ when the resolution before the BR increase has been already relatively high was also observed in \cite{Moorthy-2012}. The findings 1, 5, and 6 are new.
}
\label{tbl:key_findings} 
\end{table}
  
Our contributions are twofold: methodological innovation and observations regarding the relationship of service QoS impairments and user engagement.
(1) Methodological innovation. In addition to the abandonment ratio, we defined several new metrics to characterize the user engagement and the impact of impairments, such as the {\em video watching duration percentage}, the {\em time elapsed from the occurrence of an impairment to the end of the session} and the percentage of sessions that get abandoned within a certain time  (e.g., 60 sec) after the occurrence of an impairment. We also specified several scenarios (i.e., categories of sessions) to focus on particular types of impairments. We applied various statistical analysis methods to assess whether or not there is a {\em statistically significant relationship} between different types of impairments and QoE and user engagement metrics, such as Kolmogorov Smirnov test and LASSO regression, taking into account the covariates of the session (e.g. video duration, mean datarate). We further examined the observed relation to verify that it is not simply the result of confounding with our covariates. 
To consider the lack of sufficient interest in the content as a primary reason for abandonment of the video, we defined thresholds on video watching percentage and performed a sensitivity analysis. (2) Observations regarding the relationship of service QoS impairments and user engagement. We first highlighted the relationships that emerged in the analysis of all sessions of the first dataset. 
We then
examined whether or not they persist under specific conditions with respect to the covariates.
The examination of the aforementioned user engagement metrics and the consideration of several
covariates are some of the innovative aspects of this analysis. 
These results can guide the design of the video streaming adaptation as well as the performance of controlled field studies. The results presented in this paper have been produced using the first dataset. The analysis has been repeated using the second dataset and the key findings have been validated.
%

The paper has the following structure: Section \ref{sec:back} presents the datasets and some main trends. Section \ref{sec:perf} highlights the main findings of the analysis of the YouTube data and control study. Section \ref{sec:rel} overviews the main related research and Section \ref{sec:concl} discusses the main results and future work.

\section{Background} \label{sec:back}
This section provides a brief introduction of the datasets, the preprocessing/sanitization and treatment phase, metrics and definitions to characterize the user engagement, QoE, and impact of the application QoS on QoE, and categorization of sessions with respect to their types of impairment. It also presents an overview of the preliminary analysis of sessions. 
 
YouTube, as other increasingly popular video streaming services, including Netflix, delivers all or most of their content using HTTPS. To provide smooth streaming, they use adaptive bitrate (ABR) streaming, where a video player dynamically adjusts the video bit rate based on estimated network conditions, buffer occupancy and hardware specifications of viewers' devices.

The participants of the YouSlow study had to download the Javascript wrapper. The YouTube views were PC-based, not mobile. The two datasets collected via the YouSlow in two different time periods, namely between February 2015 and July 2016 (i.e., dataset 1, D1) and January 2017 and July 2017 (i.e., dataset 2, D2), were first pre-processed, as follows: First, sessions with missing or NaN values, or with startup delay but no RB event at time 0 of non-zero duration, or with skips (i.e. ads that allow viewers to skip the ad after five seconds) were ignored. We then removed sessions with inconsistencies that could not be resolved. Specifically, sessions with playback duration (defined as the sum of the video playing duration and rebuffering duration) equal to 0 seconds or with unknown bitrate during streaming or with a start-up delay shorter than its advertisement time, were removed. 
Finally, sessions with RB ratio (i.e., the ratio of the total RB duration over the session duration) equal to 1 were ignored. Sessions with various inconsistencies in their RB events and bitrate changes were corrected according to the rules described in our technical report. 

\subsection{Definitions and Metrics}
The QoE can be defined as the degree of delight or annoyance of a person whose experiencing involves an application, service, or system. It results from the evaluation of the user fulfillment of the expectations and needs with respect to the utility and/or enjoyment in the light of the user context, personality and current state \cite{Zhu18}. 
To characterize the user engagement and the impact of impairments, in addition to the percentage of abandoned sessions, we defined the {\em video watching duration ratio (or percentage)} and the {\em time elapsed from the occurrence of an impairment to the end of the session}. 
Specifically, the video watching duration ratio is defined as the time of pure playback duration, excluding the total time spent on rebuffering events over the total duration of the video.
The lack of sufficient interest in the content, a degraded network and/or application performance, or other contextual aspects could cause an abandonment.
To eliminate the lack of sufficient interest in the content as a primary reason for abandoning the video, we first focused on videos for which {\em more than half} of the content has been watched. This relatively strict threshold on the {\em video watching percentage (vwp)} is then relaxed to assess whether or not the main trends persist and examine the potential bias that a strong interest in the content may introduce. For that, a sensitivity analysis with video watching percentage in the interval of [20\%, 100\%] or video watching duration greater than or equal to 120 seconds was also performed. The default video watching percentage threshold is set to 50\%. 
\subsection{Categorization of Sessions}
The sessions were categorized according to the number, type, and intensity of their impairments, namely presence and type of bitrate changes (BR-, BR+), RB duration and mean data rate. The bitrate change could be of step size 1 (e.g., from medium to small) or larger (e.g., from small to large).\footnote{There are eight levels of played bitrates in the YouTube datasets, namely, tiny, small, medium, large, hd720, hd1080, hd1440, and highres, varying from 80 Kbps with resolution of 256x144 to 35-45 Mbps with resolution 3840x2160. The majority of sessions are of large or medium resolution, 36\% and 26.8\%, respectively \cite{nam16}.}
%
First, we focused on six basic categories of sessions (i.e., scenarios), namely sessions {\em without any BR change or RB} (baseline scenario), sessions with {\em only BR-}, sessions with {\em only BR+}, sessions with {\em RBs and BR+}, sessions with {\em RBs and BR-}. We then distinguished specific subgroups with respect to the RB duration or RB ratio (i.e., RB duration over the entire session duration ratio), the mean weighted data rate and the number of BR-/BR+ to highlight the impact of the intensity of the impairment.
The analysis was repeated for sessions that satisfy specific thresholds on video watching percentage (e.g., 20\%, 120 sec), mean data rate (Mbps), e.g., in the intervals [1,2], [6,8], [2.5, 8], and [1,8] to highlight the common trends, and video duration (e.g., short video clips, videos of representative duration, long videos). 
\subsection{Summary Statistics}
\begin{figure*}[t!] 
\centering
\subfloat[]{\includegraphics[width=0.35\textwidth, height=35mm]{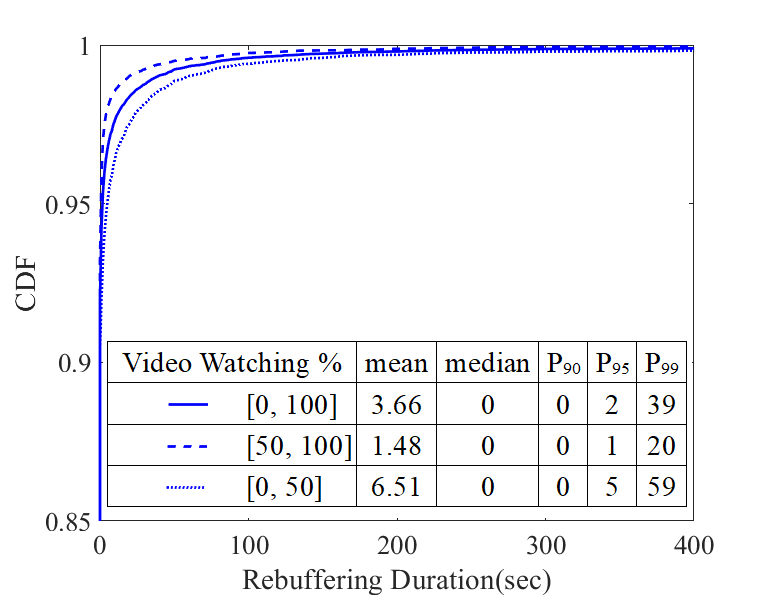}}
\subfloat[]{\includegraphics[width=0.35\textwidth, height=35mm]{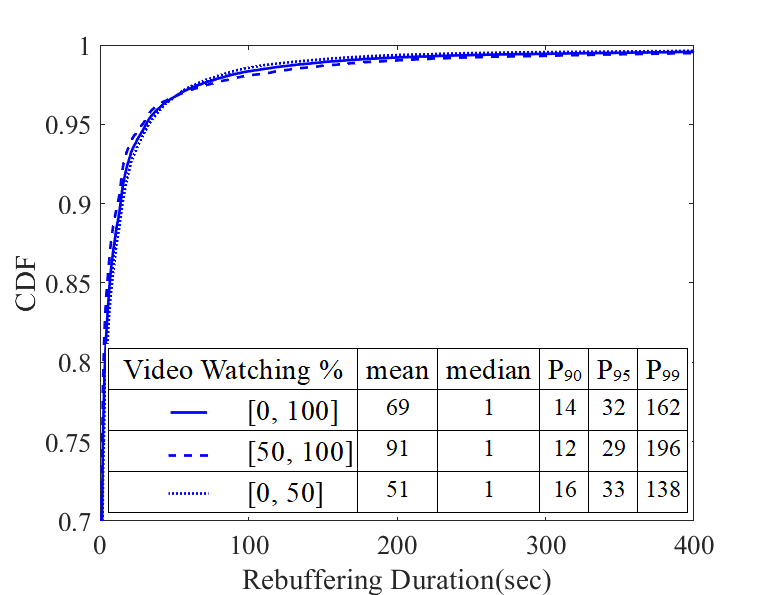}}
\subfloat[]{\includegraphics[width=0.35\textwidth, height=35mm]{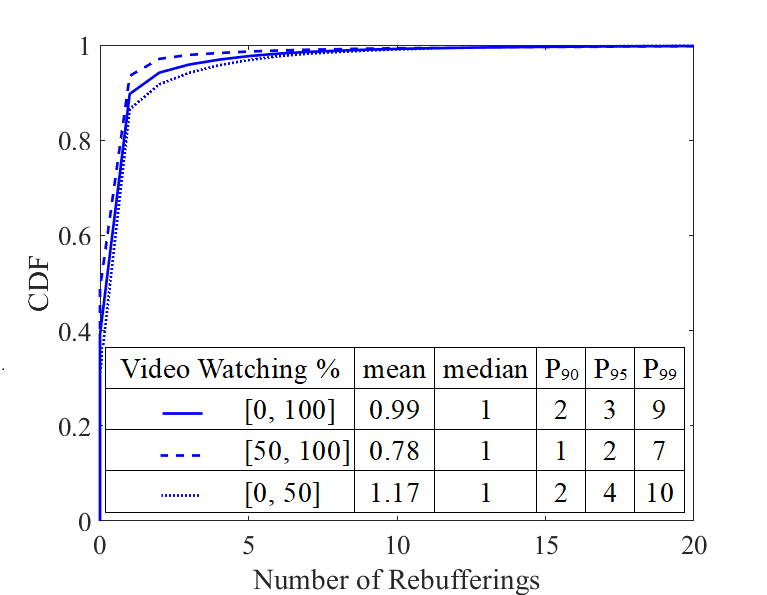}}\\
\caption{Total rebuffering duration and number of RBs. Sessions with startup delay have been excluded in Fig. (a), while they are included in Figures (b),(c).}
  \label{fig:RB_background}
\end{figure*}

To better understand the intensity and types of impairments in the collected YouTube sessions, we performed a preliminary analysis.
In general, most of sessions have no impairments and the majority of sessions with impairments have only one impairment (Table \ref{tbl:num_imp_stats}). The mean total rebufferring duration is 79 sec and the median 1 sec, while the 90\% of sessions have total rebufferring duration less than or equal to 11 sec. 
The 49\% of sessions have startup delay and only a 3\% of sessions have advertisements. In the case of sessions with video watching percentage in [50,100], the 38 \% of sessions have startup delay and a 3\% of them have advertisements.
When we focus on sessions with mean data rate (Mbps) in [2.5, 8], the 38\% of sessions have a startup delay and 3\% sessions have advertisements. Without a threshold on video watching, the 48\% of sessions have a startup delay and 3\% have advertisements. These statistics refer to the first dataset.
For sessions in which more than half of the video has been watched, the percentage of sessions with a RB duration of 10 sec or more is very low, as it is also the percentage of sessions with multiple RB and negative BR changes (0.1\%), as shown in Fig. \ref{fig:RB_background}. 

\begin{table}[h]
\centering 
\begin{tabular}{c | c | c | c | c  }
\hline
Impairments   &Min & Max& Mean & Median\\\hline
Number of RB  & 0  &153   &0.67  & 0\\  \hline
Number of BR- & 0  &8     &0.12  &0 \\\hline
Number of BR+ & 0  &8 &0.17&0\\ \hline 
\end{tabular}
\caption{Main statistics on the number of impairments for sessions with video watching percentage in [50,100]. The corresponding statistics for sessions with weighted mean datarate (Mbps) in the interval of [2.5,8] are lower, while the corresponding values increase when the video watching percentage is relaxed.}
\label{tbl:num_imp_stats} 
\end{table}
80\% of sessions have no BR changes, while 14\% of sessions include exactly one BR change and 5\% of sessions exactly two BR changes. 
For sessions with video watching percentage above 50\%, the median mean weighted data rate of the main scenarios with no BR is 2.5Mbps. For sessions with No RB/BR+ and No RB/BR-, the median mean weighted data rate is 4,9Mbps and 5Mbps, respectively, while for sessions with RB/BR+, RB/Early BR-, and RB/Other BR-, the median mean weighted data rate is 4.9Mbps, 1.6, and 1.1Mbps, respectively. The RB/Early BR- corresponds to the scenario in which the BR- occurs within 30 sec after the end of a RB, while the RB/Other BR- indicates the scenario with RB, in which the BR- occurs much later after the end of a RB. 

\section{Performance Analysis}\label{sec:perf}
This section aims to infer the impact of the various application QoS metrics on user engagement, considering different impairments with respect to their type and intensity. In the next subsections, we try to answer the following questions:
Does a user tolerate more a certain type of impairment than another? Do we observe different trends for different video duration? How do multiple impairments of different intensities impact the user engagement? Does the time elapsed between impairments impact the user engagement? 
Does a user become even less tolerant when facing a second impairment?
To assess whether or not a difference is significant, we employed the Kolmogorov-Smirnov test.

\subsection{The prominent negative impact of BR changes on video watching percentage and abandonment ratio}\label{sec:perf:neg}

We first focused on the six main scenarios: (1) no RB events and no BR changes ({\em baseline}), (2) only RBs but no BR changes (3) only BR+ changes but no other RB or BR- impairments, (4) only BR- changes but no other RB or BR+ impairment, (5)  BR+ and RB events, and (6) BR- and RB events. Sessions with startup delay are part of the sessions with RB. Sessions with BR- exhibit the lowest video watching percentage and larger abandonment ratio. This persists for different weighted mean data rate and video watching percentage thresholds. The baseline scenario (without any BR changes or RBs) exhibits prominent differences from the others. 
This also persists for all video watching percentage and mean weighted bitrate thresholds we considered. Compared to RB, BR- causes a larger number of abandonments. Actually, the differences in the number of abandonments and video watching percentage in the scenarios BR-, with vs. without RB are not significant. 
%
%
In general, for different weighted mean data rate and video watching percentage thresholds, the BR- has the smallest video watching percentage and largest abandonment ratio of all scenarios. 

%
An interesting trend regarding BR+ occurs with respect to the initial video resolution: Users with relatively low initial resolution (i.e 1 Mbps) who then experience a positive BR change, appear not as tolerant as the ones that have been watching the video at a constant data rate of 1 Mbps, without experiencing any other impairment. As the initial data rate increases (e.g., to 2.5 Mbps or more), the user becomes more "open" to BR+ changes, as reflected by the smaller abandonment ratio and larger mean video watching percentage compared to the corresponding number of the baseline scenario. 

Although there are no significant statistical differences between the RB/BR+ vs.  RB/BR- scenarios, in terms of total RB duration as well as number of RB events, these two scenarios are statistically significant different in terms of video watching percentage (for different video watching percentage and mean data rate interval thresholds).

For large mean data rates (Mbps), e.g., in [6,8] Mbps, sessions with only BR+ have the smallest abandonment ratio compared to all other scenarios, even lower than the baseline. Moorthy \textit{{\em {\em et al.}}} \cite{Moorthy-2012} had also observed that users tolerate the BR+, especially if the resolution before the bit rate increase has been already relatively high.  

\newgeometry{top=1.2in, left=0.8in, right=0.8in, bottom=1.2in}
\begin{figure*}[t!] 
\centering
\subfloat{\includegraphics[width=0.33\textwidth, height=30mm]{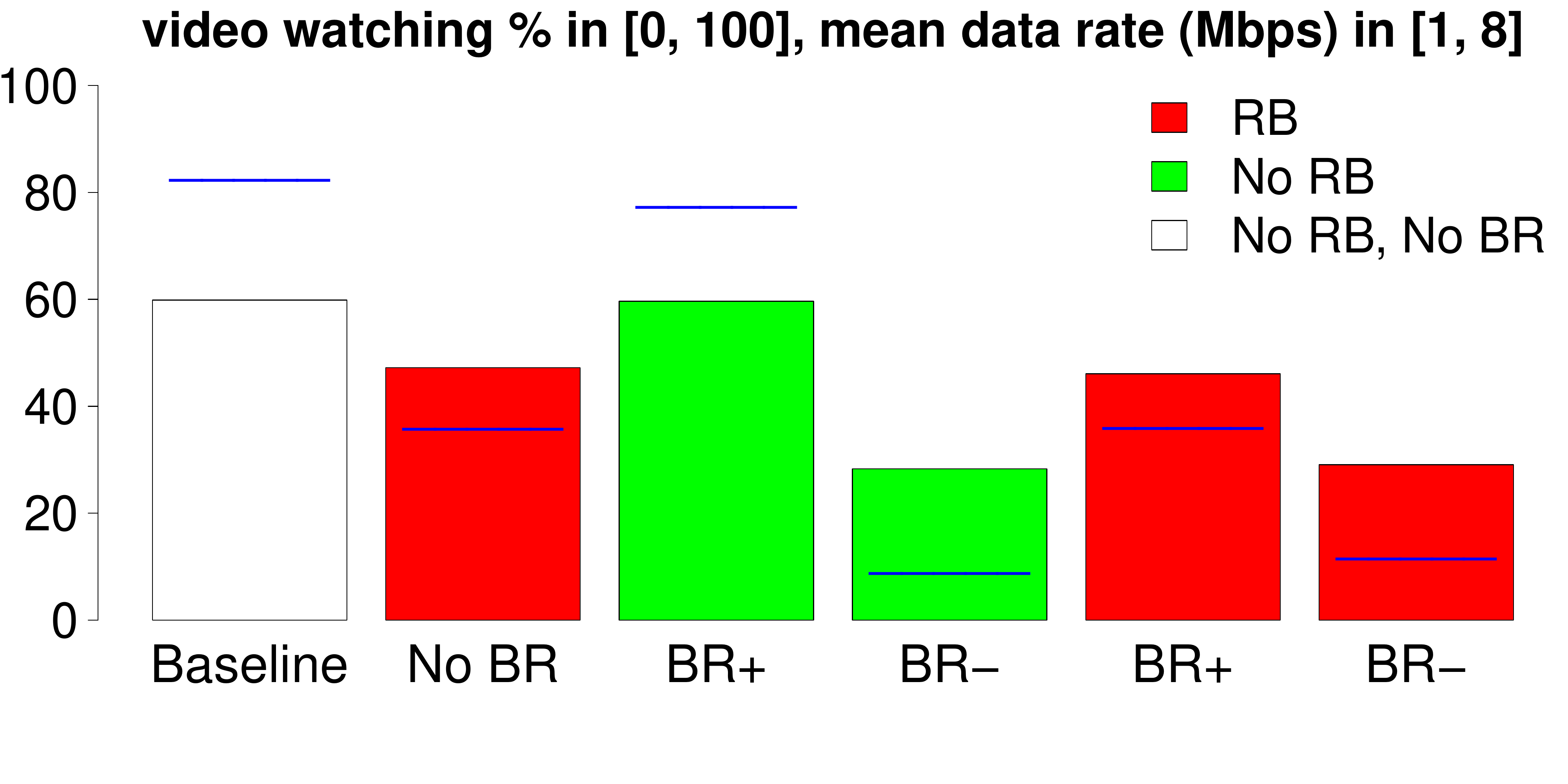}}
\subfloat{\includegraphics[width=0.34\textwidth, height=30mm]{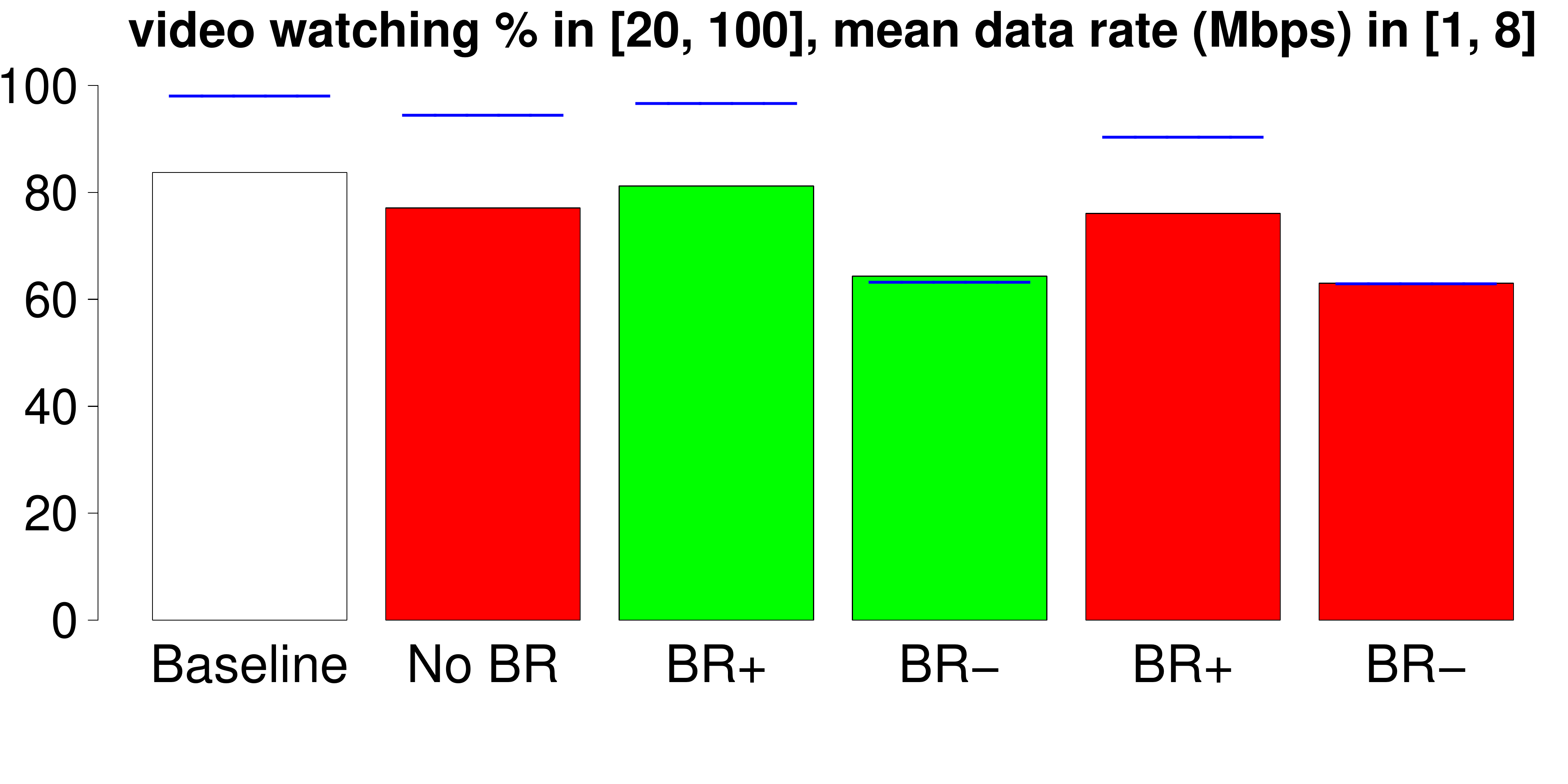}}
\subfloat{\includegraphics[width=0.33\textwidth, height=30mm]{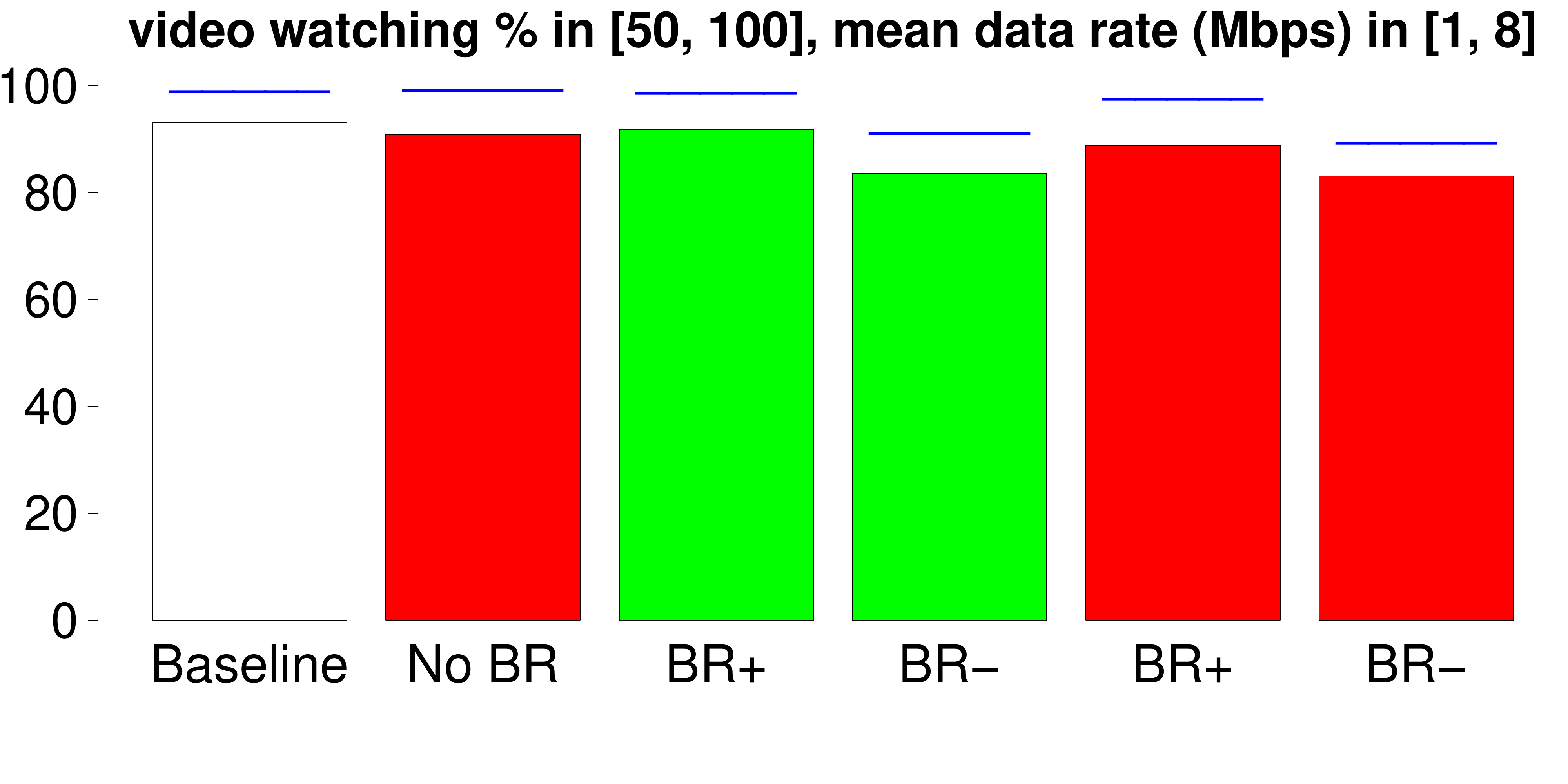}}

\subfloat{\includegraphics[width=0.33\textwidth, height=30mm]{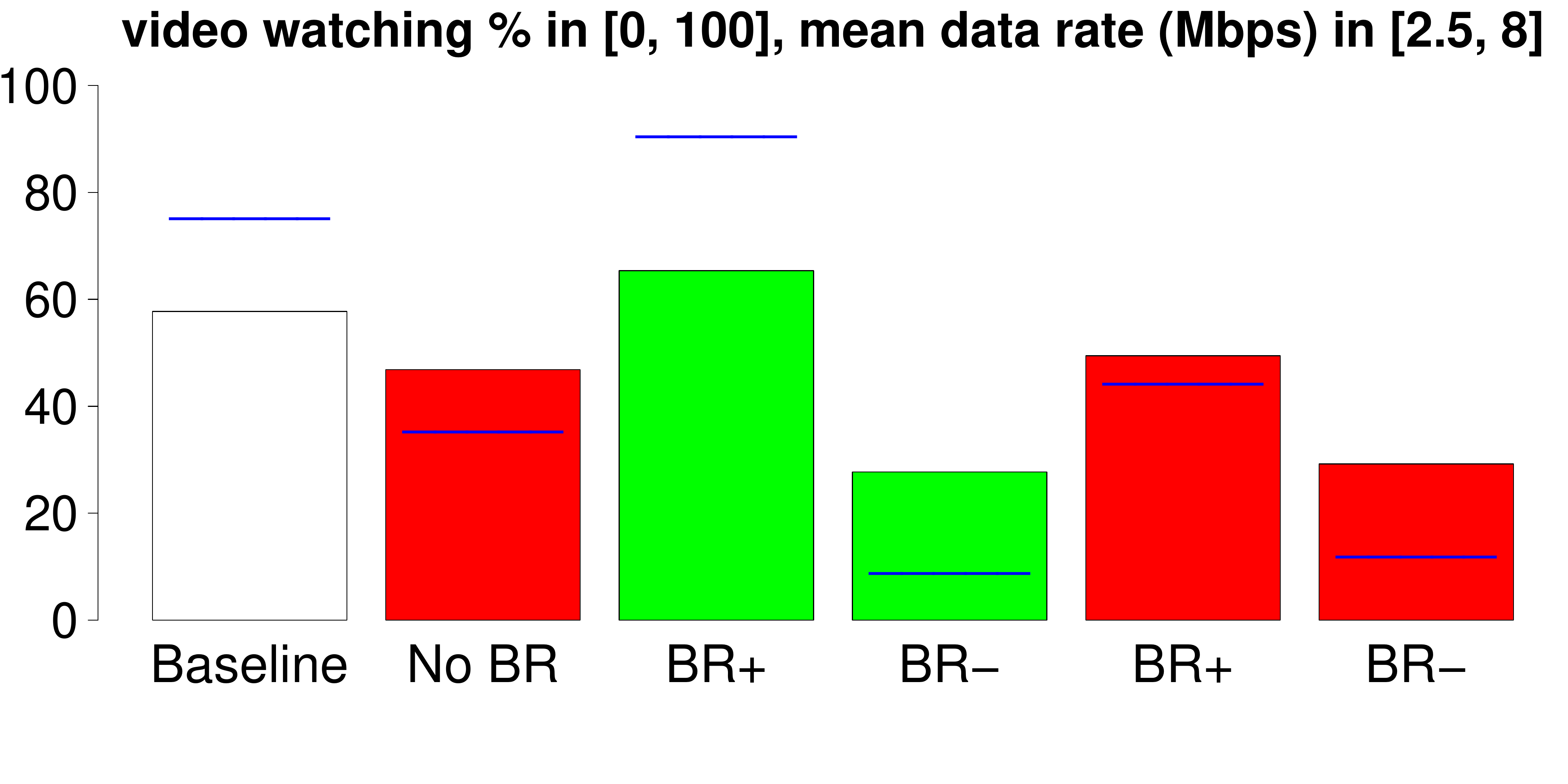}}
\subfloat{\includegraphics[width=0.34\textwidth, height=30mm]{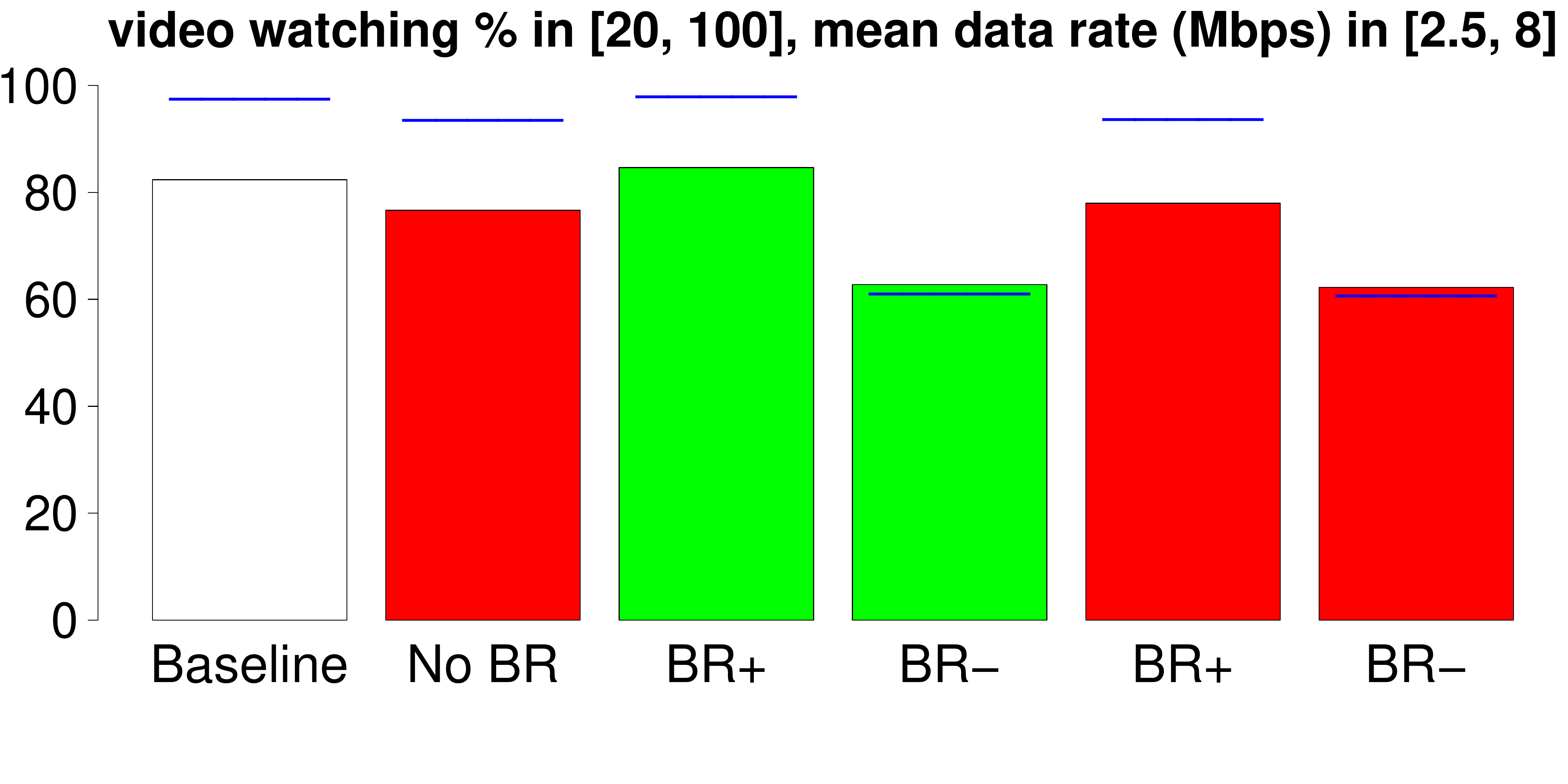}}
\subfloat{\includegraphics[width=0.33\textwidth, height=30mm]{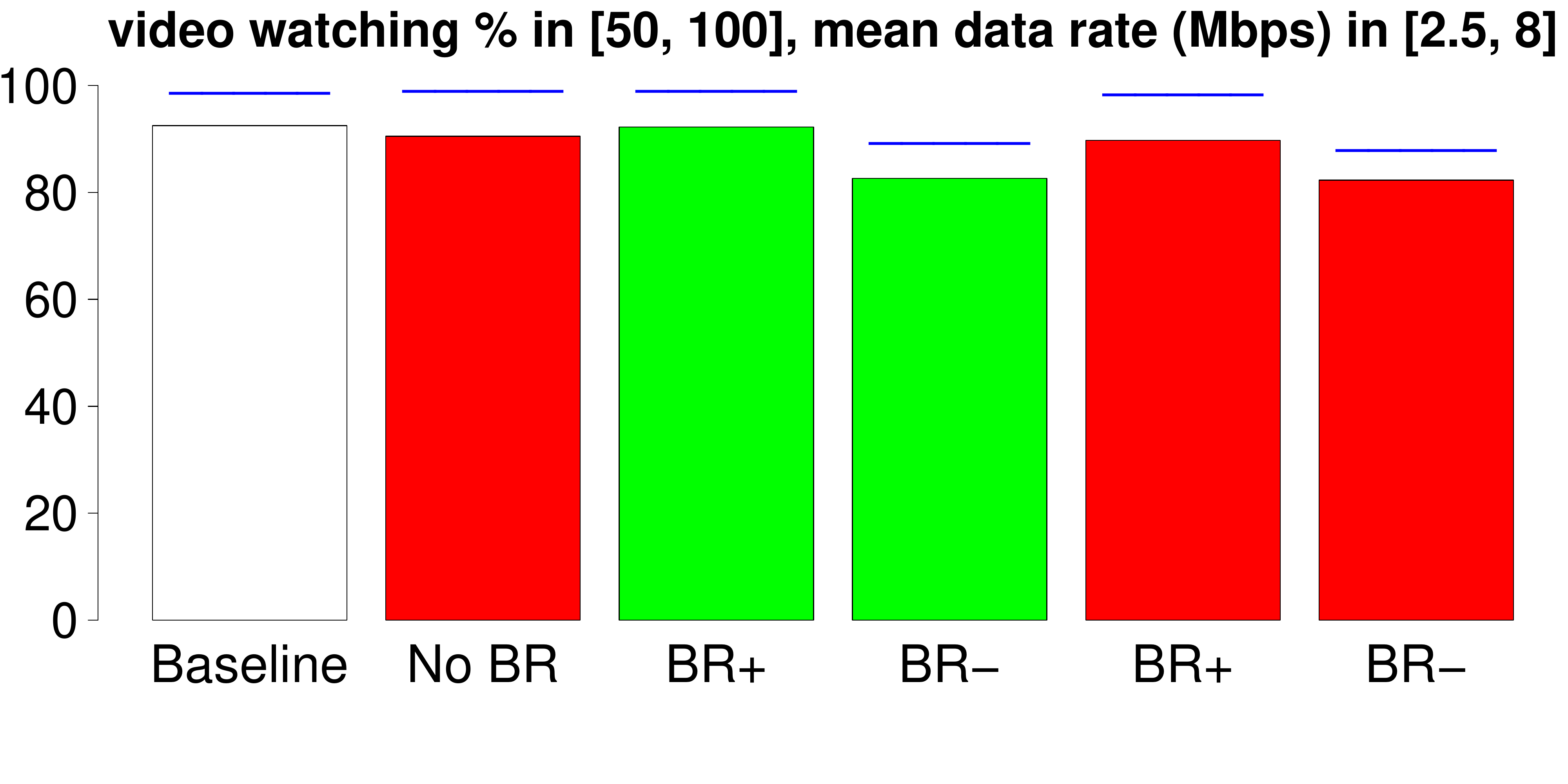}}

\subfloat{\includegraphics[width=0.33\textwidth, height=30mm]{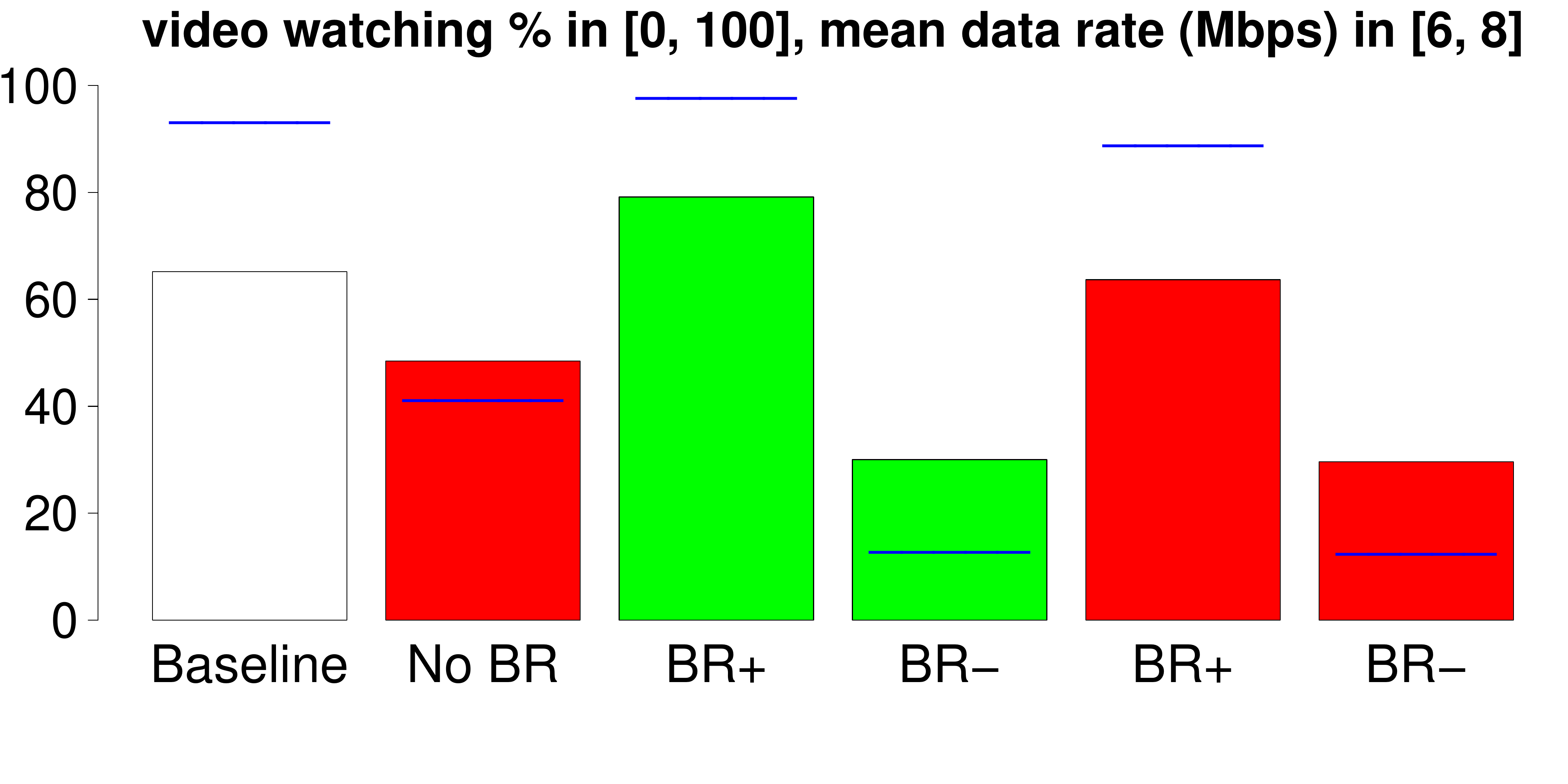}}
\subfloat{\includegraphics[width=0.34\textwidth, height=30mm]{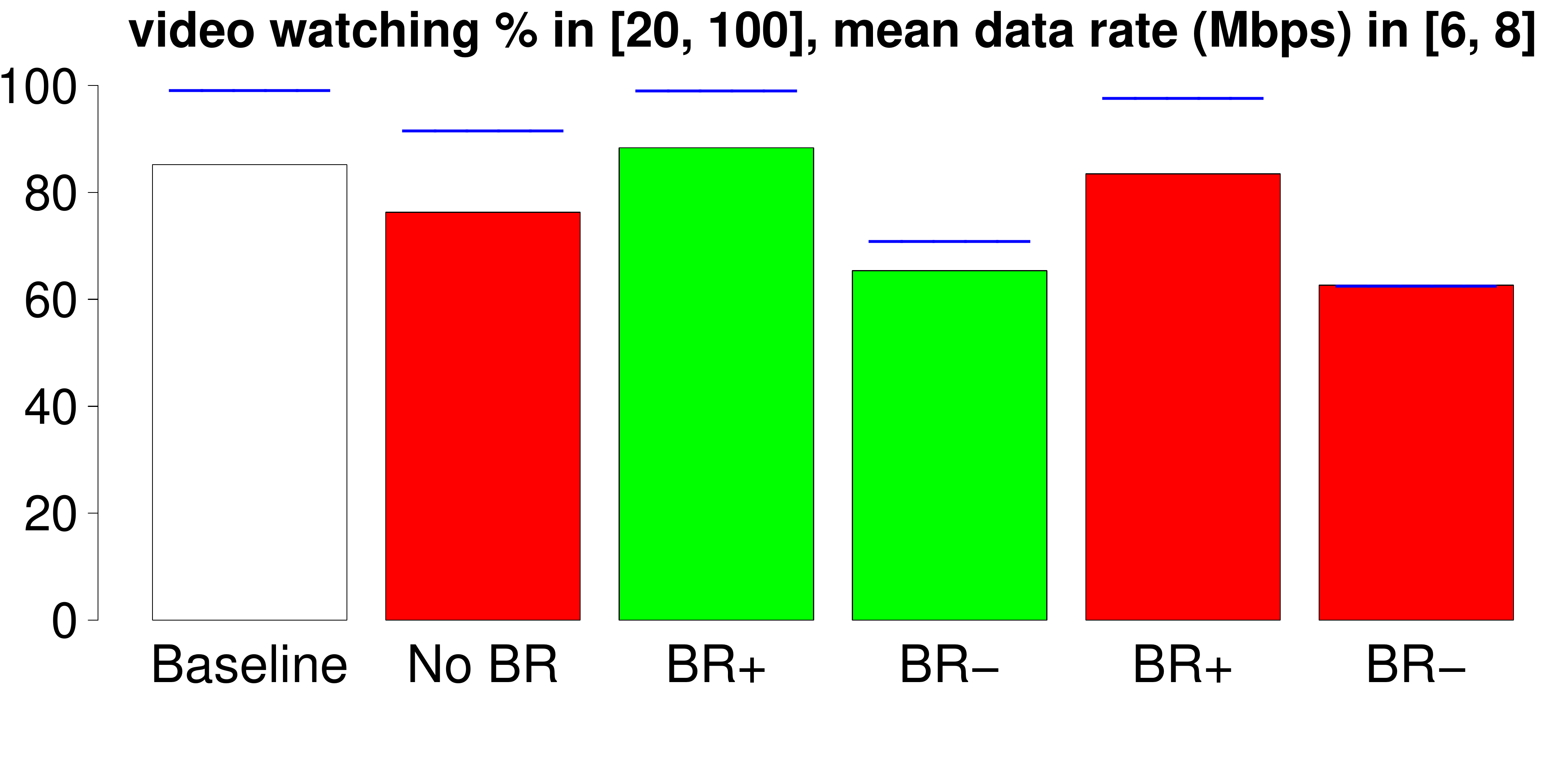}}
\subfloat{\includegraphics[width=0.34\textwidth, height=30mm]{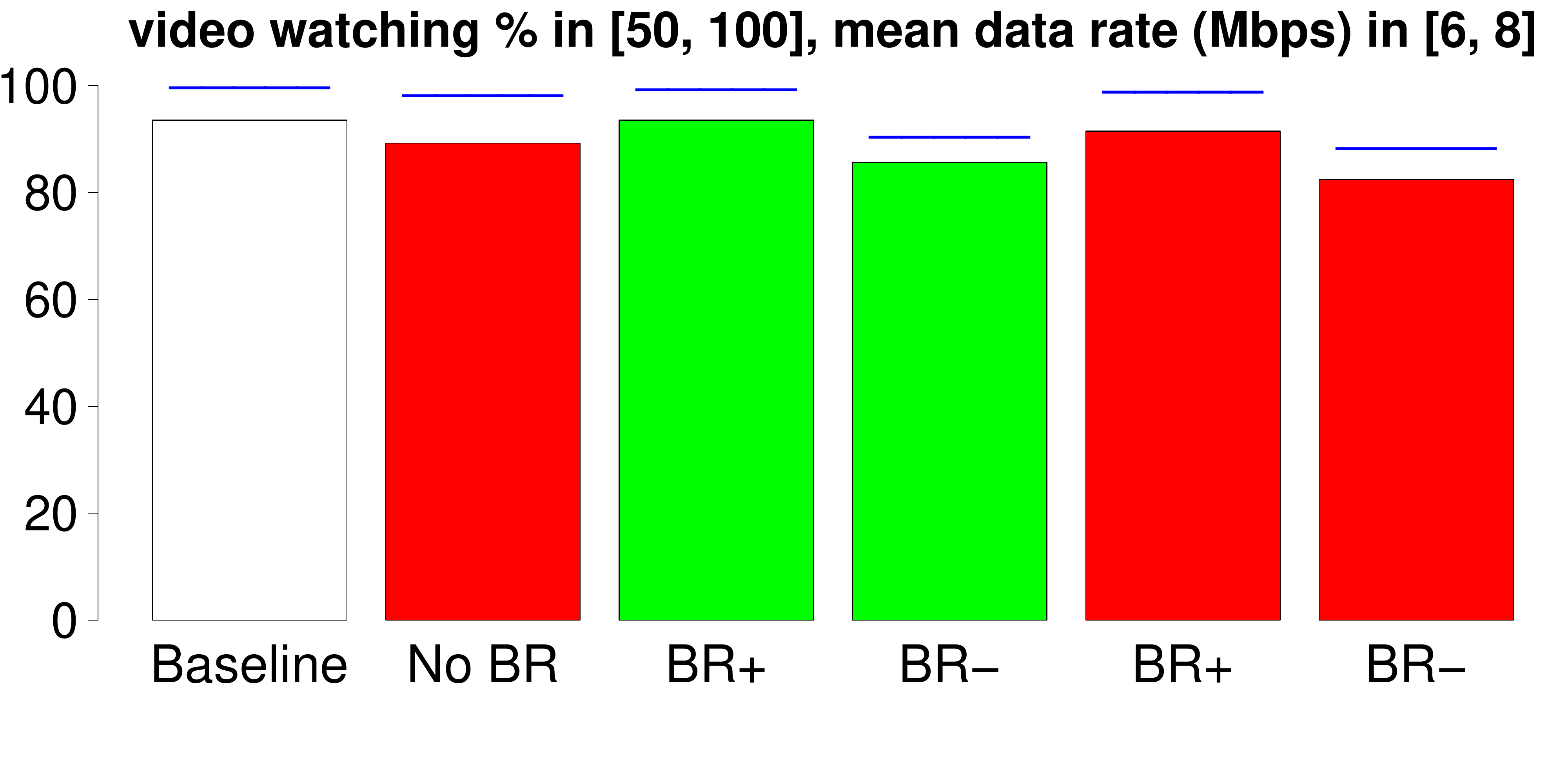}}

\caption{Video watching percentage of the main scenarios for different video watching \% and mean data rates. The blue horizontal line indicates the median.}
\label{fig:vwp_all_main6sc}
\end{figure*}

\begin{figure*}[t!] 
\centering
\subfloat{\includegraphics[width=0.33\textwidth, height=30mm]{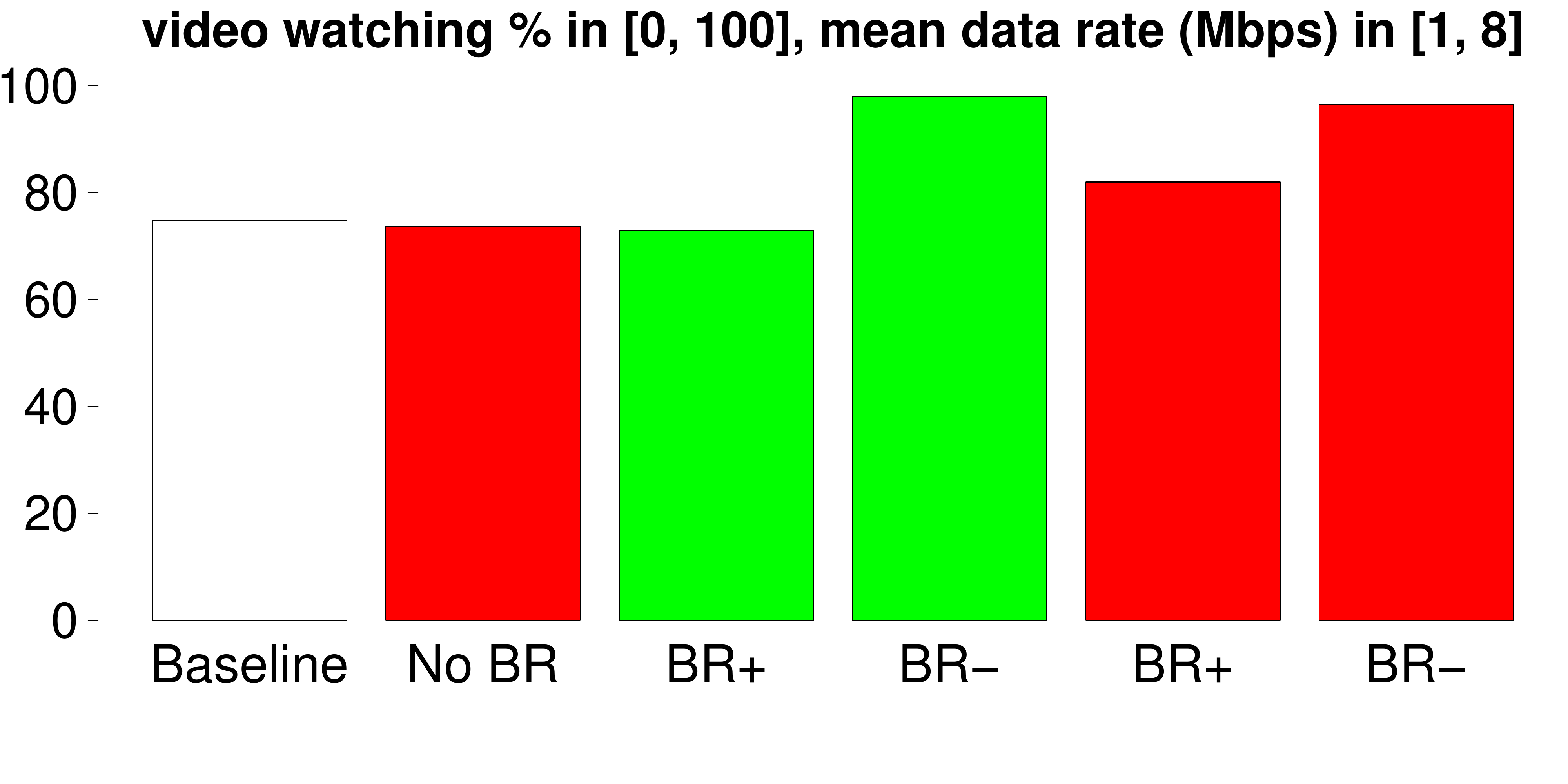}}
\subfloat{\includegraphics[width=0.34\textwidth, height=30mm]{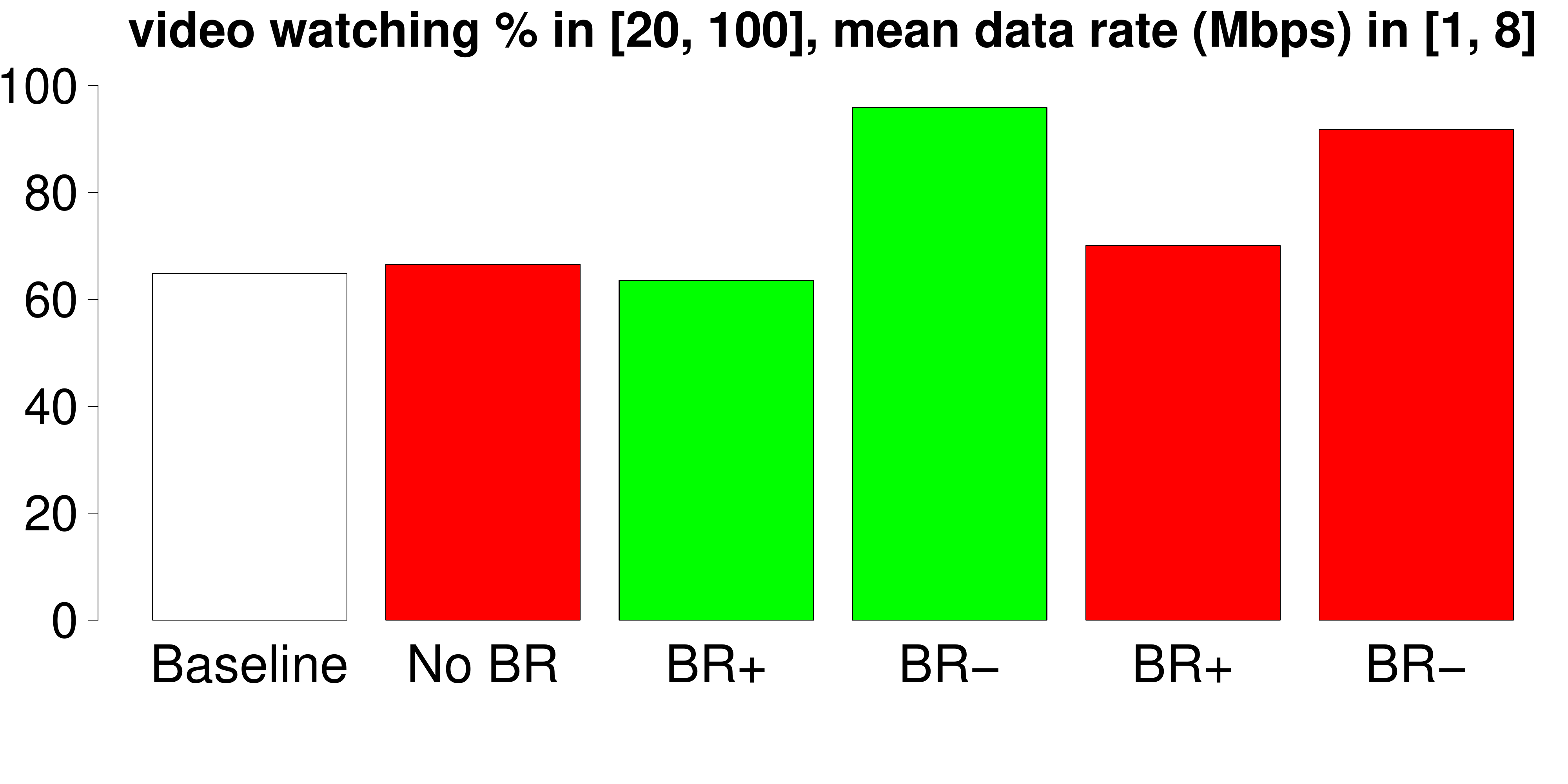}}
\subfloat{\includegraphics[width=0.33\textwidth, height=30mm]{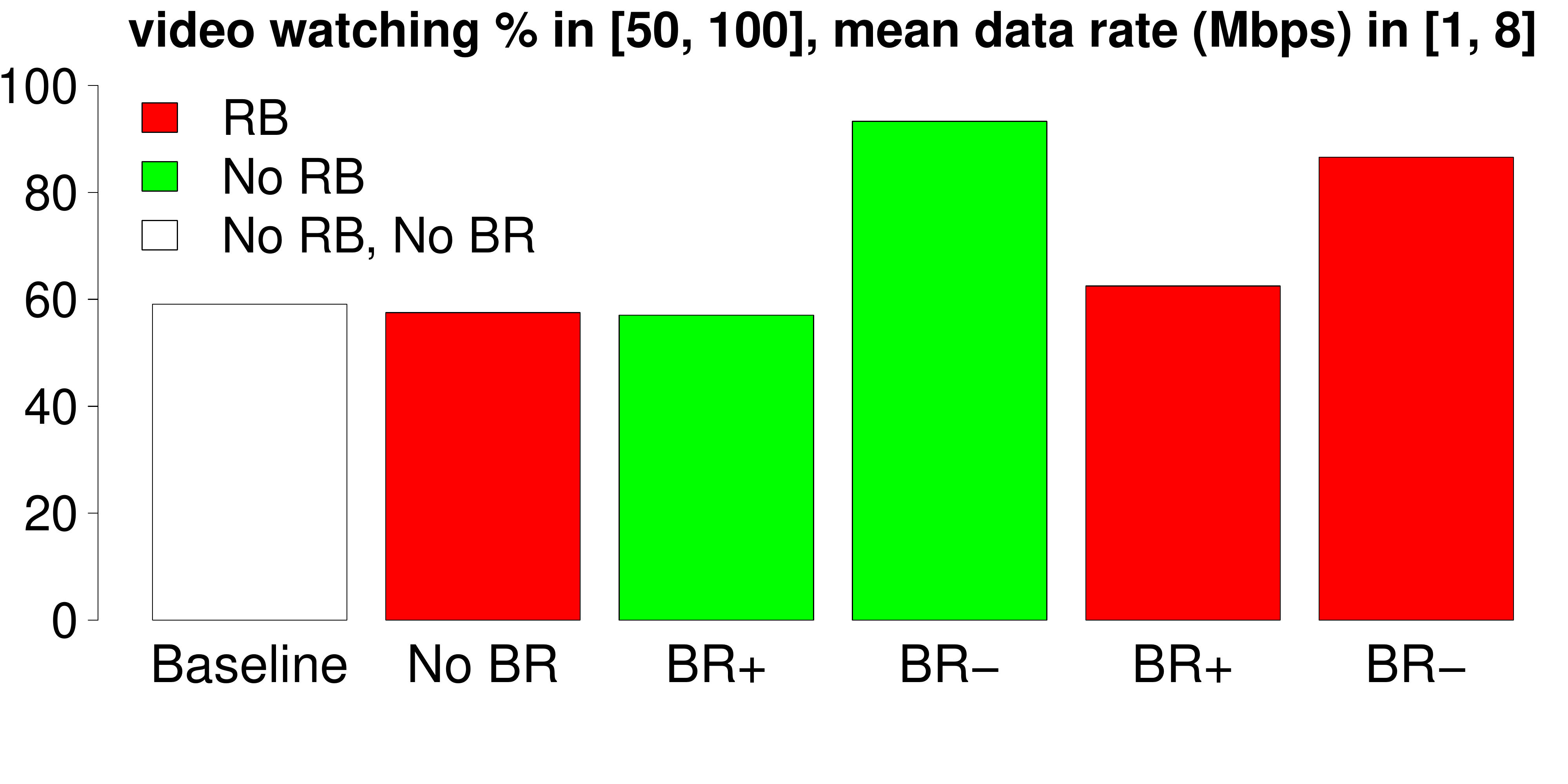}}

\subfloat{\includegraphics[width=0.33\textwidth, height=30mm]{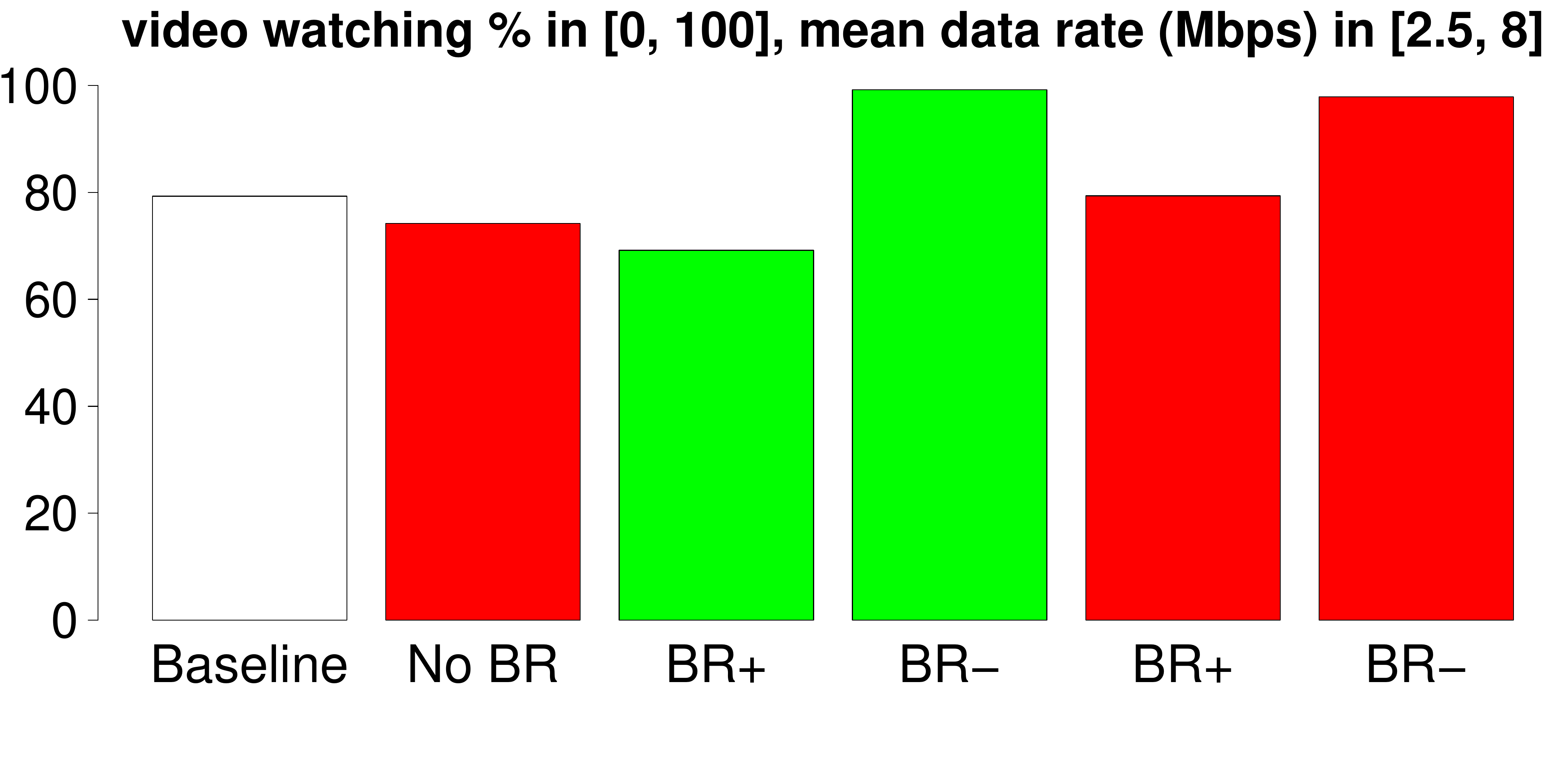}}
\subfloat{\includegraphics[width=0.34\textwidth, height=30mm]{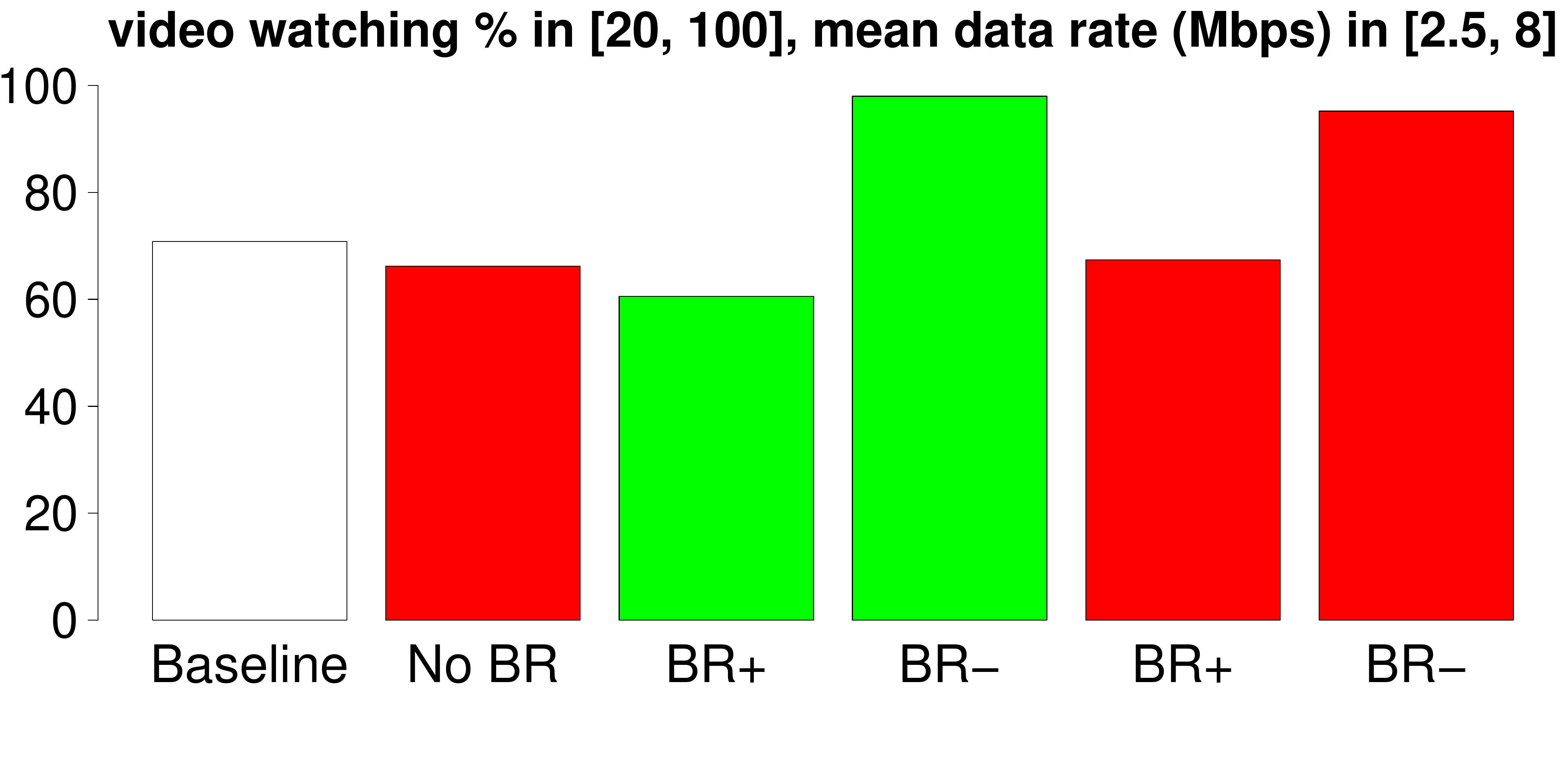}}
\subfloat{\includegraphics[width=0.33\textwidth, height=30mm]{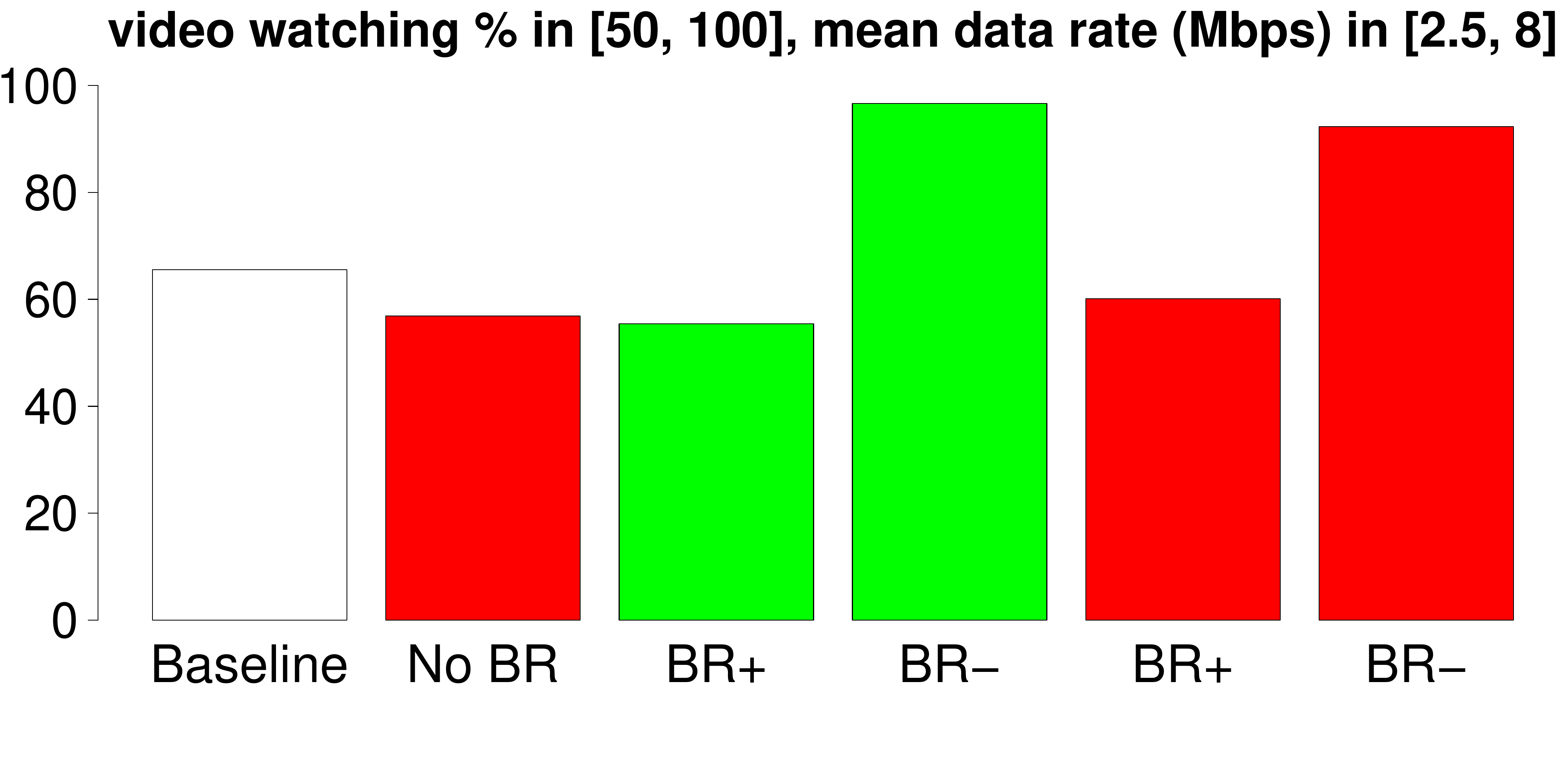}}

\subfloat{\includegraphics[width=0.33\textwidth, height=30mm]{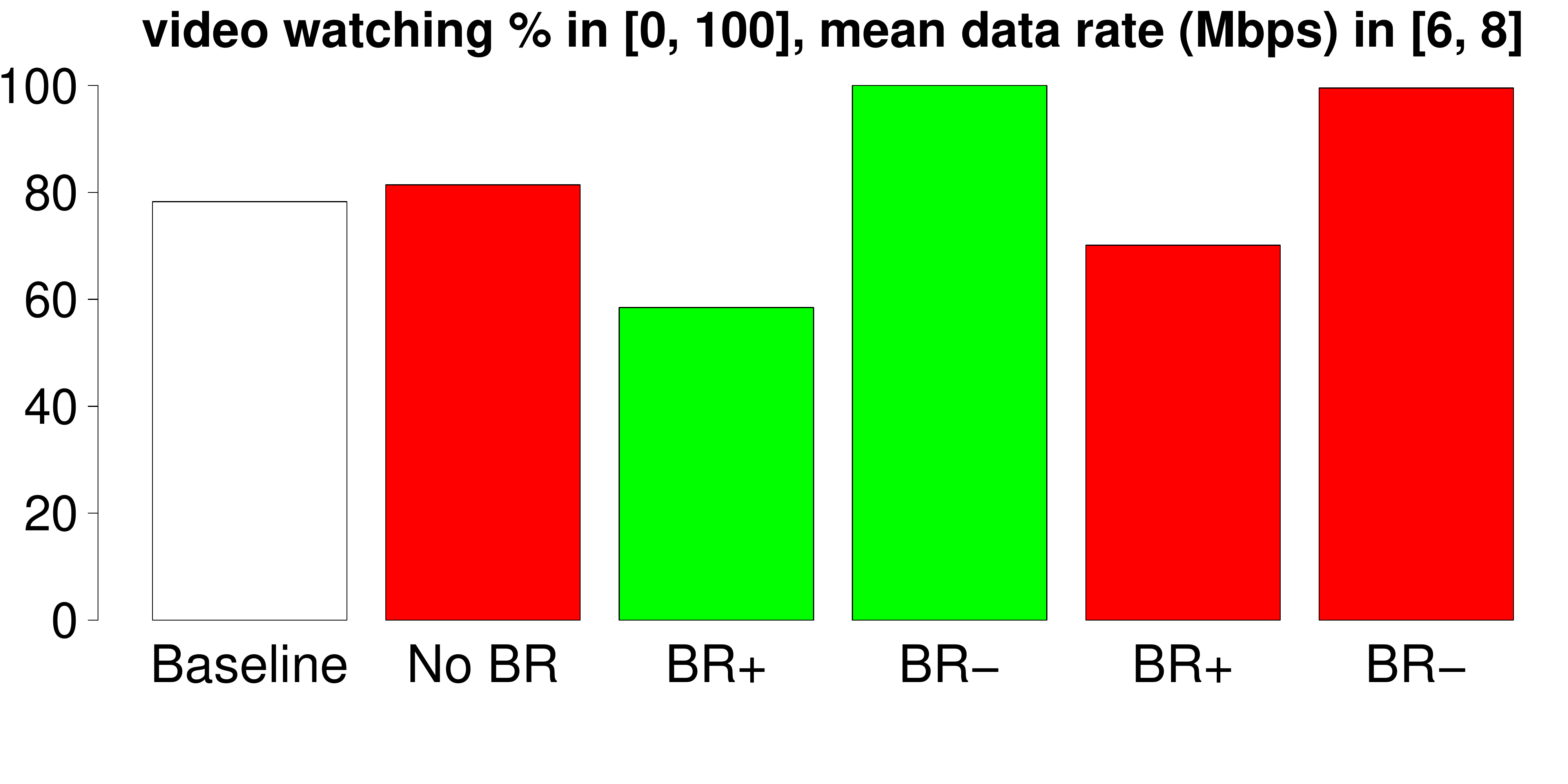}}
\subfloat{\includegraphics[width=0.34\textwidth, height=30mm]{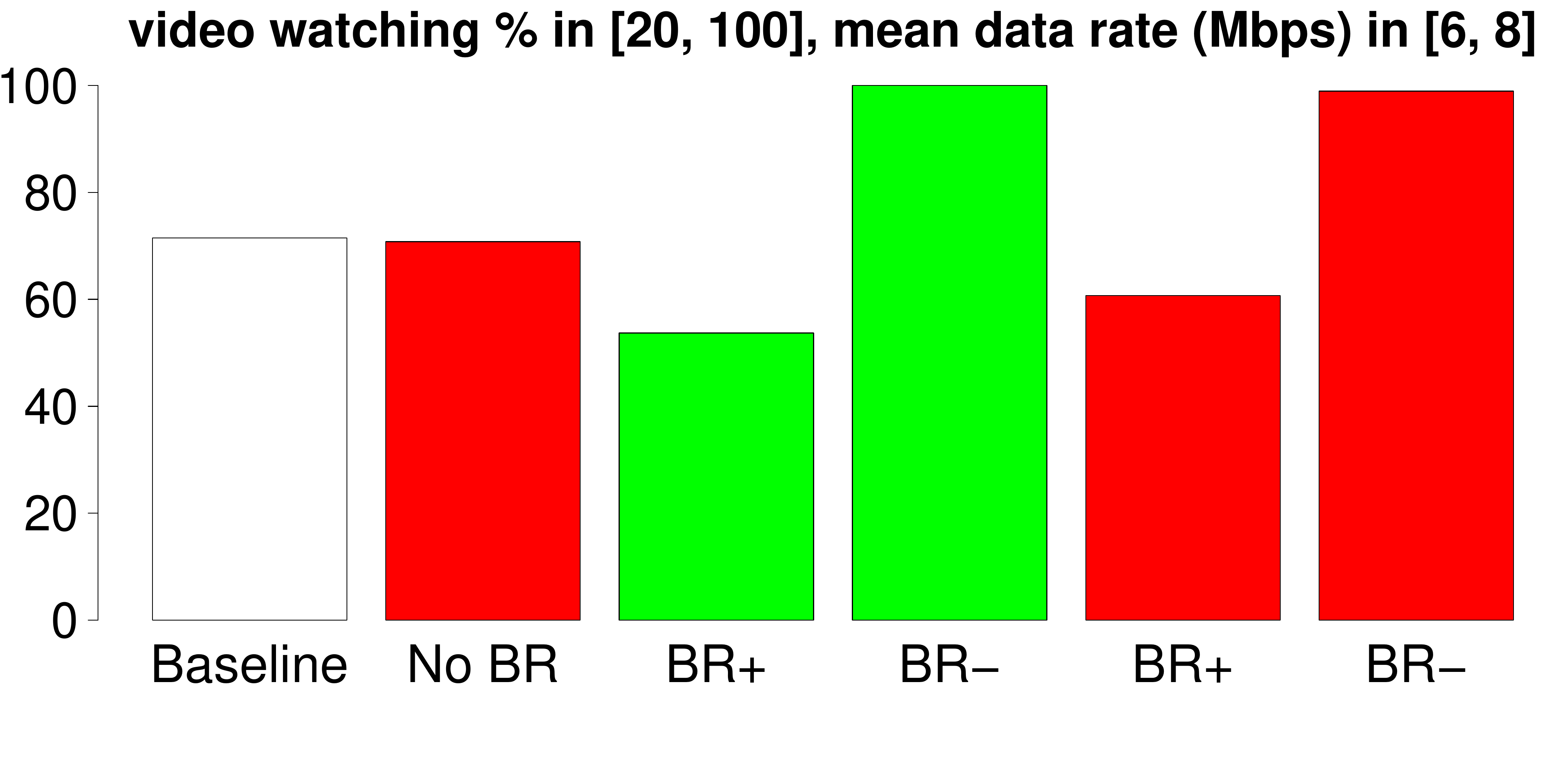}}
\subfloat{\includegraphics[width=0.33\textwidth, height=30mm]{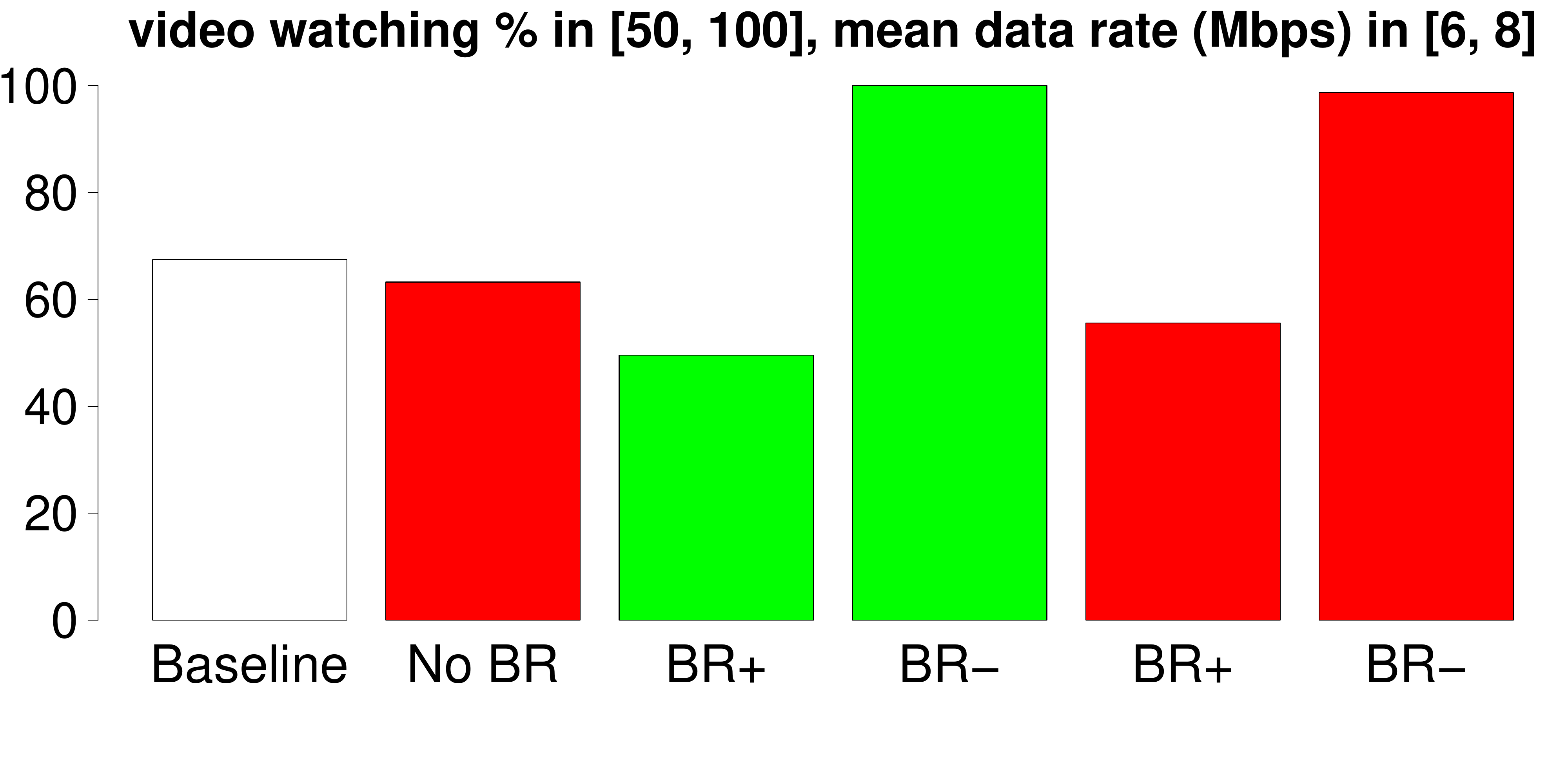}}

\caption{Abandonment ratio of the main scenarios, considering all abandoned sessions with different video watching percentage thresholds and mean data rates.}
\label{fig:abandonment_ratio_all_main6sc}
\end{figure*}
\restoregeometry
%

As expected, the lower the user interest in the content (i.e., the more relaxed the video watching percentage threshold), the smaller the tolerance for an impairment.
%
The more relaxed the video watching percentage threshold, the more prominent the differences among the six main scenarios (as shown in Figure \ref{fig:vwp_all_main6sc}).
\subsection{Sensitivity analysis on the duration of video content}
\begin{figure*}[t!] 
  \centering
\subfloat{\label{abnd_vwp_50_main6sc_vdur}\includegraphics[width=0.9\textwidth, height=60mm]{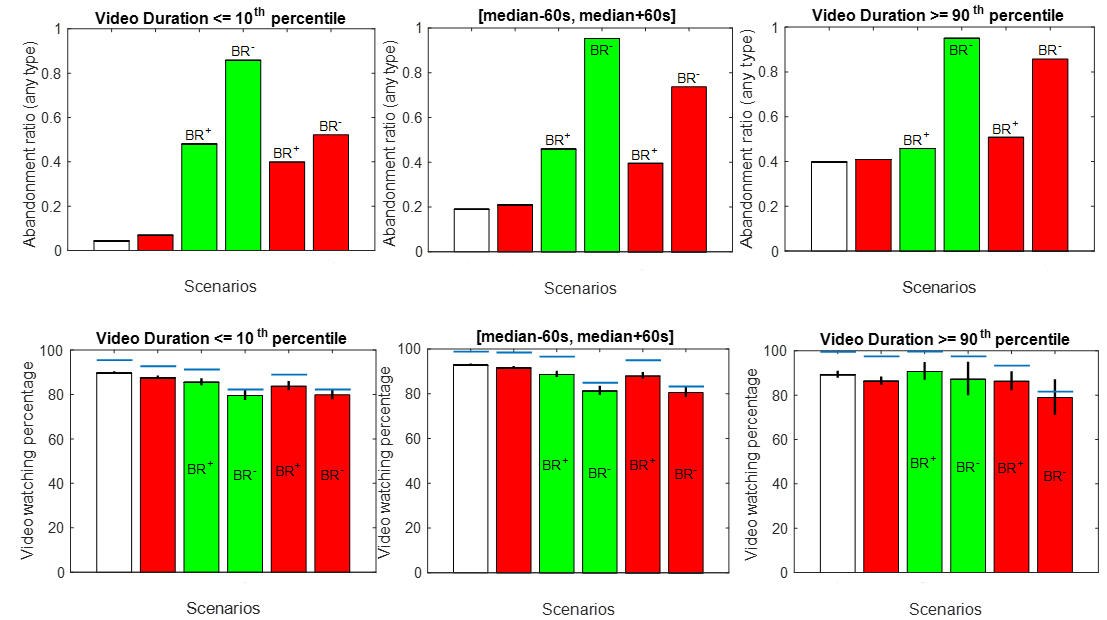}}
\caption{Abandonment ratios and watching percentage for different types of impairments and video duration, namely, short videos (less or equal to the 10th percentile), videos with typical durations (i.e., within 60 sec from the median video duration, which is about 5 min) and long videos (larger than the 90th percentile). The description of each scenario (1-6) can be found in Section \ref{sec:perf:neg}. All sessions correspond to videos that more than half of the content has been watched. The horizontal line above each bar indicates the median video watching percentage.}
\label{fig:abnd_vwp_50_main6sc_vdur}
\end{figure*}

To examine the user engagement patterns under different video content duration, we formed the following groups of sessions: (1) the representative ones with video duration in [4 min, 6 min] and mean video duration of about 5 min, (2) the short videos, i.e., sessions of video duration within the 10th percentile (which is about 2 min), and (3) the relatively long videos, with duration equal or greater than the 90th percentile of video duration (which is about 38 min). The trends discussed in the previous Section persist also in these groups. In all groups, sessions with only BR- exhibit the lowest engagement (in abandonment ratio and video watching percentage), followed by sessions with both BR- and RBs. Sessions with BR+ have also high abandonment. Same trends persist also for the video watching percentage metric, with the exception of long videos, where sessions with rebufferings have lower video watching percentage than sessions with only positive or negative bitrate changes. Surprisingly, sessions with only rebufferings exhibit similar abandonment ratio to sessions without RB and BR changes. The no BR and no RB change scenario has significant statistical differences from all others in terms of the video watching percentage (Fig. \ref{fig:abnd_vwp_50_main6sc_vdur}). Interestingly compared to rebuffering, the BR- appears to have more prominent impact on user engagement in the context of YouTube. This trend persists also when we employ other user engagement metrics, as shown in the next subsections. 
\begin{figure*}[t!] 
\centering
\subfloat{\label{vwp_0_Res_RBdur_sc}\includegraphics[width=0.35\textwidth, height=35mm]{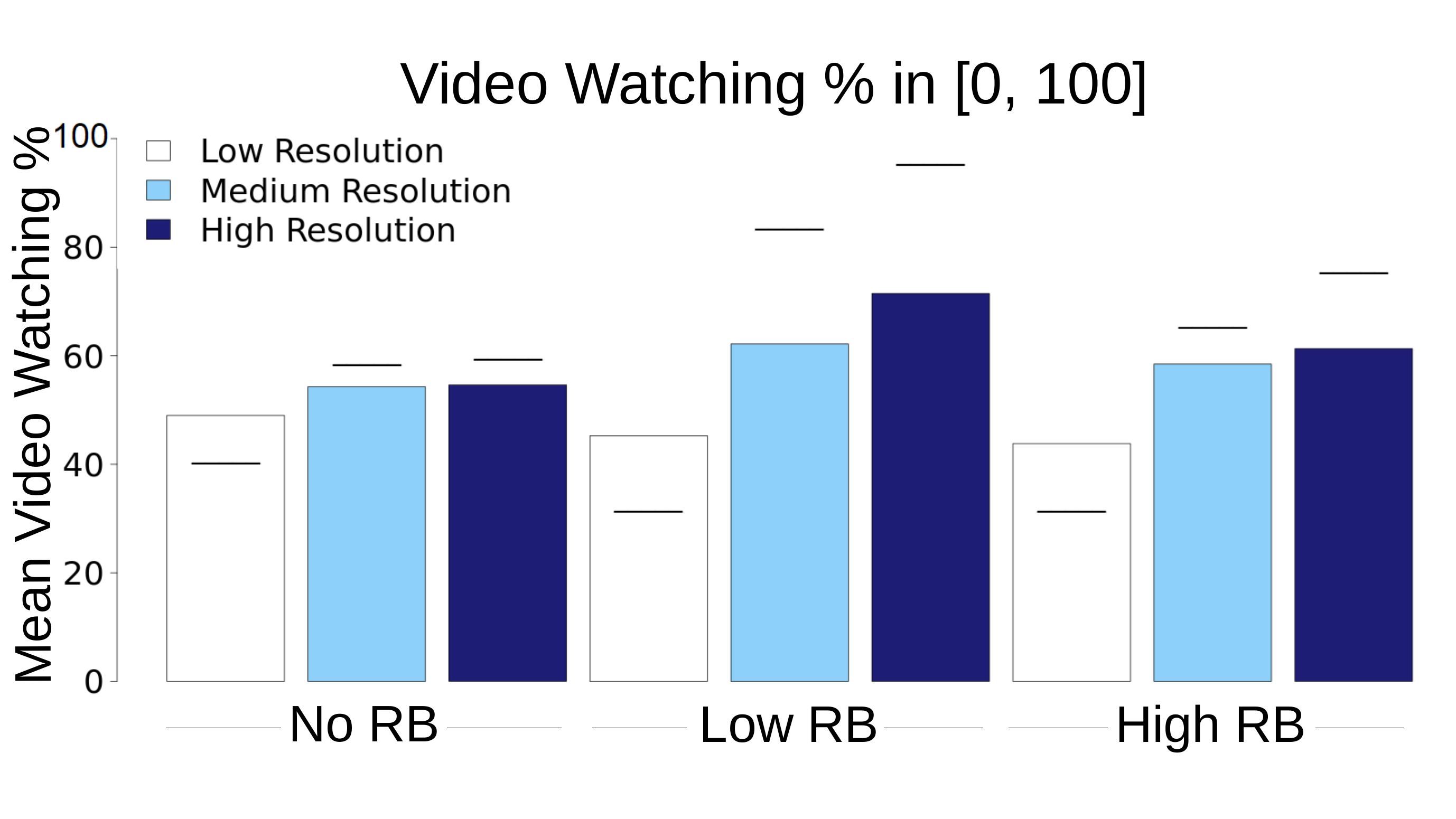}}
\subfloat{\label{vwp_20_Res_RBdur_sc}\includegraphics[width=0.35\textwidth, height=35mm]{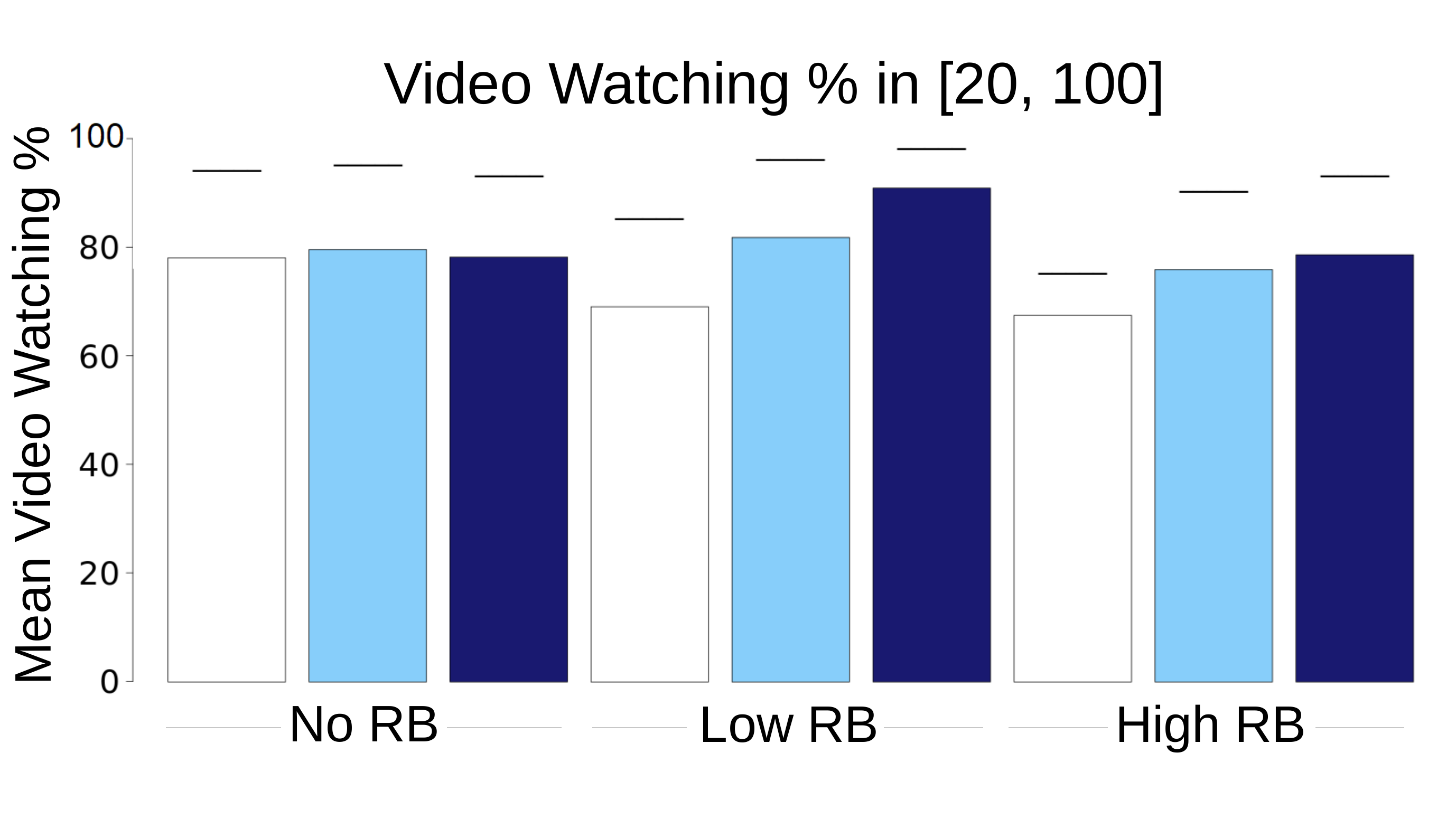}}
\subfloat{\label{vwp_50_Res_RBdur_sc}\includegraphics[width=0.35\textwidth, height=35mm]{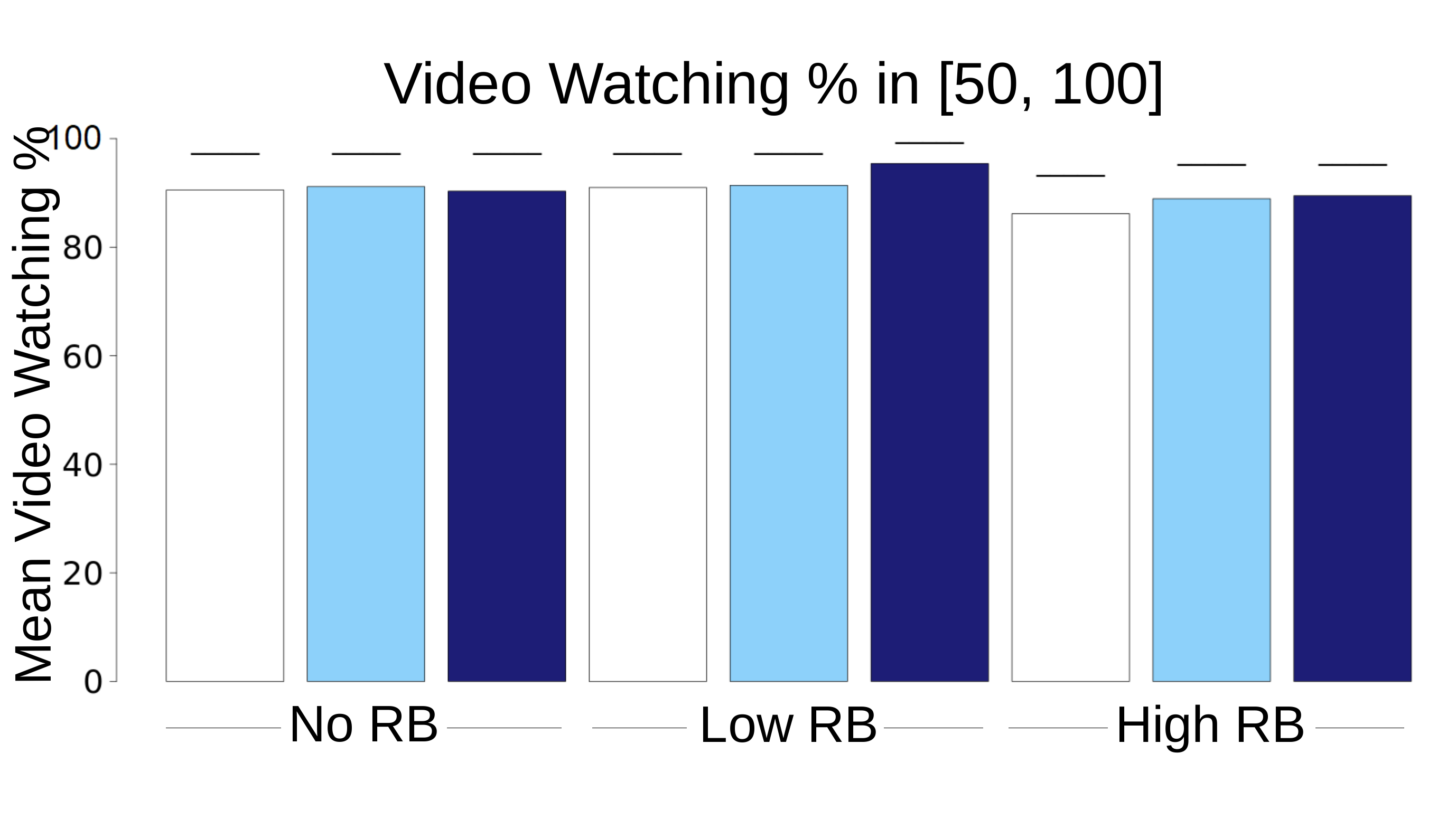}}

\subfloat{\label{mean_res_0_Res_RBdur_sc}\includegraphics[width=0.35\textwidth, height=35mm]{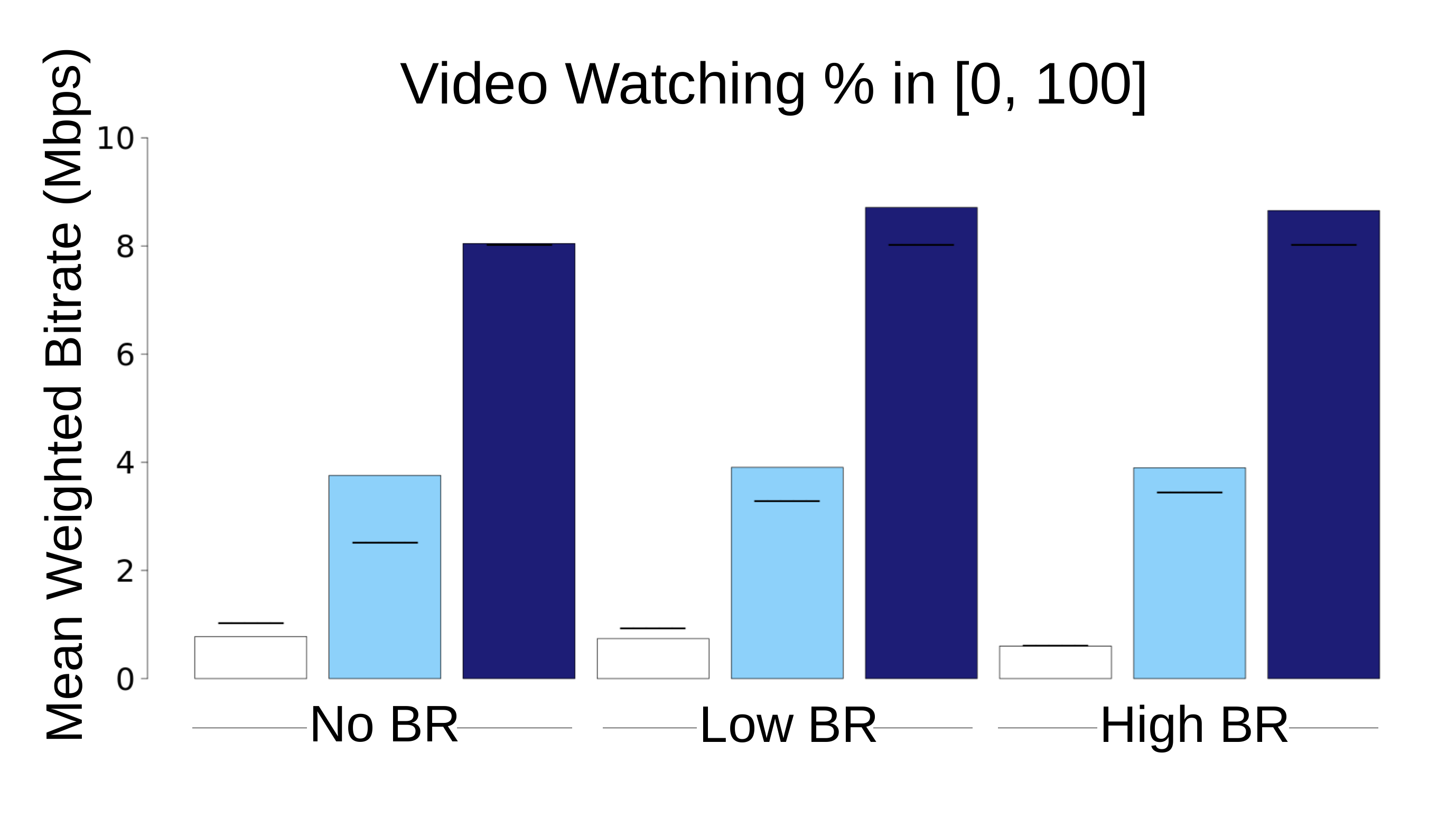}}
\subfloat{\label{mean_res_20_Res_RBdur_sc}\includegraphics[width=0.35\textwidth, height=35mm]{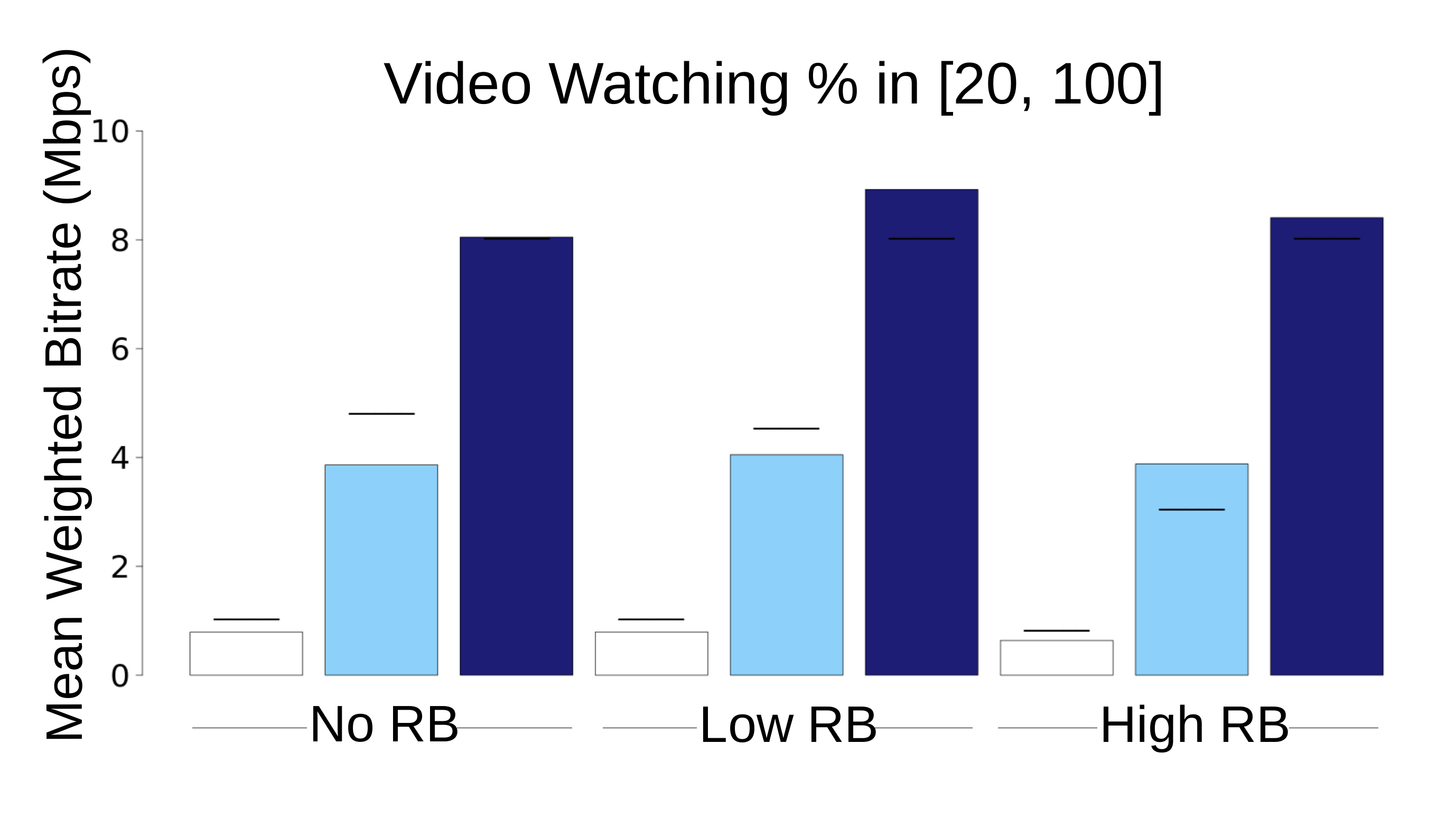}}
\subfloat{\label{mean_res_50_Res_RBdur_sc}\includegraphics[width=0.35\textwidth, height=35mm]{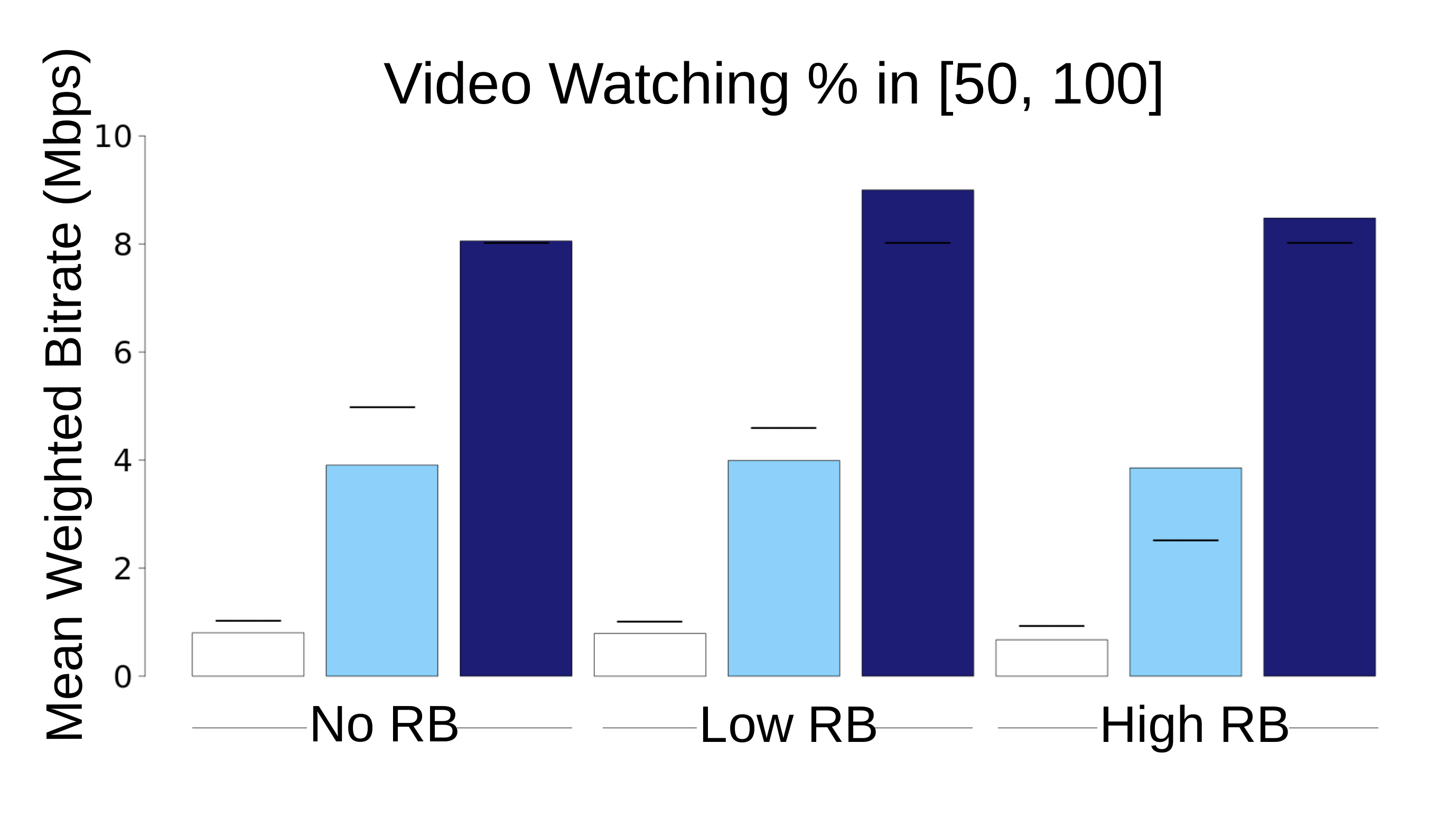}}
\caption{Video watching percentage (top) and mean weighted resolution (bottom) for different RB durations. The threshold for the video watching threshold is 0 (left), 20 (middle), and 50 (right). The thresholds in the above scenarios are as follows: Low RB $\le$ 2 sec, the High RB>2 sec, Low Resolution $\le$ 1Mbps, Medium Resolution in (1Mbps, 8Mbps), and High Resolution $\ge$ 8Mbps.
The blue horizontal line indicates the median.}
\label{fig:Res_RBdur_sc}
\end{figure*}
\subsection{Propensity in abandonment after an impairment}
Both the video watching percentage and abandonment ratio can provide an indication of the user engagement and the impact of the various impairments.
To further examine the level of severity of an impairment on the user engagement, we focus on
{\em how promptly a user abandons the session after the occurrence of an impairment}. We expect that, in general, the more severe the impact of an impairment, the more likely the abandonment of the session within a short interval after the occurrence of the impairment. To eliminate the impact of low resolution and multiple impairments, we decided to focus on sessions with {\em exactly one impairment}, namely a BR- or a RB, and relatively high mean weighted data rate (i.e., in [2.5Mbps, 8Mbps]). 
Interestingly, 91.49\% of the sessions get abandoned as soon as a BR- occurs, compared to 44.65\% when a RB occurs. This likelihood increases to 94.32\% and 55.17\% for sessions that were terminated within 60 sec from the occurrence of a BR- and RB, respectively (Table \ref{tbl:1impairnment:abnd:prob}). This trend persists for other video watching percentage thresholds and tighter (higher) mean weighted data rate.
Considering sessions with a RB, that satisfy the video watching percentage and weighted mean data rate thresholds, it is interesting that sessions with larger RB duration have lower likelihood to be abandoned. We speculate that when a RB occurs the viewer may expect that after its completion, the quality of the session will improve and thus remains longer in the session.

\begin{table}[t!]
\centering 
\begin{tabular}{| c | c | c | c | c | }
\hline
Scenario & Number of abandoned sessions (\%) & $\mathbb{P}(\mathrm{T} = $ 0) & $\mathbb{P}(\mathrm{T}$ $\leq$ 60)  & $\mathbb{P}(\mathrm{T}$ $>$ 60) \\ 
\hline
BR-&  141 (96.58\%)& 91.49\%  & 94.32\% & 5.67\% \\
\hline
BR+& 437 (56.16\%) & 17.8\% & 29.3\% & 70.7\% \\ \hline
RB&  589 (76.59\%) & 44.65\% & 55.17\%  & 44.82\% \\ \hline
\end{tabular}
\caption{All {\em abandoned} sessions with {\em exactly one} impairment (namely, BR-, BR+, or RB), mean weighted resolution in the interval [2.5Mbps, 8Mbps] and video watching percentage in [50,100]. The time T (in sec) is the time from the beginning of the impairment to end of session.}
\label{tbl:1impairnment:abnd:prob} 
\end{table}

We then compared the three populations of abandoned sessions, namely the ones with a single BR-, the ones with a single BR- which were abandoned within 60 sec from the occurrence of the BR-, and the baseline (no RB/no BR), with respect to the video watching percentage. 
The sessions with BR- that were abandoned within 60 sec have a statistically significant difference in video watching percentage from the ones without RB or BR.
Moreover, all abandoned sessions with BR- have statistically significant difference from all the baseline sessions (regardless the time when the abandonment occurred).
Similarly, sessions with a RB that were abandoned within 60 sec from the time that the RB occurs have a statistically significant difference from the baseline scenario. On the other hand, we cannot reject the null hypothesis, in the case of sessions with RB that were not abandoned within 60 sec from the time the RB occurred.

The negative BR change seems to have a more direct impact on the user-engagement than RB: 90\% of users that experience a single BR-, abandoned the session within 60 sec after the occurrence of the bitrate change, as opposed to 66\% in the case of single RB, for the video watching percentage in [20\%, 100\%]. Moreover, the 95\% of the sessions were abandoned within 185 sec after the occurrence of the BR-, while the same percentage of sessions got abandoned within 696 sec, in the case of RB. Fig. \ref{fig:ECDF_for_2vs1_impairments_start_time_until_end_session} shows this comparison for different thresholds.
Specifically, the second column of Table \ref{table:8} was calculated without any constraint on the Abandonment Type (that is, it includes sessions that ended normally). The third column considered only sessions that were abandoned with abandonment type 1 or 2. 

Interestingly, even the BR+ events may result in abandonments:
e.g., from sessions with exactly one impairment, namely a BR+,  the 27\% of them are abandoned within 60 sec from the time when a BR+ occurred, for sessions with video watching percentage in [50, 100], while this percentage changes to 19\% for sessions with video watching percentage in [20, 100]. 
\begin{table}[t!]
\centering 
\begin{tabular}{| c | c | c | c | c | c | c |} 
\hline
Impairment &  $\mathcal{P}_{95}$ & $\mathbb{P}$(T $\leq$ 60)  &  Abandonment \% &$\mathcal{P}_{95}$ & $\mathbb{P}$(T $\leq$ 60)  &  Abandonment \%\\
\hline
BR- &  185 &  0.90 & 96.42 &  77  &  0.95 & 97.98\\
\hline 
RB&    696 &  0.66 & 84.17 &  764 & 0.55 &  79.04\\
\hline
0 $\leq$ RB $\leq$ 2&   603  & 0.73 & 85.84& 737 & 0.61&  80.23\\
\hline
RB$>$2& 978   & 0.30& 76.82 & 1038   & 0.30 & 73.56\\
\hline
** RB$>$20 &  1442   & 0.35 & 85.00 &  1456   &  0.37& 87.88\\
\hline
\end{tabular}
\caption{The time elapsed from the start of the RB or the occurrence of the BR- to the end of the sessions (T), for sessions with exactly one impairment, RB or BR-. The threshold for the video watching percentage is 20\%. Similar trends are observed for video watching percentage threshold of 50\%.
The $\mathcal{P}_{95}$ corresponds to the 95th percentile of T distribution. The two stars indicate that the number of observations is less or equal to 40. The columns 2-4 (5-7) correspond to sessions with weighted mean data rate (Mbps) in [1,8] ([2.5,8]), respectively.}
\label{tbl:1impairment_abnd_within60_part_02} 
\end{table} 
Users seem to be more tolerant on RBs than on BR- (Table \ref{tbl:1impairnment:abnd:prob} \& \ref{tbl:1impairment_abnd_within60_part_02}). These tables considers pre-processed sessions without startup delay.  The third column applies the following equation for all abandoned sessions, while the fourth estimates the abandonment ratio (considering all pre-processed sessions).
\begin{equation}
\mathbb{P}(\mathrm{T} \leq 60 \text{ sec}) = \frac{\# \text{ of abandoned sessions that have T} \leq 60 \text{ sec}}{\# \text{ all abandoned sessions}}
\end{equation}

\begin{table}[h]
\centering
\setlength\tabcolsep{4pt}
\begin{minipage}{.48\textwidth}
\centering
\begin{tabular}{|c | c | c|} 
\hline 
Scenario & $\mathcal{P}_{95}$ (sec)& $\mathbb{P}$(T $\leq$ 60 sec) \\
\hline 
BR-  & 185 & 0.90\\
\hline
RB   & 690 & 0.67\\
\hline
\textbf{RB$>$2}  & \textbf{942}  & \textbf{0.37}\\
\hline
** RB$>$20 & 615  & 0.67\\
\hline
BR-  & 77 & 0.95\\
\hline 
RB &  756 & 0.56\\
\hline
RB$>$2 & 952 & 0.37\\
\hline
** RB$>$20 & 622 & 0.68\\
\hline
\end{tabular}
\end{minipage}
\hfill
\begin{minipage}{.48\textwidth}
\centering
\begin{tabular}{|c | c | c|} 
\hline 
Scenario & $\mathcal{P}_{95}$ (sec)& $\mathbb{P}$(T $\leq$ 60 sec) \\
\hline 
BR-  & 262 & 0.88 \\
\hline
RB   &  698 & 0.67 \\
\hline
\textbf{RB$>$2}  & \textbf{972}  & \textbf{0.22}\\
\hline
** RB$>$20 & 963 & 0.43 \\
\hline
BR-  & 96 & 0.94 \\
\hline 
RB &  761 & 0.55 \\
\hline
RB$>$2 & 972 & 0.25 \\
\hline
** RB$>$20 & 1113 & 0.50 \\
\hline
\end{tabular}
\end{minipage}
\caption{Exactly one impairment, namely RB or BR-,  for different cases, in terms of video watching percentage criterion and mean weighted datarate. 
T is the time elapsed from the end of the RB or the occurrence of the BR- to the end of the session. The two stars indicate that the number of observations is smaller than 40. 
The threshold for the video watching percentage is 20 (left) and 50 (right). 	
The first(last) four rows correspond to sessions with weighted mean data rate (Mbps) in [1,8] ([2.5,8]), respectively.}
\label{table:8} 
\end{table}

\clearpage
\subsection{On the impact of rebufferring duration and resolution}
Krishnan {\em {\em et al.}} \cite{krishnan2013video} reported that startup delay of 2 sec or more causes more viewers to abandon the video. We speculate that users could be even more annoyed by a rebuffering event of similar duration during playback.  In a field study performed by Hossfeld {\em {\em et al.}} \cite{hossfeld2008testing} with 1 min video sessions, single rebufferings of 2 sec or more have significant impact on the MOS. This is also consistent with the ITU-T G.247 model. Similar observations were made for the case of single or multiple rebufferings in short videos (duration of 30 sec). Given these findings, we selected the threshold of 2 sec to distinguish the short (low) rebuffering duration events from the high (long) and classified the sessions according to their total rebuffering duration into new scenarios. The majority of sessions (80\%) have total rebuffering duration less than 2 sec (Fig. \ref{fig:RB_background}).

YouTube sessions with RBs larger than 2 sec, but no BR change have higher video watching percentage than the no RB with BR- scenario (Figure \ref{fig:Res_RBdur_sc}). As expected, the higher the RB duration, the lower the video watching percentage. For video watching percentage in [50,100], the differences in the video watching percentage across scenarios of different RB duration (for mean data rate within specific interval) are small. The differences become more prominent for video watching percentage in [20,100] and in [0,100]. However, the differences in the mean data rate across scenarios with RB duration in the same range are significant.
Under large RB duration (e.g., 15 sec or more), the video watching percentage is higher than in sessions with BR- but no RB, though the difference is not statistically significant. Similar results are observed for low and medium RB duration (with a median RB duration of 2 sec and 14 sec, respectively). 

\begin{figure*}[t!] 
\hspace*{-1.0cm}
\subfloat{\label{vwp_20_RBratio_startup_sc}\includegraphics[width=0.60\textwidth, height=60mm]{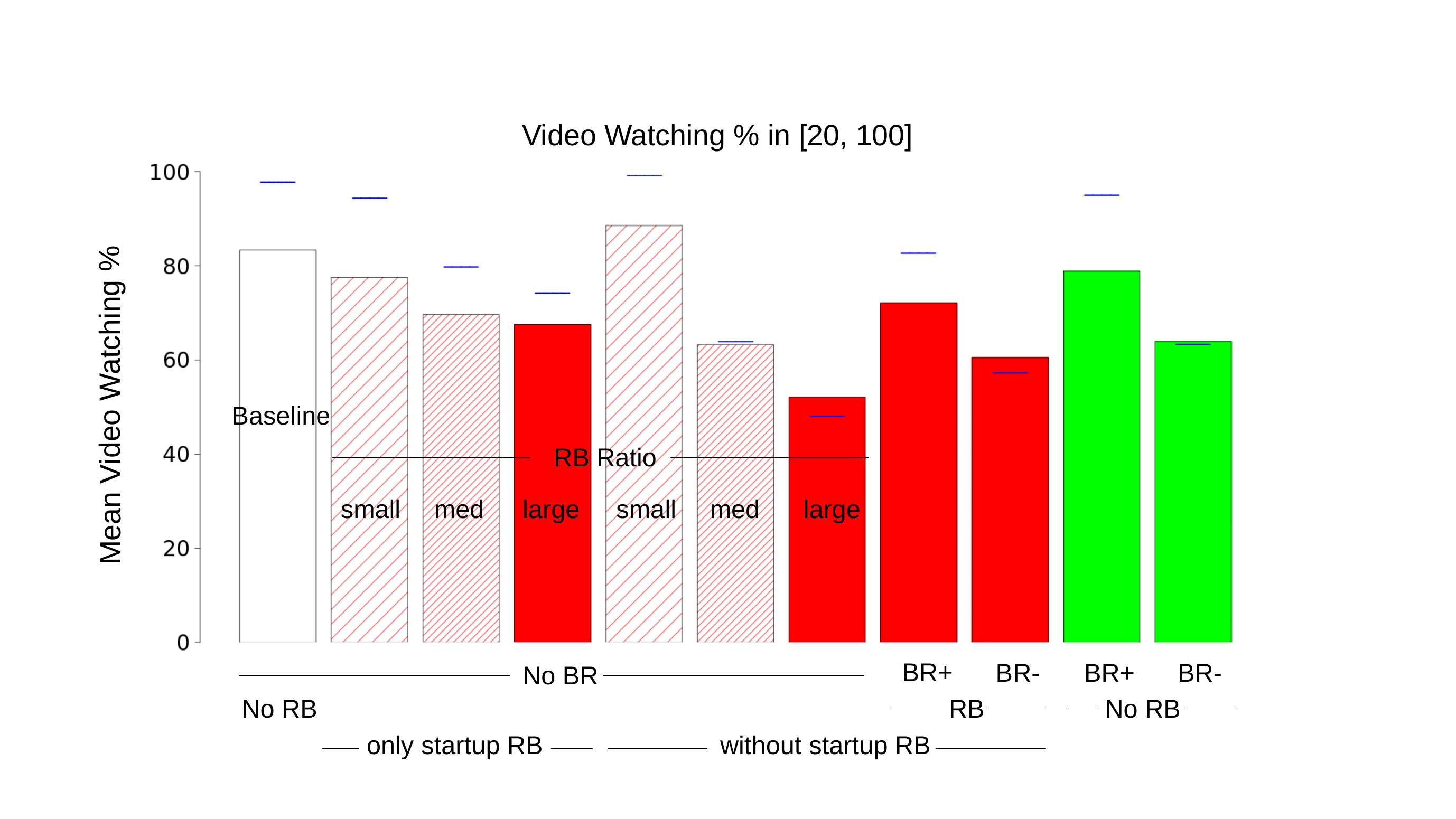}}
\hspace*{-1.0cm}
\subfloat{\label{mean_weighted_res_20_RBratio_main6sc}\includegraphics[width=0.60\textwidth, height=60mm]{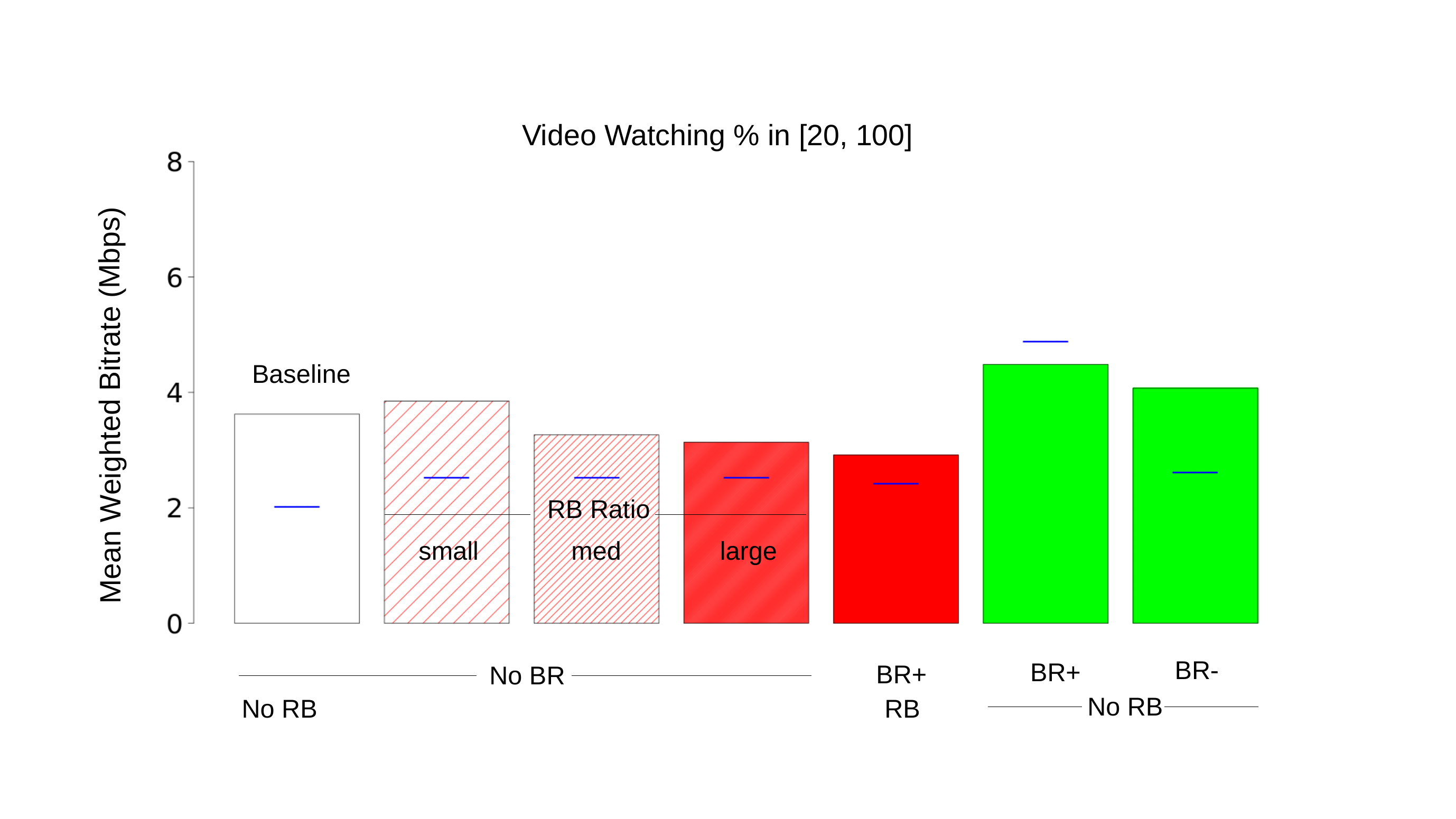}}
\caption{The video watching percentage (right) and mean weighted resolution (left) for different RB ratios. The small RB ratio corresponds to $\le 0.05$, the medium RB ratio to (0.05, 0.2], while the large RB ratio of 0.2 or more. 
The blue horizontal line indicates the median.}
\label{fig:vwp_and_mwr_50_RBratio_startup_sc}
\end{figure*}

For sessions with exceptionally {\em high RB ratio}, larger than 0.2 (that corresponds to a median RB duration of 111 sec), the mean video watching percentage becomes lower than this of the BR- scenarios. In general, only when RB ratio is very large, its impact becomes more noticeable than BR- impact (e.g., Fig. \ref{fig:vwp_and_mwr_50_RBratio_startup_sc}).
All scenarios with low or no RB have significant statistical differences with all others. Interestingly, the scenarios with high RB do not exhibit significant statistical differences with each other, even in the case of low vs. high resolution, in which the mean weighted resolution and the RB duration are statistically significant higher and lower, respectively. This could be due to the {\em manual} user selection of lower resolution. Similar observation can be made for the high RB/lower resolution vs. high RB/medium resolution scenarios. The high RB/low Res sessions include a relatively large percentage of sessions with BR changes that occur shortly after the end of RB. We speculate that the viewer may manually request the BR adaptation after such long RB, especially in cases of high interest in the content (as reflected by the large video watching percentage). Interestingly, although the high RB/high Resolution sessions have statistically significant higher mean weighted resolution and lower mean/median RB duration than the High RB/low resolution, their difference in video watching percentage is not sTtatistically significant. As in the case of high RB/low resolution vs. high RB/medium resolution, this could also be the result of the manual user selection of a lower resolution.

\subsection{On the time of the occurrence of a rebufferring}
\begin{figure*}[t!] 
\centering
\subfloat{\label{vwp20_starting_time_RB_Scatterplot_RBdur}\includegraphics[width=0.34\textwidth, height=45mm]{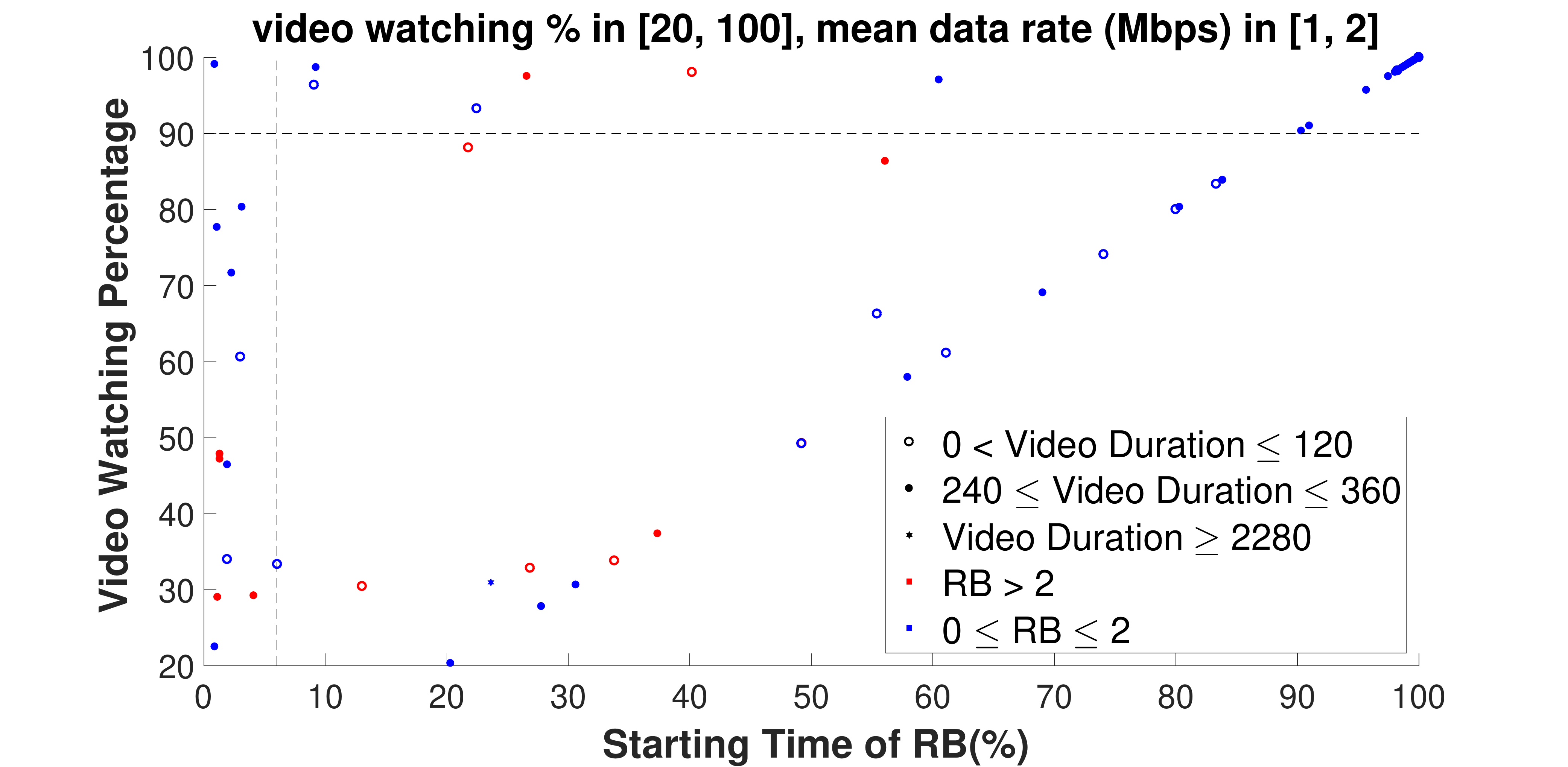}}
\subfloat{\label{vwp20_starting_time_RBdur_with_startup_delays}\includegraphics[width=0.34\textwidth, height=45mm]{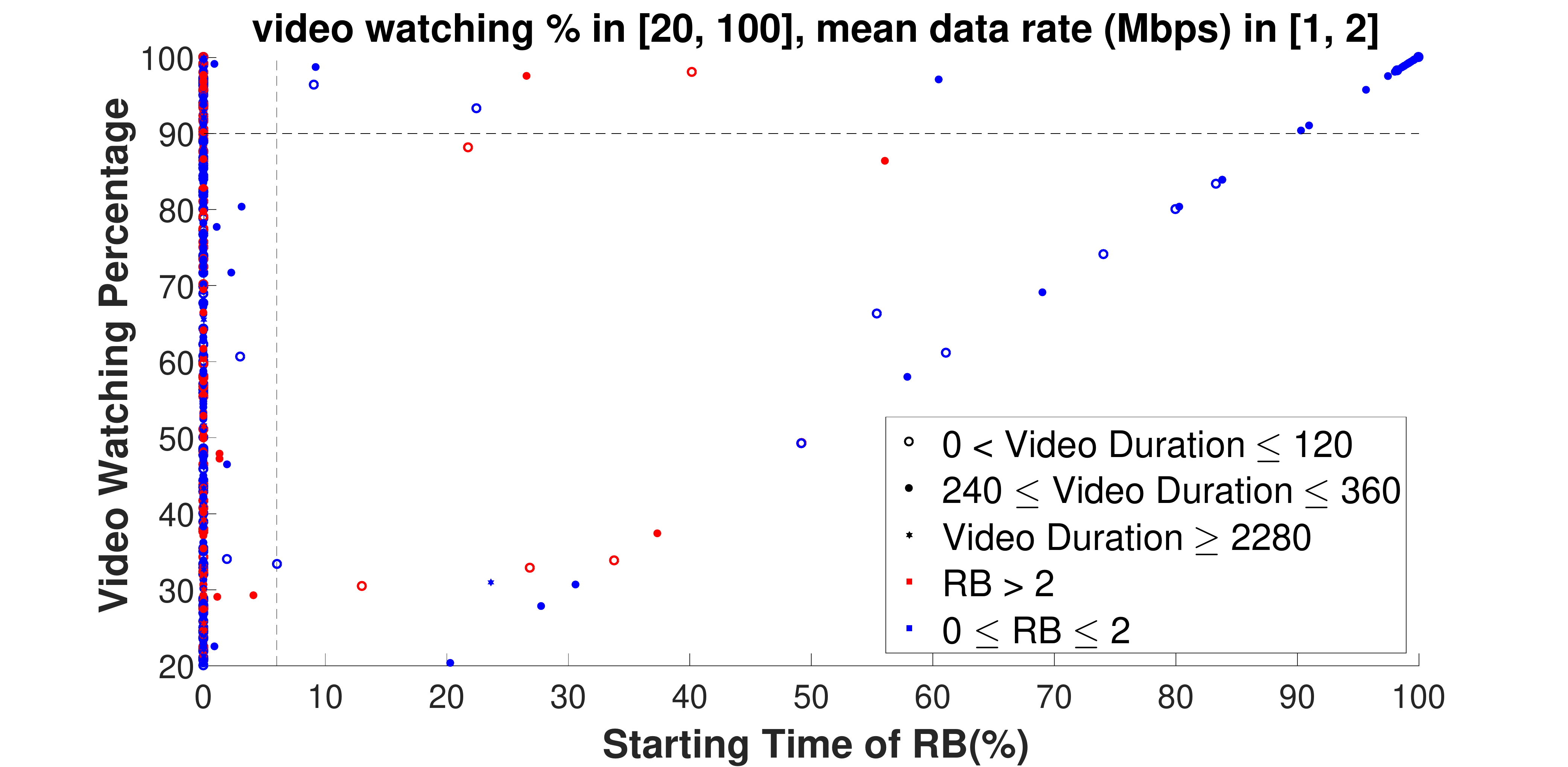}}
\subfloat{\label{vwp20_starting_time_BR_Scatterplot}\includegraphics[width=0.34\textwidth, height=45mm]{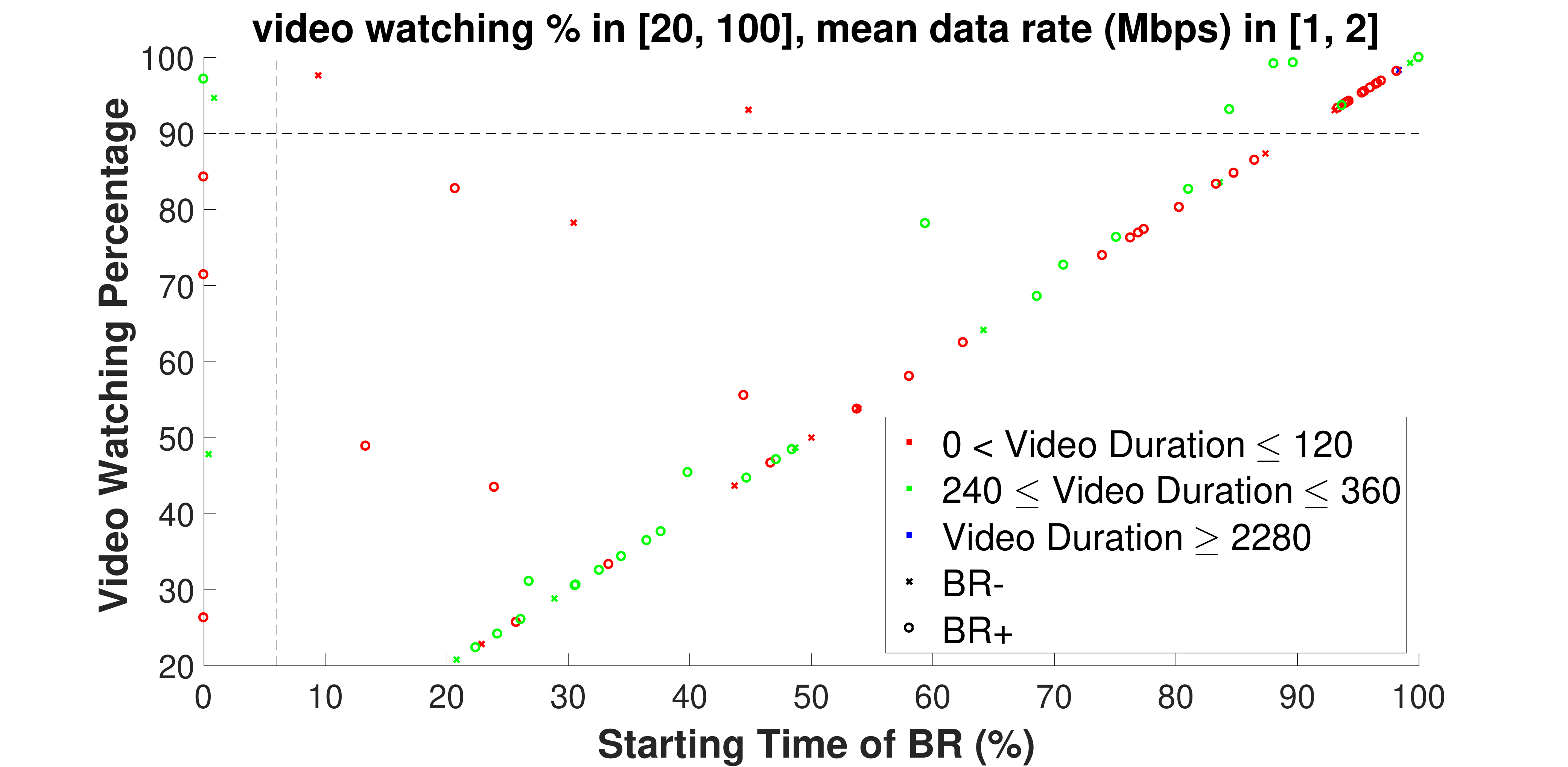}}
	
\subfloat{\label{vwp50_starting_time_RBdur_with_startup_delays}\includegraphics[width=0.34\textwidth, height=45mm]{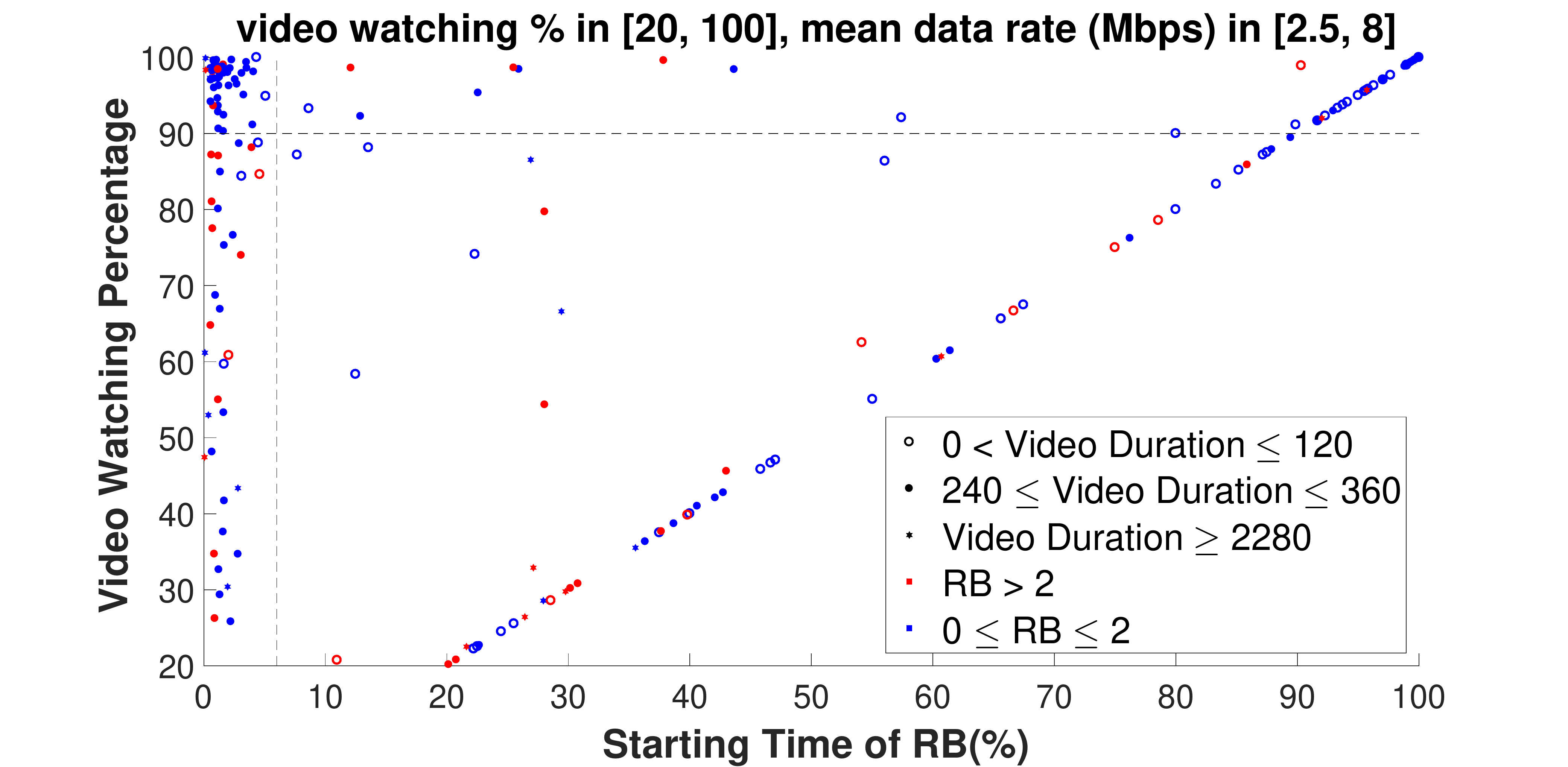}}
\subfloat{\label{vwp50_starting_time_BR_Scatterplot}\includegraphics[width=0.34\textwidth, height=45mm]{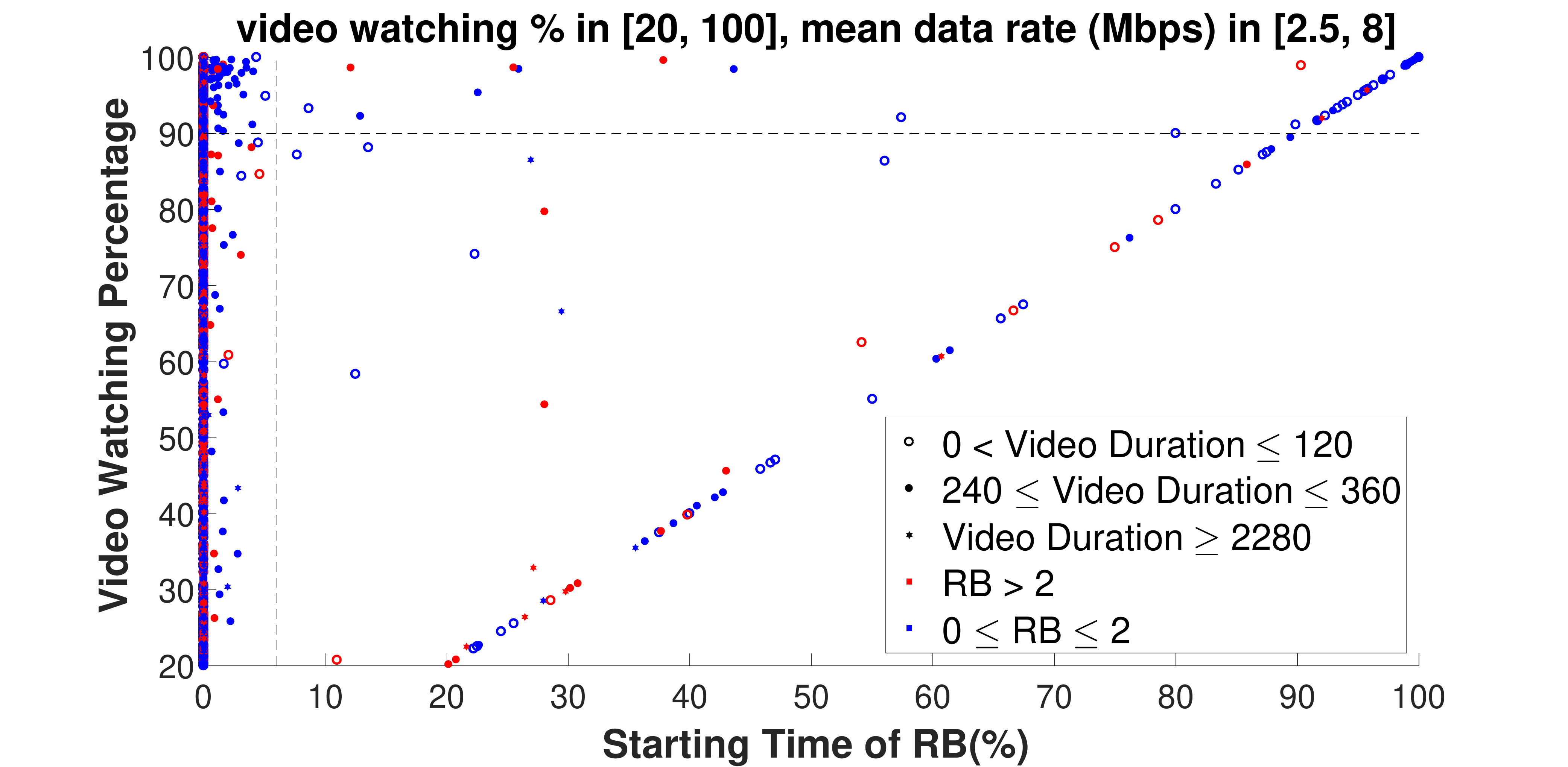}}
\subfloat{\label{vwp50_starting_time_RB_Scatterplot_RBdur}\includegraphics[width=0.34\textwidth, height=45mm]{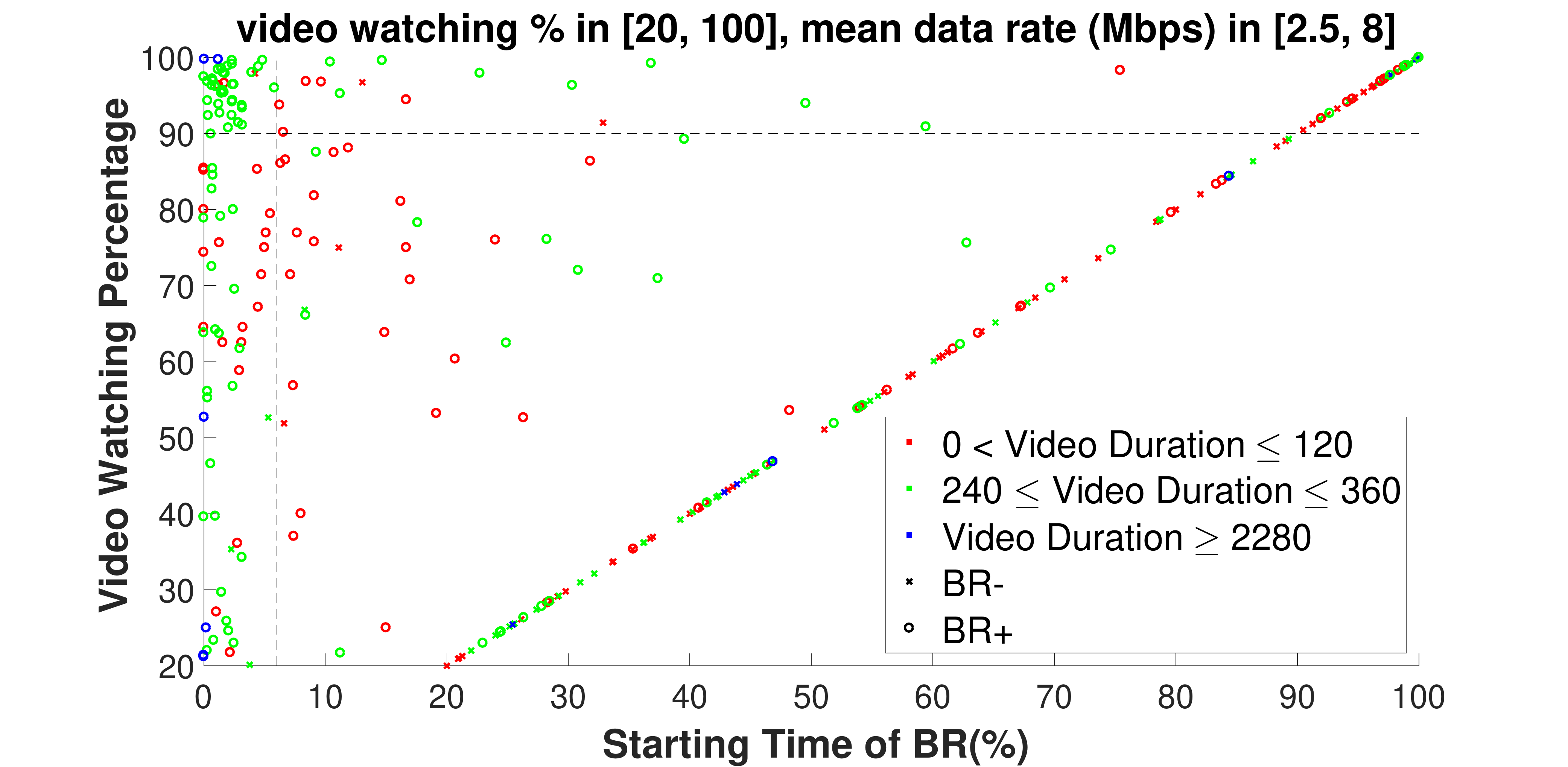}}

\subfloat{\label{vwp50_starting_time_RBdur_with_startup_delays}\includegraphics[width=0.34\textwidth, height=45mm]{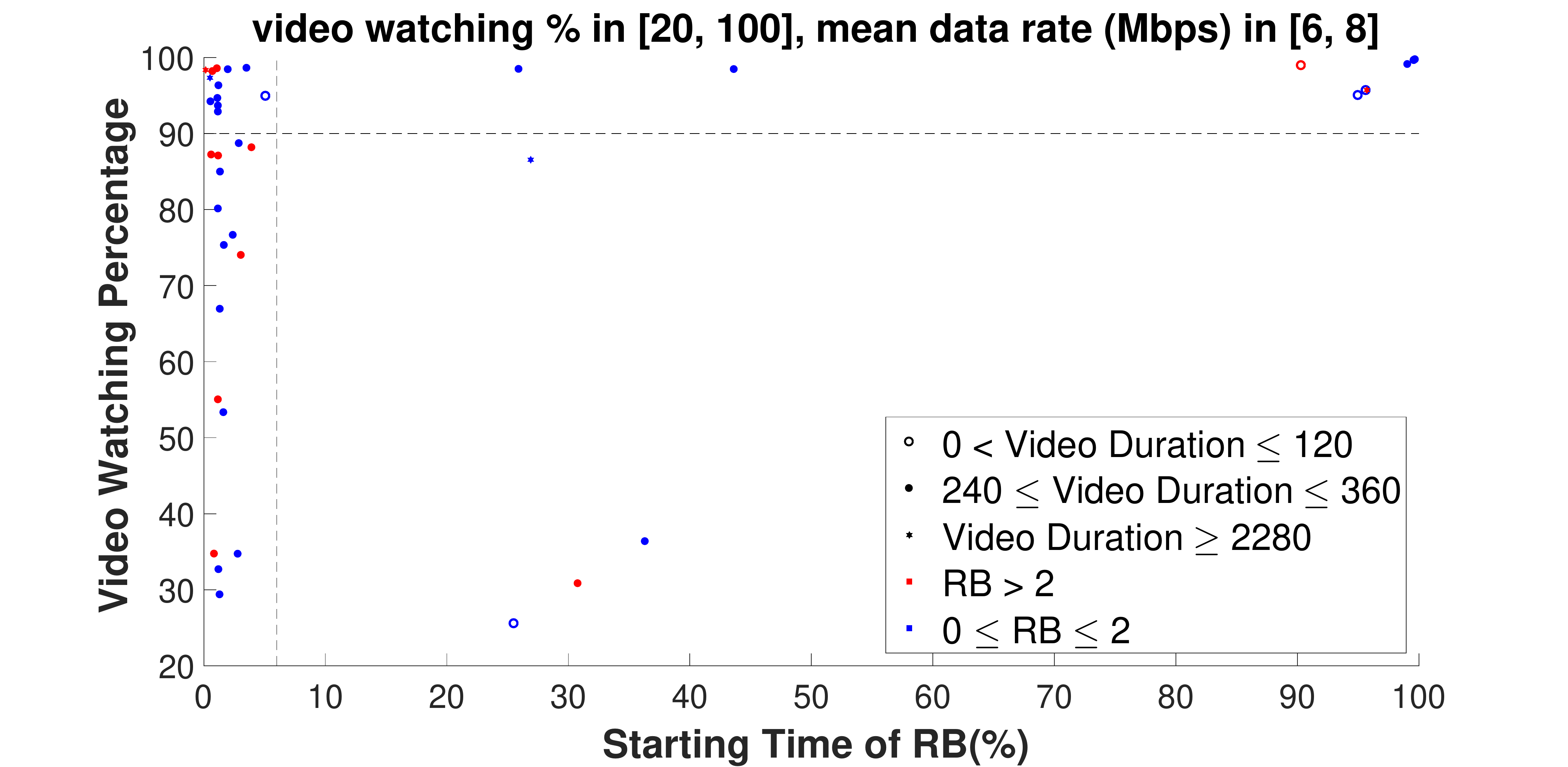}}
\subfloat{\label{vwp50_starting_time_BR_Scatterplot}\includegraphics[width=0.34\textwidth, height=45mm]{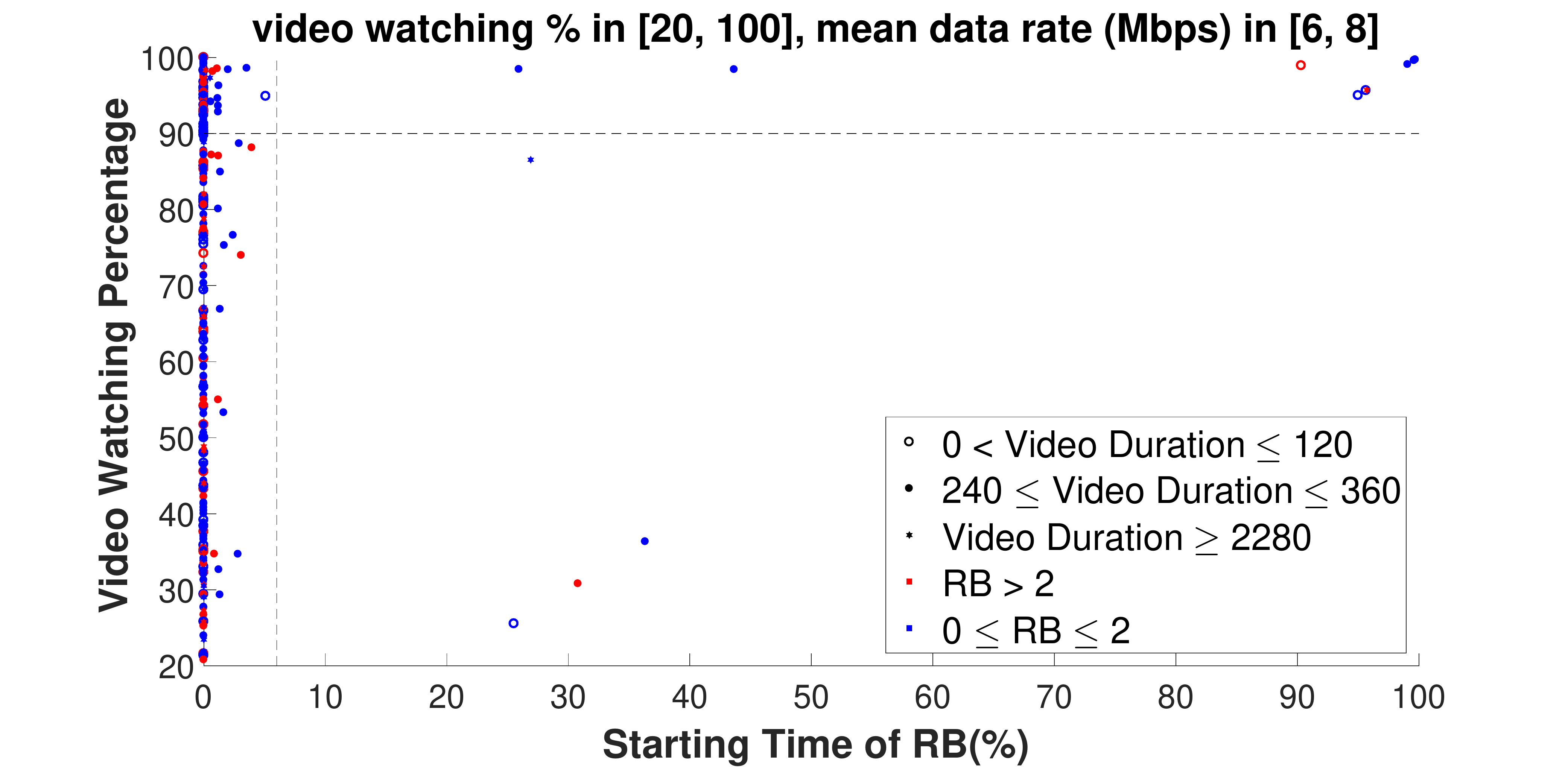}}
\subfloat{\label{vwp50_starting_time_RB_Scatterplot_RBdur}\includegraphics[width=0.34\textwidth, height=45mm]{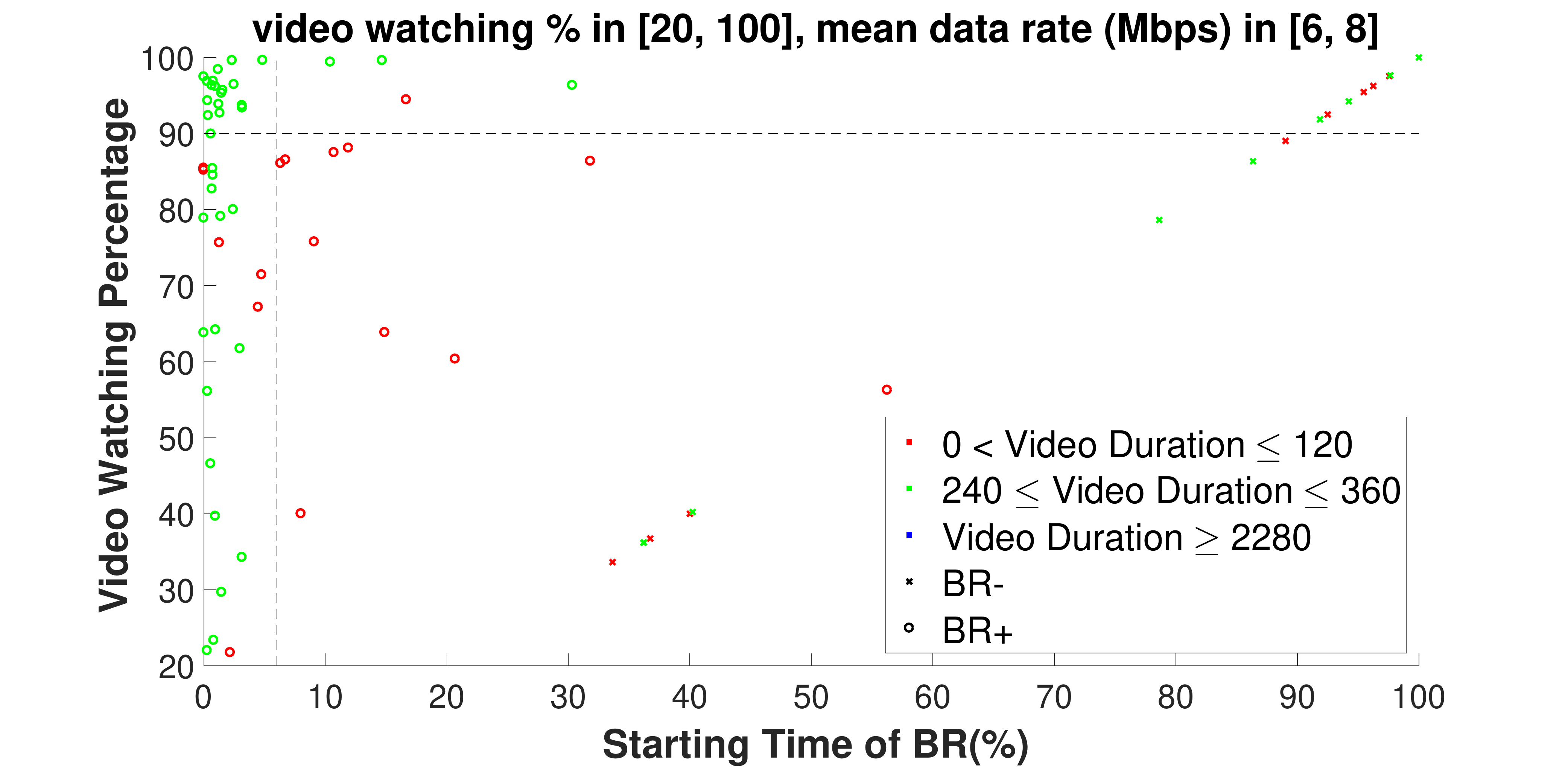}}
\caption{The video watching percentage and {\em starting time} of the impairment (RB or BR), for all {\em abandoned} sessions with ($2^{\mathrm{nd}}$ column) and without ($1^{\mathrm{st}}$ and $3^{\mathrm{rd}}$ columns) startup delay, a {\em single impairment}, and video watching percentage more than 20\%.}
\label{fig:vwp_starting_time_p_RB_scatterplot_RBdur}
\end{figure*}

Compared to startup delays, RBs have a larger impact on the video watching percentage (as shown in Fig. \ref{fig:vwp_and_mwr_50_RBratio_startup_sc}). Other studies had also reported similar trends (e.g., Hossfeld {\em {\em et al.}} \cite{hossfeld2008testing}). 
For video watching \% in [20,100], sessions with large RB ratios in which the RB occurs early have higher mean video watching percentage (67\%) than sessions with large RB ratio that occur later during the session (52\%). 
Two interesting populations of sessions are present in the dataset with respect to the time that a RB occurs and the video watching percentage: one in which the RB occurs very early on (e.g., startup delay), without affecting the video watching percentage and a second one in which users are really sensitive to RBs and abandon the session within 60 sec from its occurrence (following a linear trend, Fig. \ref{fig:vwp_starting_time_p_RB_scatterplot_RBdur}). 
The majority of sessions in the second population are of small video duration. The difference between these two scenarios is more prominent in the case of video watching percentage in [50,100] (95\% vs 74\%). Furthermore, as expected, the higher the mean data rate, the thinner the linear trend (i.e., the smaller the size of the second population). Most of the RBs are startup delay events with small impact on the video watching percentage.
The reverse occurs for low mean data rates: The majority of the RB events occur at the start of the sessions and their video watching percentage varies.
The mean weighted mean data rate of the first population is slightly higher than the one with the linear trend, while the median is the same (2.5Mbps).  Approximately 76\% of the sessions with a single impairment, a RB, are abandoned, and a little more than half of them are abandoned within the first minute (Table \ref{tbl:1impairnment:abnd:prob}).


\subsection{Features with large predictive power on video watching percentage} 
\begin{figure*}[t!] 
  \centering
\subfloat{\label{abnd_vwp_50_main6sc_vdur}\includegraphics[width=0.9\textwidth, height=60mm]{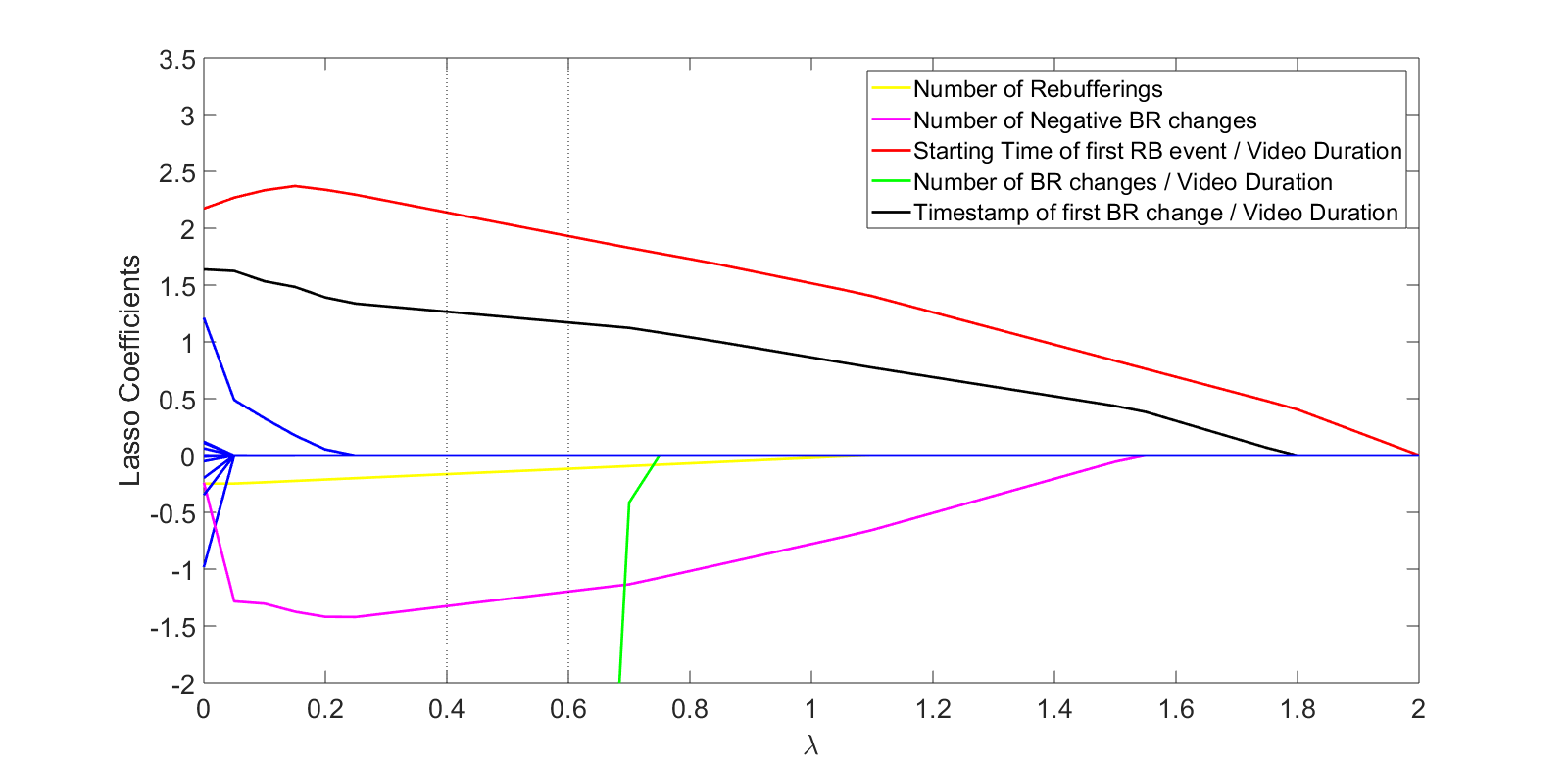}}
\caption{Fitted lasso coefficients against $\lambda$. The vertical dotted lines at $\lambda$ = 0.4 and 0.5 indicate the interval of $\lambda$ values that yield the smallest cross-validation error. The blue lines, that are not described in the legend, represent the parameters with zero LASSO coefficients for $\lambda < 0.4$.}
\label{fig:lasso}
\end{figure*}

We considered a set of 37 parameters based on RB duration, number of RB events, data rate, number of BR- and BR+ changes that characterize the sessions. In order to identify the parameters with the largest predictive power on video watching percentage, we applied 10-fold cross-validation LASSO regression and estimated their coefficients. Specifically, for a wide range of values of $\lambda$ in [0,2]\footnote{For $\lambda$=2, all LASSO coefficients become zero.}, we ran the 10-fold cross-validation and estimated the value of $\lambda$ that yields the smallest cross-validation error. We repeated this process several times and each time obtained a $\lambda$ in [0.4, 0.6]. After that, we examined the derived plot (Fig \ref{fig:lasso}) for $\lambda$ in [0, 2] with step size of 0.05. The dominant parameters are the ones with non-zero LASSO coefficients for $\lambda$ in [0.4, 0.6]. For $\lambda$  in the aforementioned interval, the dominant parameters, namely the number of BR changes per video duration, starting time of first RB event per video duration, timestamp of first BR change per video duration, number of negative BR changes, number of rebufferings (RBs), remain the same.

%

\subsection{Interplay of rebuffering events and bitrate changes}
Rebufferings and BR changes in sessions are not independent but may occur jointly in sessions as a result of the Adaptive Bitrate (ABR) algorithm or a viewer-initiated request for resolution change. 
%
The ratio of sessions with a positive BR change that is followed by a rebufferring within 30 sec over the total number of sessions with a positive BR change is lower than the number of sessions in which a positive BR change follows a RB within 30 sec of its occurrence over the total number of sessions with at least a positive BR change (3.7\% vs. 8.4\%). Similar trend exists for BR-. The number of sessions with a RB followed by a BR- within 30 sec is larger than the number of sessions with at least a BR- followed by a RB within 30 sec.\footnote{5.7\% of sessions with a RB followed by a BR- within 30 sec over the total number of sessions with a BR- compared to the 1.7\% of sessions with at least a BR- followed by a RB within 30 sec over the total number of sessions with BR-.}

We define the early negative bitrate change as the BR- that occurs {\em shortly after the end of a rebuffering}, while late bitrate changes follow {\em much later} (i.e., 30 sec or more) the end of a rebuffering event.
Note that the total number of such sessions is small: only 39 sessions of early BR- and 11 sessions of late BR- exist in the dataset. 
We speculate that a late bitrate change is perceived as a more severe impairment than an early bitrate one, which could be viewed as a direct consequence of the rebuffering event that just occurred. That is, users may perceive the rebuffering and the negative bitrate change that follows shortly after as a single impairment, instead of two distinct impairments. On the other hand, it is more likely that a late bitrate change will be perceived as a new impairment. It may be also more likely that the network impairment will cause the user to manually downgrade the resolution immediately after a rebuffering (through which the impairment becomes apparent) than much later after a RB. Given that the participants of the YouSlow study had to download and run a Chrome plugin, we speculate that most of them were computer savvy. As an attempt to improve their viewing experience, at the face of the aforementioned network impairments, these users might have decided to manually downgrade the resolution.

To assess the aforementioned hypothesis, the user engagement metrics have been computed for sessions with early bitrate changes and late bitrate changes. As before, the analysis focuses on sessions where at least half of the video has been watched. Indeed early bitrate changes have lower abandonment ratios than late bitrate changes. Similarly, the video watching percentage is higher for sessions with early bitrate changes than for late ones (for all different video durations).  Note that the total number of sessions with early or late BR change and RB is relatively small. Similar trend exist also in the second dataset. We noticed that for the sessions with a rebuffering duration of 10 sec or more (332 sessions), only the 6\% of them had negative BR change. We speculate that the automatic bitrate streaming adaptation was disabled in these sessions.

Compared to the early BR change sessions, the late ones have a significantly smaller percentage of sessions that ended normally (i.e., not abandoned). The majority of abandoned sessions with early bitrate changes ended with user-initiated abandonment, while most of the abandoned sessions with late bitrate changes were abandoned within 5 sec after a BR change. This is consistent with our speculation that the late bitrate changes are perceived as a more severe impairment than early bitrate changes. Users seem to be more tolerant with the early BR- change than the late one,
even though these sessions end with lower bitrate and slightly larger BR change 
than the late BR- one
. We speculate that users perceive the rebuffering event and the early negative BR change as a single degradation.

\subsection{Memory of the impairments, when the last impairment is BR-}
Does an earlier impairment affect how the viewer perceives a followup impairment during a session? To answer this question, we focused on two scenarios: the first scenario included all sessions with {\em exactly one} impairment, specifically, a BR-. The second scenario includes all sessions with {\em exactly two impairments}, where the second impairment is exactly a BR-, while the first impairment can be a RB or BR (positive or negative), but without a startup delay. A time margin between the occurrence of two impairments greater than or equal to 30 sec was also employed for ensuring that the user perceives them as two distinct impairments. 

\begin{figure*}[t!] 
\centering
\subfloat {\includegraphics[width=0.5\textwidth, height=45mm]{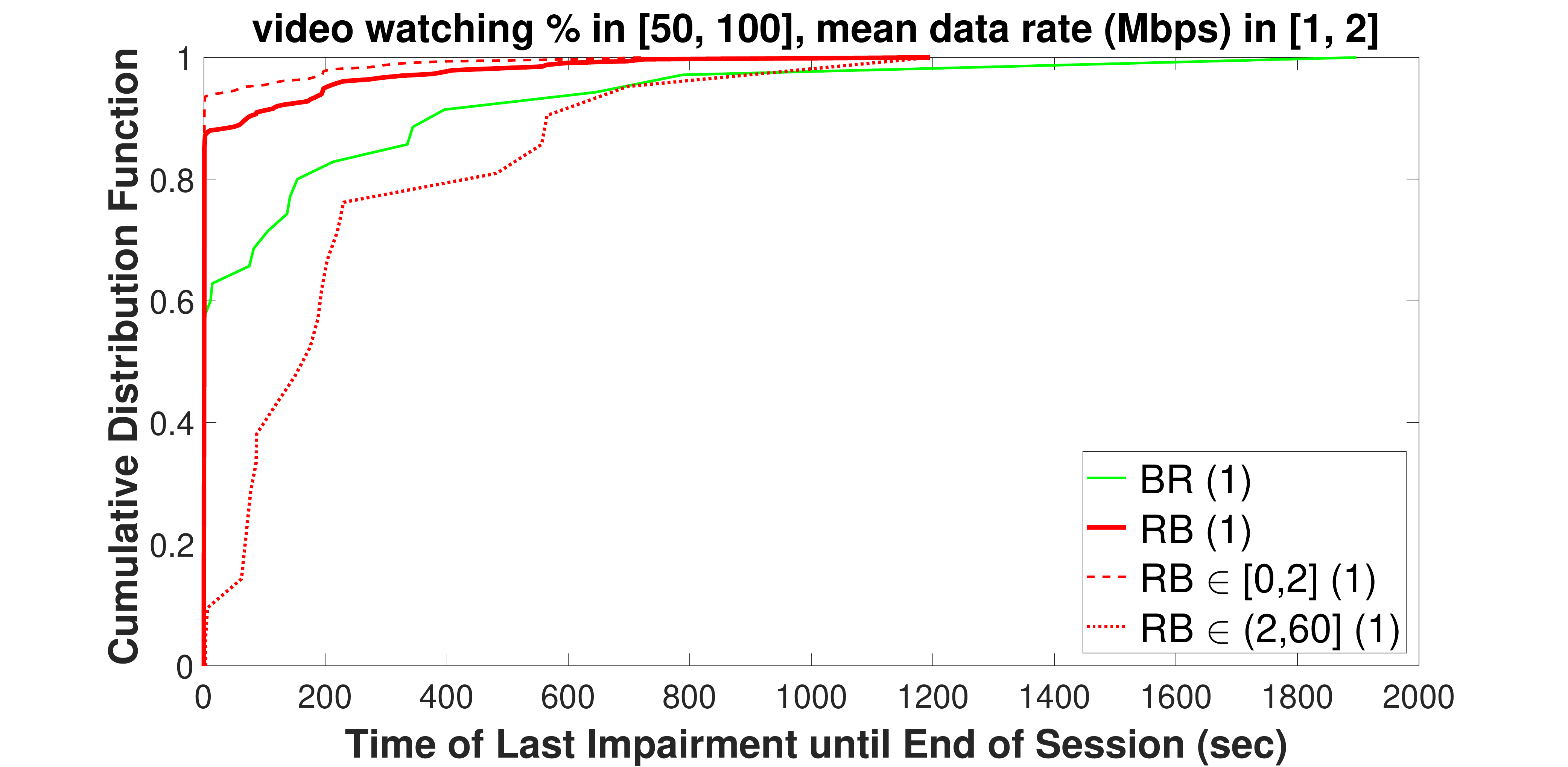}}
\subfloat {\includegraphics[width=0.5\textwidth, height=45mm]{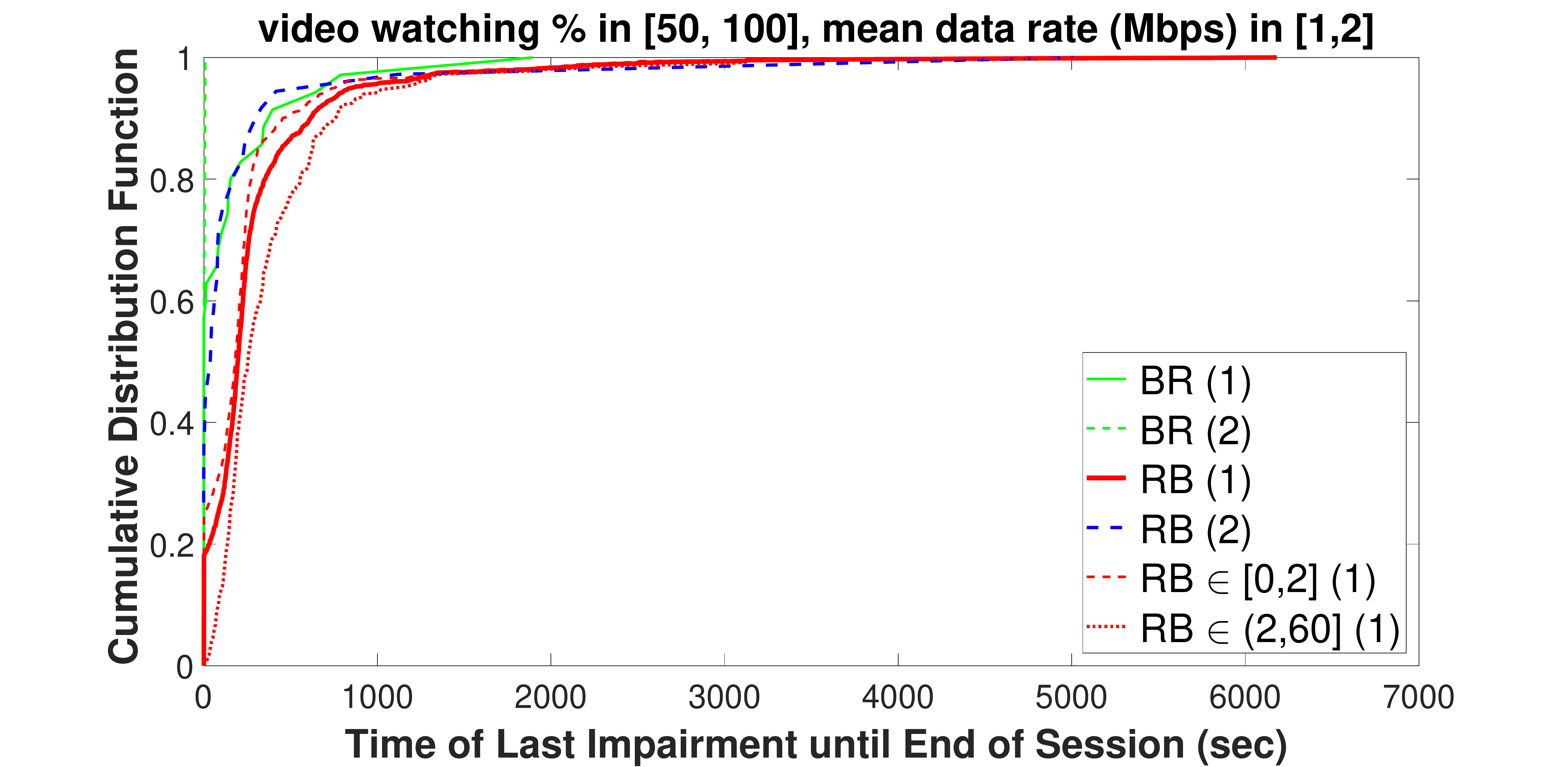}}
  
\subfloat{\label{ECDF_for_2vs1_impairments_start_time_until_end_session_without_startup_delays}\includegraphics[width=0.5\textwidth, height=45mm]{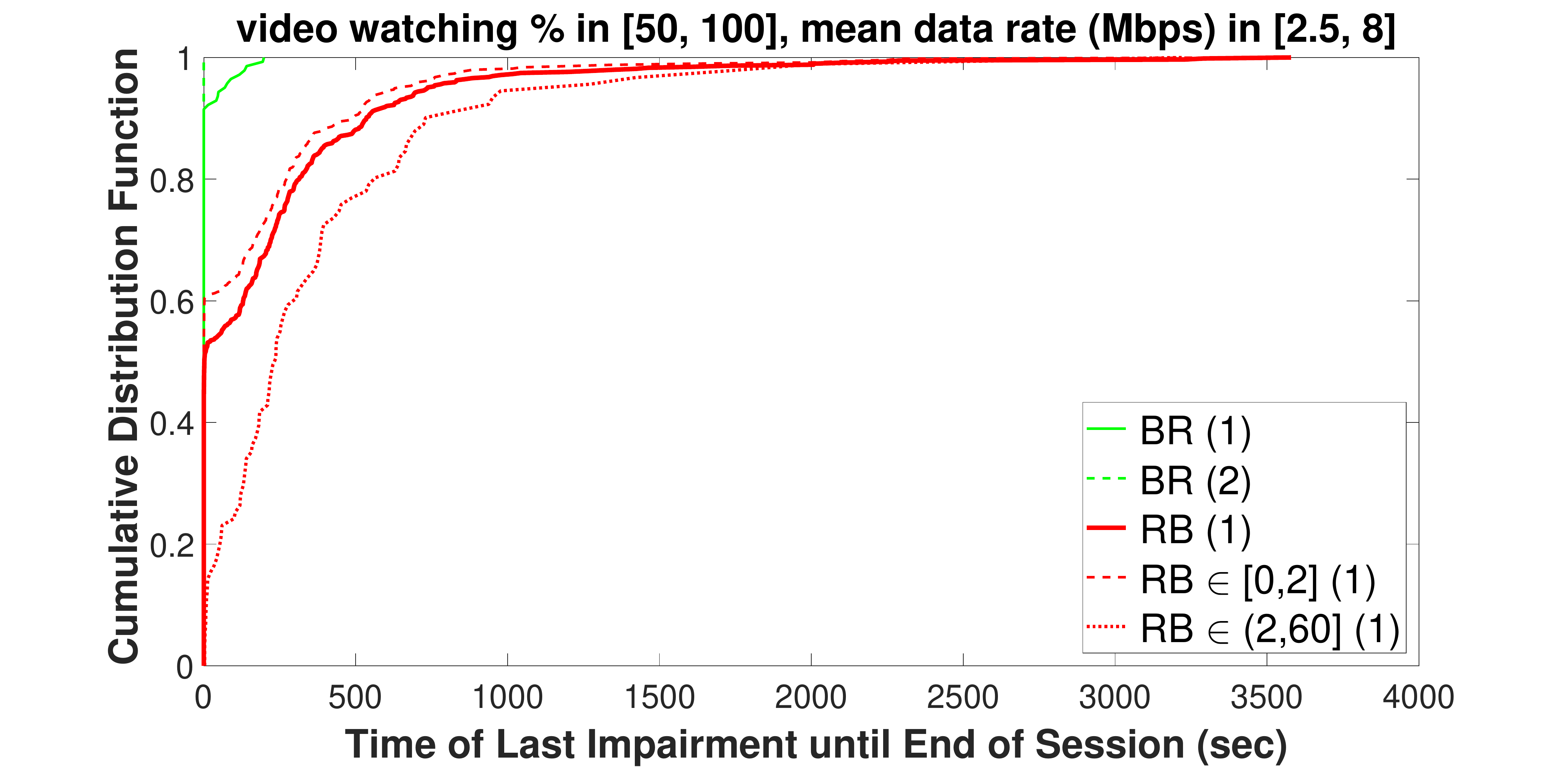}}
\subfloat{\label{ECDF_for_2vs1_impairments_start_time_until_end_session_with_startup_delays}\includegraphics[width=0.5\textwidth, height=45mm]{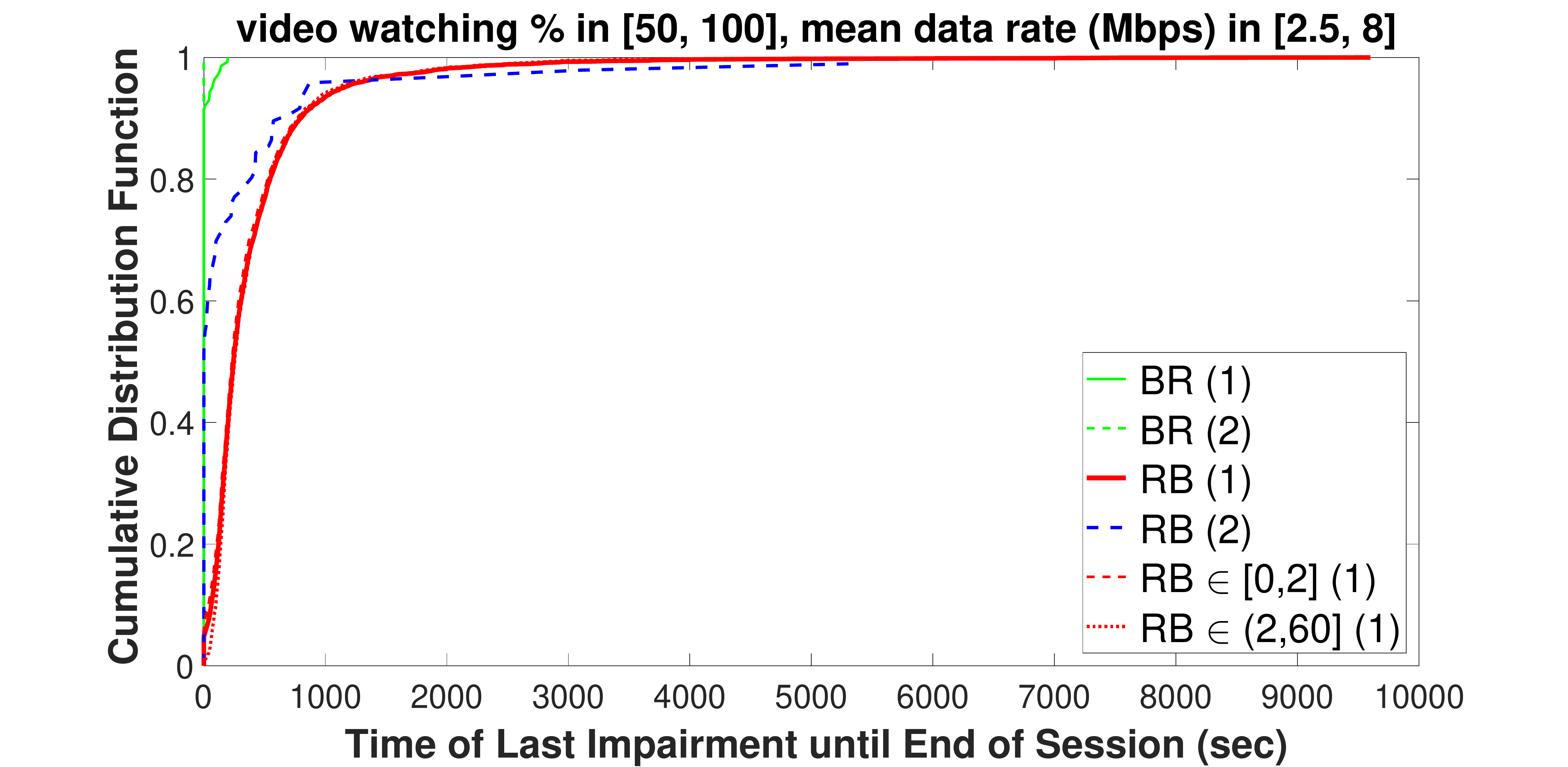}}\\
\addtocounter{subfigure}{-4}
\subfloat[]{\includegraphics[width=0.5\textwidth, height=45mm]{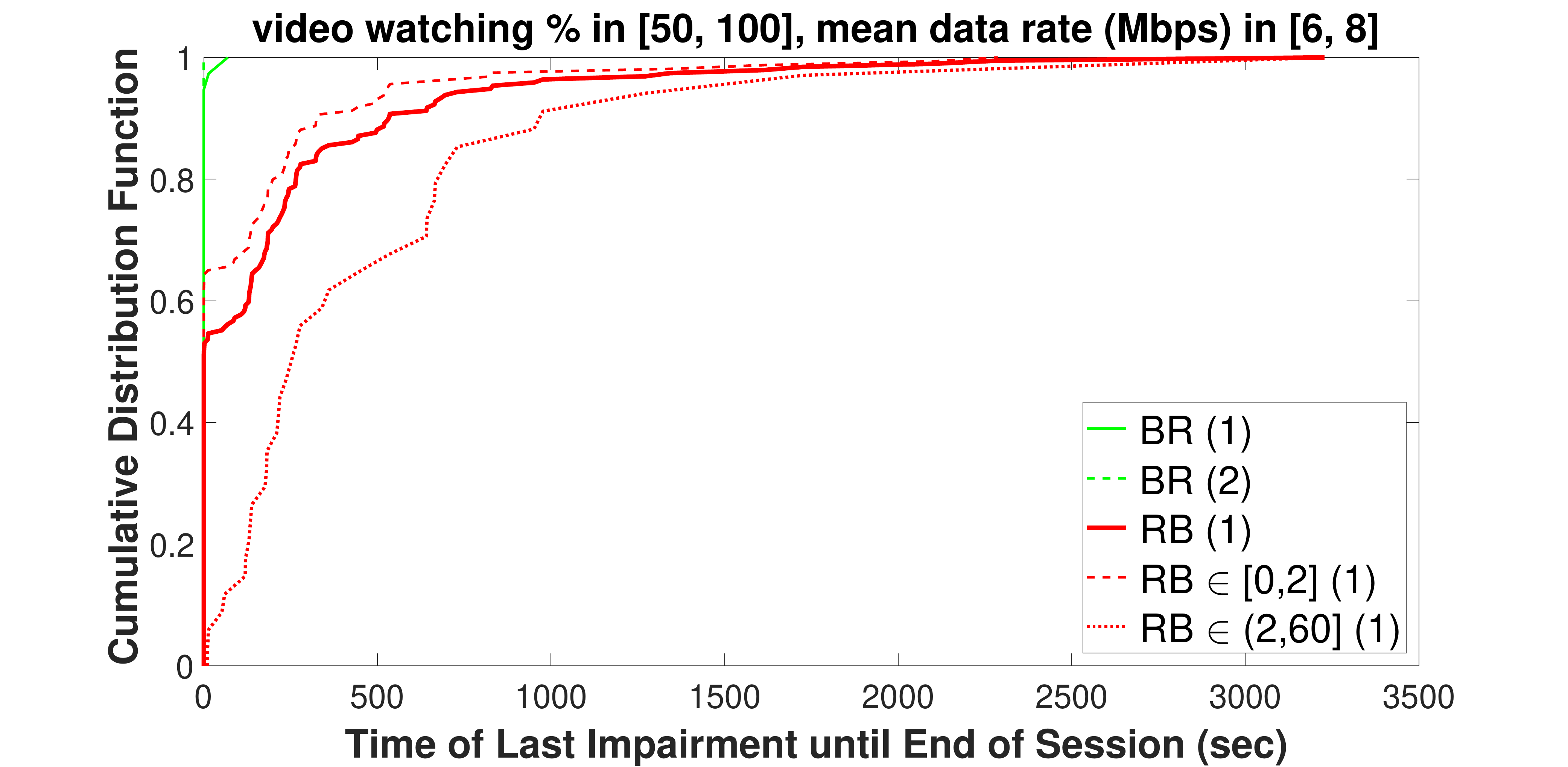}}
\subfloat[]{\includegraphics[width=0.5\textwidth, height=45mm]{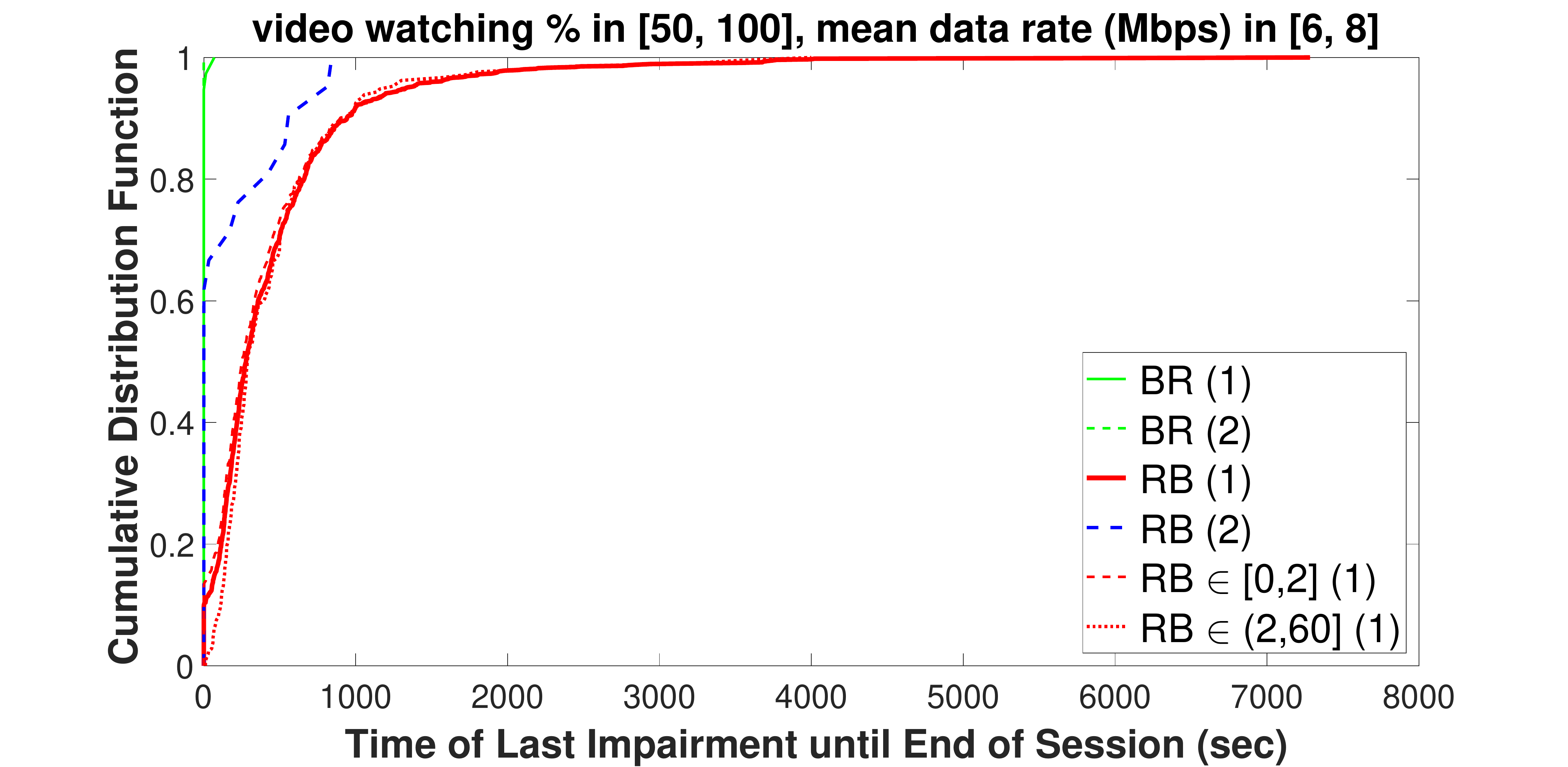}}  
\caption{The time lag between the {\em start} of the impairment until the end of session, considering a time margin of at least 30 sec between impairments, for all abandoned sessions, without and with startup delay in (a) and (b), respectively. The number in parenthesis next to the impairment type (RB, BR) indicates the number of impairments, while the interval corresponds to the RB duration.}
\label{fig:ECDF_for_2vs1_impairments_start_time_until_end_session}
\end{figure*}

We first considered sessions with weighted mean data rate in the interval of [1Mbps, 8Mbps]. It does appear that there is memory in how the BR- impairments are perceived: a user that experiences a BR-, after experiencing an impairment earlier in that session, tends to become more prone to abandon the session, than the user that experienced a BR- for the first time, without also having experienced any other impairment during that session. The time elapsed from the occurrence of the second impairment (the BR-) until the abandonment of the session is smaller than the time elapsed from the time when the BR- occurred to the abandonment of the session for sessions with exactly one impairment. Moreover, the probability of an abandonment is 94.12\% and 96.95\% for the sessions with one and two impairments, respectively. Sessions with two impairments tend to get abandoned earlier, after the occurrence of the second impairment, than sessions with only one BR- (Fig. \ref{fig:ECDF_for_2vs1_impairments_start_time_until_end_session}). 
Their difference was statistical significant according to the non-parametric two-sample Kolmogorov-Smirnov test for video watching percentage above 50\% and 20\%, while for the 120 sec threshold, we can not reject the null hypothesis for a significance level $\alpha=0.05$.

Interestingly, in the case of sessions with mean weighted data rate in the interval of [2.5Mbps, 8Mbps], abandoned or not abandoned, there is no statistically significant difference from the time elapsed from the occurrence of the last (second) BR- until the end of the session between sessions with exactly two impairments with the last one to be a BR- and sessions with exactly one impairment, which is a BR-.  In all the above cases, the percentage of sessions abandoned by users when there are exactly two impairments is greater than when there is only one. The number of sessions with two impairments, where the last one is RB is small to be able to make conclusive statements.

\subsection{Multiple positive bitrate changes}
It is evident that the No RB/BR+ with two or more bitrate changes has significantly larger mean weighted data rate than the scenarios with only one BR+. The mean weighted data rate in the case of multiple BR+ is 5650 kbps, while in the case of only one BR+, it is 4206 Kbps. In the baseline scenario (no RB/no BR), it is 4630 Kbps. The differences are statistically significant (as reported by the Kolmogorov-Smirnov test).
We speculate that there are two underlying populations of sessions: (1) sessions generated by users with high speed connections, during which the player starts with relatively low default video quality and dynamically adapts to higher data rates, and (2) the ones with lower speed (even lower than the sessions of the baseline scenario, i.e., of no RB and no BR change).

Users from the first population face more positive bit rate changes but are less irritated by them, given that they tolerate (or are accustomed to) the data rate adaptation of the player, resulting to lower abandonment rate, than the users of the second population. 

To examine the abandonment under BR+ and the influence of the mean data rate, we focused on sessions with mean weighted data rate above 4640 Kbps (mean weighted data rate of the baseline scenario) as well as the ones with lower data rate, namely in the interval [2.5, 4.64) Mbps. For sessions with one BR+, the lower the mean data rate, the higher the probability of abandonment within 60 sec from the occurrence of the BR+. Specifically, for sessions of mean date rate in [2.5-4.64] Mbps,  this probability is 0.71, while for sessions of mean data rate larger than 4.64 Mbps, this probability becomes 0.18. For sessions with {\em multiple BR+} and mean date rate in [4.640-8] Mbps, the probability of abandonment within 60 sec from the occurrence of the BR+ is 0.54.
Note that there are very few sessions with multiple BR+ and mean data rate in the interval [2.5, 4.64M) Mbps.
In addition, we examined the distribution of the time when the BR+ occurred in the cases of multiple vs. single BR+. Even though, in the single BR+ change scenario, the distribution is almost uniform, the multiple bitrate changes scenario seems to have a larger number of BR+ changes in the first half of the video. The Kolmogorov-Smirnov test for the two distributions rejected the null hypothesis.
\begin{figure*}[t!] 
  \centering
  \subfloat[]{\label{vwp_bsl_BR-_BR+_ecdf}\includegraphics[width=0.5\textwidth, height=50mm]{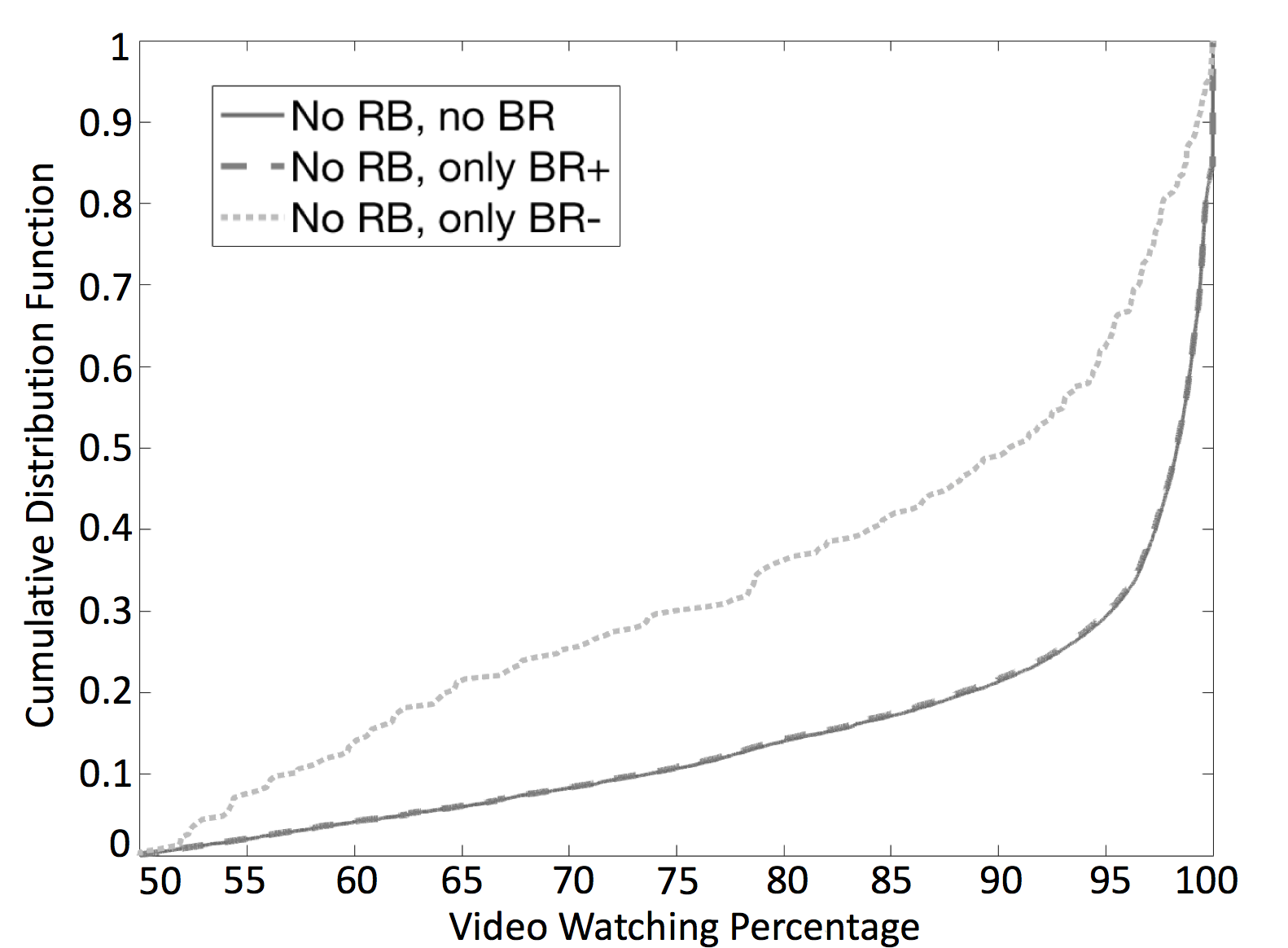}}
\subfloat[]{\label{1_multiple_BR+_starting}\includegraphics[width=0.5\textwidth, height=50mm]{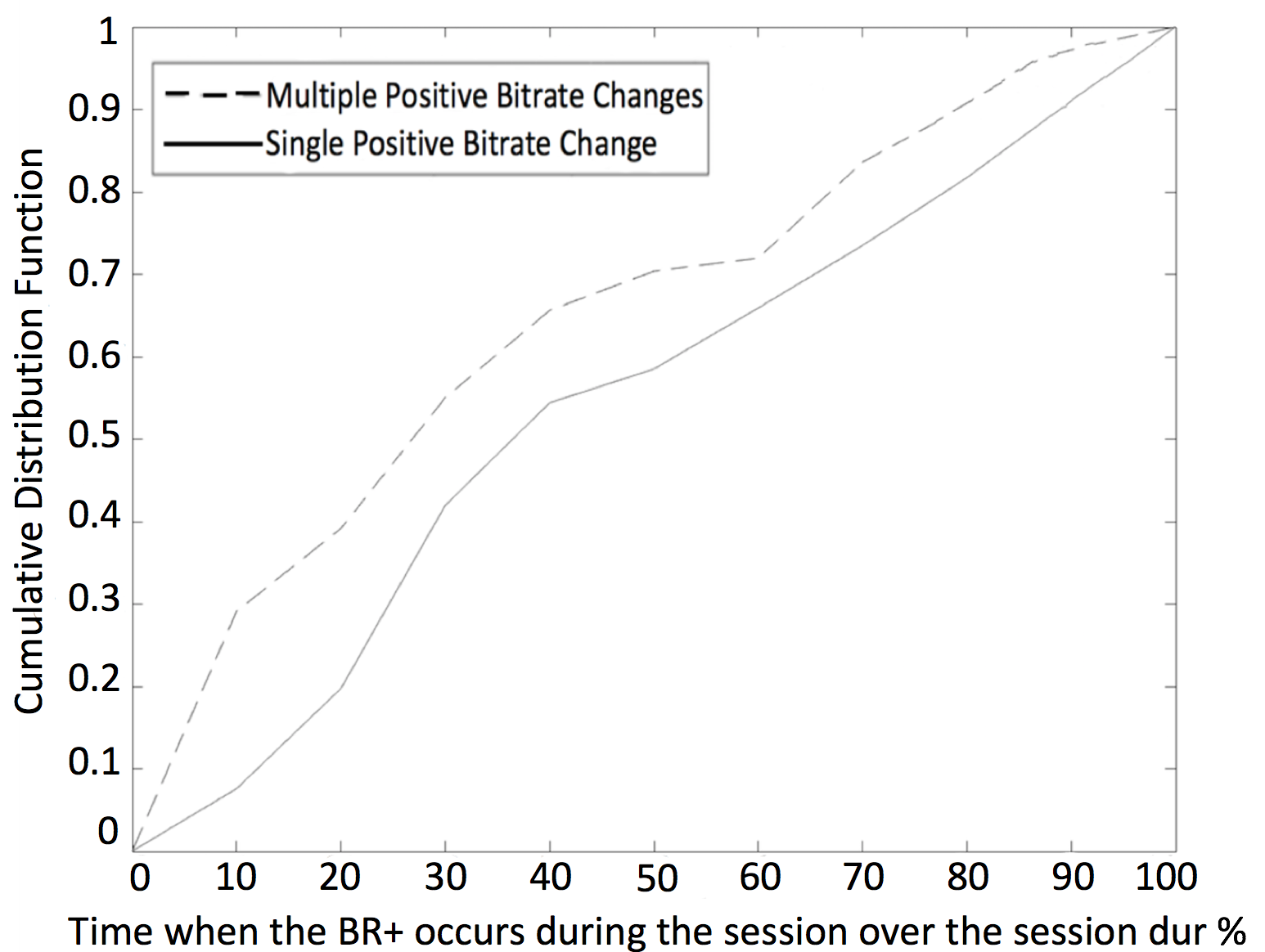}}
\caption{Left: ECDF of video watching percentage for sessions with no RB and no BR vs. no RB and only BR+ vs. no RB and only BR-. Right: Time when the BR+ occurs during the session, in sessions with exactly one impairment, a BR+ vs. multiple BR+, considering sessions with video watching \% in [50,100] and mean weighted data rate (Mbps) in [2.5, 8].}
\label{fig:1_multiple_BR+_starting}
\end{figure*}

\section{Related Work}\label{sec:rel}
Monitoring tools that collect various QoS measurements, such as player state, statistics of buffering events, and video quality level have been developed \cite{chen2014qoe, wamser2015yomoapp,nam16,nam-rep16,plakia2016videoqoe}. The analysis of collected QoS measurements to infer the QoE of video streaming, and specifically of YouTube, a major source of traffic worldwide, has been presented in \cite{nam16, nam-rep16, hoss13, smodovan16, Orsolic-2017, Casa17}. The impact of various mechanisms and infrastructures for optimizing the QoE has been the focus of \cite{smodovan16,Argy18}. 
In general, the ground-truth for the QoE has been formed based on the explicit opinion scores reported by users (e.g., in the context of audiovisual tests or at the end of their service via a GUI), based on measurements collected using physiological metrics~\cite{wilson2000users, mandryk2006using}, according to specific mathematical models proposed by ITU (e.g.,ITU-T Rec. P.1203)
\footnote{The ITU-T Rec. P.1203 model predicts the impact of
audiovisual quality variations and stalling events on quality
experienced by the end user in multimedia mobile streaming
and fixed network applications using Adaptive Bitrate streaming,
based on a previous estimation of audio and video quality
and information on startup delay and stalling events during the
media session. The model predicts the Mean Opinion Score
(MOS) on a 5-point Absolute Category Rating (ACR) scale as
a final audiovisual quality MOS score.}, or using implicit QoE metrics (e.g., abandonment ratio, video watching \%, revisits).

The closest to our work are the papers by Nam and Schulzrinne \cite{nam16,nam-rep16}, which 
highlighted the role of RBs, BR changes and startup delay on the abandonments in YouTube. 
%
Hossfeld {\em {\em {\em et al.}}}\cite{hoss13} showed that
the total RB duration is not sufficient for properly modeling YouTube QoE, indicating the impact of the number of RBs. 
Often the player performs RBs per two minutes to avoid the playback of the video at the lowest quality level \cite{smodovan16}. Instead of downloading the same segment multiple times, the wasted traffic could be used to download segments in a higher quality. Assuming that the available bandwidth can be predicted, Moldovan {\em {\em {\em et al.}}}\cite{smodovan16} proposed a technique that can be employed to prevent a large number of RBs.


Diallo {\em {\em {\em et al.}}} \cite{diallo2014impacts} reported that the RB is a critical parameter, while the bitrate becomes significant in the case of popular content. Also, in the case of popular content, startup delay under a certain threshold is less significant. Krishnan and Sitaraman \cite{krishnan2013video} aimed to establish a causal relationship between video quality and viewer behavior. They found that viewers are more likely to abandon a video for startup delay larger than 2 seconds. The larger the startup delay, the higher the likelihood of abandonment. Furthermore, sessions with RB duration more than 1\% of the video duration exhibit a reduced average playback time. Finally, they observed that an unsuccessful experience may result to a lower frequency of user revisits.

Apart from the aforementioned "out-in-the-wild" studies, several controlled field studies have been performed, in which users viewed short artificially-impaired videos and provided feedback about the quality of the video they watched, during the viewing session as well as in the end. Based on these continuous-time scores, an overall opinion score is estimated.  
Moorthy {\em {\em {\em et al.}}} \cite{Moorthy-2012} focused on {\em mobile users} and found that BR- results to poor QoE, after a long period of high quality. Moreover, users seem more tolerant to BR+, especially when the resolution before the bitrate change is relatively high. This is consistent with our observation in the case of sessions with mean weighted data rate in [2.5,8]~Mbps in terms of abandonment ratio and video watching percentage. However, for low mean weighted resolution, BR+ become the most annoying impairment.
Moreover, a large RB results in better QoE than a large number of smaller ones. 
%
Wamser {\em {\em {\em et al.}}} \cite{wams16} performed a YouTube QoE study based on monitoring in smartphones. They comparatively analysed the QoE experienced by the participants using the fixed HD quality configuration against the DASH configuration. DASH changes the video quality without incurring in playback stallings, whereas the fixed quality configuration results in video stallings. They reported that the DASH approach results in a nearly optimal QoE, while the fixed HD quality results in poor QoE for downlink bandwidth under 4Mbps. 
However it is difficult to compare the aforementioned results with our work, since they focus on mobile phones in the context of a field study, and use opinion scores.

The relation of QoS and QoE parameters has been examined in various papers, such as in \cite{shaikh2010quality,wu2009quality}, specifically
by applying the Weber Fechner Law and IQX \cite{hossfeld2008testing,hossfeld2012initial}, simple regression models\cite{dobrian2011understanding, chen2009oneclick}, full-reference signal processing techniques (e.g., VQM \cite{pinson2004new}), or  machine-learning algorithms \cite{menkovski2009optimized, menkovski2009predicting, menkovski2010online, joumblatt2013predicting, shafiq2014understanding,plakia2016videoqoe,Casa17}.

Hands and Wilkins \cite{hands1999study} examined the quality and acceptability for video streaming under different network conditions and showed that the burst size (number of consecutive dropped packets) has a considerable impact on QoE and acceptability.
 The role of the context on QoE for various streaming services has been highlighted in \cite{hecht2014all,jumisko2008does,xue2012study,pinson2012influence}. 
Specifically, Jumisko-Pyykk{\"o} and Hannuksela  \cite{jumisko2008does} focused on people watching mobile television on specific mobile devices in different usage contexts (like waiting on the railway station, travelling by bus and spending time in a cafe) and evaluated their experience in terms of satisfaction and acceptance of quality, acceptability, entertainment and information. They found that the acceptance and satisfaction of different contexts does not impact the ratings, while the environmental context is important for entertainment and information ratings.
 Xue and Chen \cite{xue2012study} evaluated the influence of display size, viewing distance, ambient luminance and user movement on subjective perceived quality. Wu \textit{{\em {\em et al.}}}~\cite{wu2009quality} characterized the QoS based on interactivity, vividness and consistency and the QoE using as metrics the concentration, enjoyment, telepresence, perceived usefulness, and perceived easiness of use and applied Pearson correlation to map the QoS to QoE. 

Joumblatt {\em {\em {\em et al.}}}\cite{joumblatt2013predicting} designed predictors of user dissatisfaction by combining user-level feedback with low level machine and networking performance metrics, such as RTT, Jitter, resets, retransmissions, per-connection data rates, host data rates, CPU, and SNR. Predictors were then trained based on the user feedback (namely satisfied or annoyed) and performance data.
Shafiq {\em {\em et al.}} \cite{shafiq2014understanding} analyzed the impact of cellular network performance on mobile video user engagement from the perspective of a network operator. The network performance metrics included TCP flow throughput, flow duration, handover rate, signal strength, and the physical location's land cover type. They observed that many network features exhibit strong correlation with abandonment rate and skip rate: Network load and handovers increase the abandonment rate and higher throughput does not always mean lower abandonment. Moreover, skip rate has a direct relationship with maximum flow inter-arrival time and number of flows. They emphasized the need for network operators to continuously monitor and proactively alleviate the factors that adversely impact user engagement. Casas {\em {\em {\em et al.}}} \cite{Casa17} analyzed YouTube measurements from cellular networks and also highlighted the positive correlation of the average and maximum session throughput as well as of the signal strength with QoE. 

Our earlier work \cite{plakia2016videoqoe} had also examined the impact of the startup delay and buffering ratio on the QoE using measurements collected in smaller-scale field studies of mobile users. Sessions with startup delay higher than 10 sec obtain lower QoE scores. Sessions with poor network performance during the last 15 sec are likely to be terminated with poor connectivity.  The parameters with a dominant impact on the QoE include the frequency of buffering events, weighted mean video resolution ratio, and packet loss. We then proposed machine-learning models for predicting the QoE with a median error less than 0.1. The analysis of measurements collected from a followup, more controlled, field study, indicated a strong impact of the number of buffering events and buffering ratio on QoE for each user. The sensitivity of users to the different types of impairment varies across users. Lenient and strict users were also present, further motivating the need for personalized adaptation in video streaming.

%
%

The impact of the YouTube infrastructure and the distributed CDN has been also investigated. For example, Argyropoulos {\em {\em {\em et al.}}}\cite{Argy18} examined the impact of the host infrastructure as well as the cache server on the delivered quality, showing that the distance of the cache server and its capacity affect
the delivery to a multitude of users and may result to higher
startup delay, stalling events and switching to a lower quality
during playback. 

\section{Conclusions and Future Work}\label{sec:concl}
In general, most of sessions have no impairments and the majority of sessions with impairments have only one. Sessions with BR- exhibit the lowest
video watching percentage, largest abandonment ratio and highest likelihood to get abandoned within a few seconds after the occurrence of the impairment. This trend persists for different weighted mean data rate and video watching percentage thresholds. Interestingly, 91.49\% of the sessions get abandoned as soon as a BR- occurs, compared to 44.65\% when a RB. This likelihood
increases for sessions that were terminated within 60 sec from the occurrence of a BR- and RB, respectively. Sessions with two impairments, in which the last one is a BR-, tend to get abandoned earlier, after the occurrence of the second impairment than sessions with only one BR-.

BR+ is not "well-received" in scenarios with low data rates.
Furthermore, although there are no significant statistical differences between the RB/BR+ vs. RB/BR- in terms of total RB duration and number of RB events, these two scenarios are statistically significant different in terms of video watching percentage. Only when the RB duration is very large, its impact becomes more prominent than BR-. RBs have also more noticeable impact on user-engagement than startup delay events. Two populations were present: one with RBs that occur early during sessions, in which, their impact on user engagement varied, and another one with RBs throughout the session, in which users abandoned the video promptly after its occurrence. According to Lasso regression the dominant features include the number of BR changes over video duration, starting time of first RB event over video duration, timestamp of first BR change over video duration, the number of RBs, number of BR changes, number of negative BR changes, mean weighted bitrate, median BR change level, max BR change level and ads duration over video duration.

Monitoring the rebufferring events and bitrate changes is important to quantify and improve the user engagement. We suggest the implementation of a user-centric ABR player that takes into consideration the impact of the bitrate changes and rebbuferings for optimizing the adaptation process. Moreover, per-user statistics about the revisit and viewing duration per video, taking into consideration the user device, context (e.g., time-of-the-day, position), content type can be also incorporated in the user-engagement model to improve not only the adaptation process but also the prefetching/caching.

\bibliographystyle{plainnat}
\medskip
\bibliography{references18}
\end{document}